\documentclass[aps,pra,reprint,twocolumn,superscriptaddress,nofootinbib,floatfix]{revtex4-2}

\usepackage{graphicx}
\usepackage{amsmath,amssymb,amsthm,mathrsfs,amsfonts}
\usepackage{mathtools}
\usepackage{bm}
\usepackage{xcolor}
\usepackage{hyperref}
\usepackage{placeins}
\hypersetup{hidelinks,
  pdftitle={A parity obstruction to degree-optimal syndrome extraction in bivariate bicycle codes},
  pdfauthor={Guangyi Li, Yu Gan, Zeguan Wu, Xueyue Zhang, Zheshen Zhang, Junyu Liu}}
\usepackage{booktabs}
\usepackage{enumitem}
\usepackage{longtable}

\newcommand{\kd}[1]{\mbox{$(#1)$}}
\newcommand{\paramcell}[1]{\parbox[t]{\linewidth}{\raggedright #1}}

\renewcommand{\topfraction}{0.95}
\renewcommand{\bottomfraction}{0.9}
\renewcommand{\textfraction}{0.05}
\renewcommand{\floatpagefraction}{0.8}
\renewcommand{\dbltopfraction}{0.95}
\renewcommand{\dblfloatpagefraction}{0.8}
\theoremstyle{plain}
\newtheorem{theorem}{Theorem}

\newcommand{\Zq}{\mathbb{Z}_q}
\newcommand{\F}{\mathbb{F}_2}
\newcommand{\col}{\mathsf{d}}          
\newcommand{\colbar}{\bar{\mathsf{d}}} 
\newcommand{\wA}{w_A}                  
\newcommand{\wB}{w_B}

\begin{document}

\title{A parity obstruction to degree-optimal syndrome extraction in bivariate bicycle codes}

\author{Guangyi Li}
\email{guangyi@umd.edu}
\affiliation{Joint Center for Quantum Information and Computer Science, University of Maryland, College Park, MD 20742, USA}
\affiliation{Department of Computer Science, University of Maryland, College Park, MD 20742, USA}
\author{Yu Gan}
\affiliation{Department of Computer Science, University of Pittsburgh, Pittsburgh, PA 15260, USA}
\author{Zeguan Wu}
\affiliation{Department of Computer Science, University of Pittsburgh, Pittsburgh, PA 15260, USA}
\author{Xueyue Zhang}
\affiliation{Department of Applied Physics and Applied Mathematics, Columbia University, New York, NY 10027, USA}
\author{Zheshen Zhang}
\affiliation{Department of Electrical Engineering and Computer Science, University of Michigan, Ann Arbor, MI 48109, USA}
\author{Junyu Liu}
\email[Corresponding author: ]{junyuliu@pitt.edu}
\affiliation{Department of Computer Science, University of Pittsburgh, Pittsburgh, PA 15260, USA}

\begin{abstract}
Syndrome extraction for a CSS code has depth at least the Tanner-graph
degree $\Delta$, attained by edge colouring if gate order is ignored;
correctness adds a parity constraint. We prove three
general obstructions to depth-$\Delta$
extraction and show that, for every two-block group-algebra code with
weight-three polynomials whose checks overlap in at most two qubits, no
correct single-round circuit with one ancilla per check and one CNOT per
Tanner edge has depth six: the minimum is seven, so the seven-layer cycle
of Bravyi \emph{et al.} is optimal for their seven bivariate bicycle
codes. This is the first analytic, family-wide impossibility result,
extending prior instance-specific solver-based infeasibility results.
The hypothesis cannot be dropped: a
single orbit of four-qubit overlaps with a collision element of even order
and $2$-height at most eight makes depth six attainable, and a
$[[72,8,6]]$ code of this kind has a verified six-layer circuit. The
refutations carry DRAT certificates. As a further application of the same
scheduling scheme, we construct Stab-QRAM, a quantum random access memory
for affine Boolean data that uses only CNOT and $X$ gates and has the
minimum depth possible for this gate set.
\end{abstract}

\maketitle

\section{Introduction}
\label{sec:intro}

Quantum low-density parity-check codes promise fault-tolerant memories
with constant overhead~\cite{BreuckmannEberhardt2021,Panteleev2022,
LeverrierZemor2022,Dinur2023}, and the bivariate bicycle (BB) codes of
Bravyi \emph{et al.}~\cite{Bravyi2024} are among the families closest to
implementation, with architectures~\cite{Yoder2025,Xu2024} and
variants~\cite{Voss2024,WangMueller2024,Berthusen2025,LiangChen2026} built
on them. A memory measures its stabilisers once per cycle, and the depth of
the syndrome-extraction circuit sets how many idle locations accumulate
errors between measurements. The published cycle uses seven layers of CNOT
gates for every member of the family~\cite{Bravyi2024}. Each data qubit
and each check has degree six in the Tanner graph, so six layers is a lower
bound for any circuit with one CNOT per Tanner edge. One CNOT layer is one
seventh of the published cycle and one idle location per qubit per cycle.
Throughout, optimality refers to the CNOT depth of a single interleaved
round with one ancilla per check and one CNOT per Tanner edge, without
extra ancillas, staggering across rounds, or connectivity constraints; it
makes no claim about circuit distance or logical error rate. This paper is
about the gap between six and seven.

Ignoring gate order, the bound is attainable: scheduling CNOTs is an
edge-colouring problem, K\H{o}nig's theorem colours a bipartite graph with
$\Delta$ colours~\cite{Konig1916}, and this is how depth-$\Delta$
extraction circuits are obtained for CSS codes~\cite{TDB2022};
depth optimisation of linear reversible and Clifford circuits is studied in
Refs.~\cite{PMH2008,Jiang2020,MaslovZindorf2022,Kutin2007,Bataille2022,GoubaultMartiel2023,MaslovYang2023}.
If all gates commute, any such colouring gives a correct circuit. This is
the case for a quantum random access memory
(QRAM)~\cite{Giovannetti2008,Jaques2023} that stores affine Boolean data.
The lookup $|\mathbf{x}\rangle|\mathbf{0}\rangle\mapsto
|\mathbf{x}\rangle|f(\mathbf{x})\rangle$ with $f$ affine over $\F$ is a
circuit of CNOT and $X$ gates in which every CNOT is controlled by an
address qubit and targets an output qubit, and we treat it as a further
application of our scheme. In syndrome extraction the gates do not all
commute. A CNOT from an $X$-ancilla into a data qubit and a CNOT from that
qubit into a $Z$-ancilla do not commute, each $X$-first ordering on a
shared qubit adds a cross term, and a depth-$\Delta$ colouring is a correct
measurement only if the cross terms cancel. That order matters has been
known since the surface code~\cite{DKLP2002,Fowler2012,TomitaSvore2014};
for general CSS codes, Conrad \emph{et al.} observed that two CNOT pairs on
two shared qubits must agree in order~\cite{Conrad2018}, Zhang
\emph{et al.} stated the general condition, that the number of $X$-first
orderings over the shared qubits of any overlapping pair must be even, and
built a constraint solver around it~\cite{Zhang2026ASC}, as have other
automated schedulers~\cite{Liu2026AlphaSyndrome,Viszlai2026PropHunt}, and
Nguyen \emph{et al.} constructed schedules of depth $\Delta$ when one
polynomial weight is even and $\Delta+1$ when both are odd, leaving the
optimality of the latter to numerical evidence~\cite{Nguyen2026}. Bravyi
\emph{et al.} asked whether the depth of their cycle can be reduced and
reported numerical indications that the $[[72,12,6]]$ code admits no
depth-$6$ circuit~\cite[Sec.~10]{Bravyi2023arXiv}. Zhang \emph{et al.}
report CP-SAT-based certification that no depth-$6$ syndrome-extraction
circuit exists for the seven BB instances of Ref.~\cite{Bravyi2024}.
These are instance-specific solver conclusions; their paper does not
present independently checkable infeasibility certificates or a
family-wide analytic obstruction~\cite{Zhang2026ASC}. Depth-six
measurement circuits related to BB codes exist in other models: morphing
circuits change the code between measurements~\cite{ShawTerhal2025,
ShawTerhal2026}, and flag schemes change the gate
set~\cite{ChamberlandBeverland2018,ChaoReichardt2018}; neither belongs to
the published cycle's model.

We settle the question in that model: one ancilla per check, one CNOT per
Tanner edge, a single round, correctness meaning that the circuit
implements the ideal measurement. We first prove three general obstructions
to depth-$\Delta$ interleaved extraction: a mod-4 condition on the degree
profile of a CSS code (Theorem~\ref{thm:mod4}), with explicit forms for
two-block codes and for regular CSS codes; the impossibility of a common
layer set for two-block codes with both polynomial weights odd; and a
parity condition on schedules twisted by a group homomorphism. Our main
result, Theorem~\ref{thm:main}, concerns two-block codes over any finite
abelian group with $|A|=|B|=3$ and pairwise distinct sums $a_s+b_t$, a
family that contains all seven BB instances of Ref.~\cite{Bravyi2024}. For
these codes the exact minimum depth of a correct single-round interleaved
circuit is $7$. The proof forces every data qubit into a rigid pairing of
adjacent layers and then counts $X$-first corners, of which there are
$(\Delta/2)^2$, an odd number. The overlap hypothesis cannot be dropped:
for codes with one collision orbit whose collision element $\col$ has even
order and $2$-height at most $8$, i.e.\ $\col\notin512G$, the exact minimum
depth is $6$ (Theorem~\ref{thm:six}). That proof uses a group-independent
template, a lifting argument and a reduction of the template keys to a
one-parameter family with a universal half-period symmetry.

The refutations for the seven codes of Ref.~\cite{Bravyi2024} and for
eleven further instances carry verified DRAT certificates, every schedule
we report is checked by $\F$ simulation, and we map the remaining regime
computationally. A search over all two-block codes of length up to $144$
yields depth-six codes with competitive parameters, among them
$[[72,8,6]]$, $[[108,12,6]]$, $[[144,16,6]]$ and $[[144,8,\le10]]$ codes
with verified six-layer circuits (Table~\ref{tab:depth6codes}). Up to
$n=48$ they match the best depth-seven two-block codes, and from $n=72$ on
they trail them by one step in $k$ or in $d$. As a further application of
the scheduling scheme, we construct Stab-QRAM, a QRAM for affine Boolean
data built from CNOT and $X$ gates only (Supplementary Section~S9). It
uses one qubit per address bit and per output bit, no ancillas and no $T$
gates, and its depth, at most one more than the largest row or column
weight of the data matrix, is optimal among schedules of these gates. All
CNOTs of the lookup commute, so the parity constraint of syndrome
extraction does not arise. The construction uses only standard
results~\cite{DehaeneDeMoor2003,Aaronson2004,Konig1916}.
We do not claim a criterion for general odd weights, a complete
characterisation of the coinciding-sum case, or any statement about
circuit-level distance. Nor do we claim optimality in amortised or
staggered models, which allow operations from consecutive rounds to
overlap, as in left--right syndrome-extraction circuits~\cite{Strikis2026}.
Their depth convention includes ancilla preparation and measurement,
unlike the single-round CNOT depth considered here. Lower bounds for
circuits with CNOTs outside the Tanner graph or with extra ancillas are
open.

\section{Model and correctness condition}
\label{sec:model}

\paragraph*{The cross term.}
Let $H_X,H_Z$ be the check matrices of a CSS code, $\mathbf{x}$ the data
register, $\mathbf{y}$ the $X$-ancillas (prepared in $|+\rangle$) and
$\mathbf{z}$ the $Z$-ancillas (prepared in $|0\rangle$). Each Tanner edge
is executed by one CNOT, controlled by the $X$-ancilla for $X$-edges and by
the data qubit for $Z$-edges; there are no other gates and no further
ancillas, and the ancillas are measured after the last layer. A
\emph{schedule} $\pi$ assigns layers $1,\dots,D$ to the edges so that no
qubit is in two CNOTs of one layer; $D$ is its depth. The circuit implements
\begin{equation}
(\mathbf{x},\mathbf{y},\mathbf{z})\mapsto
\bigl(\mathbf{x}+H_X^{\mathsf T}\mathbf{y},\;\mathbf{y},\;
\mathbf{z}+H_Z\mathbf{x}+M_\pi\mathbf{y}\bigr),
\label{eq:se-map}
\end{equation}
where $(M_\pi)_{ji}$ is the number, modulo $2$, of shared qubits of $X_i$
and $Z_j$ on which the $X$-edge precedes the $Z$-edge. Each half of this
map, $\mathbf{z}\mapsto\mathbf{z}+H_Z\mathbf{x}$ or
$\mathbf{x}\mapsto\mathbf{x}+H_X^{\mathsf T}\mathbf{y}$, consists of
commuting CNOTs, and the cross term $M_\pi\mathbf{y}$ appears only when the
two halves are interleaved. On its own, each half is a lookup of the kind
used in the Stab-QRAM of Supplementary Section~S9.

Correctness means that for every input state of the data register the
circuit, followed by the $X$-basis measurement of $\mathbf{y}$ and the
$Z$-basis measurement of $\mathbf{z}$, realises the ideal joint projective
measurement of the commuting checks $\{X_i\}\cup\{Z_j\}$: the recorded
outcome associated with each ancilla is the eigenvalue of its designated
check, and the post-measurement data state is the projected state.
Redundancy-dependent classical recombination of the labelled outcomes is
not included in this definition.

Propagating the measured Pauli operators backwards through the CNOTs, and
using the even overlap of each commuting $X$/$Z$ check pair in the second
identity (Supplementary Section~S2), gives
\begin{equation}
\begin{aligned}
U^\dagger Z_{z_j} U
  &= Z_{z_j}\,\overline{Z}_j
     \prod_i Z_{a_i}^{(M_\pi)_{ji}}, \\
U^\dagger X_{a_i} U
  &= X_{a_i}\,\overline{X}_i
     \prod_j X_{z_j}^{(M_\pi)_{ji}} .
\end{aligned}
\label{eq:heisenberg}
\end{equation}
Here $\overline{Z}_j,\overline{X}_i$ are the checks on the data register
and $a_i,z_j$ the ancillas. If $M_\pi=0$, these are the ideal measurement
operators, since each ancilla is a $+1$ eigenstate of its own measured
Pauli.

Conversely, suppose that $M_\pi\neq0$, and let $u_j$ be a nonzero row of
$M_\pi$. Write
\[
Z_a^{u_j}:=\prod_i Z_{a_i}^{(u_j)_i}.
\]
For any input data state $\rho$, the marginal expectation of the recorded
outcome $s_j\in\{\pm1\}$ of $Z_{z_j}$ is
\[
\begin{aligned}
\mathbb{E}_{\rho}[s_j]
&=
\operatorname{Tr}\!\left[
  \bigl(\rho\otimes |+\rangle\!\langle+|_a
  \otimes |0\rangle\!\langle0|_z\bigr)
  U^\dagger Z_{z_j}U
\right] \\
&=
\operatorname{Tr}(\rho\,\overline{Z}_j)\,
{}_z\langle0|Z_{z_j}|0\rangle_z\,
{}_a\langle+|Z_a^{u_j}|+\rangle_a
=0,
\end{aligned}
\]
because $u_j\neq0$ and hence
${}_a\langle+|Z_a^{u_j}|+\rangle_a=0$. Since $s_j$ is binary, its marginal
distribution is therefore uniform. In particular, on a simultaneous
eigenstate of the commuting checks, the ideal outcome of
$\overline{Z}_j$ is deterministic whereas the recorded outcome $s_j$ is
uniform. Thus the circuit realises the stipulated ideal measurement if
and only if $M_\pi=0$.

\begin{figure*}[t]
\centering
\includegraphics[width=\textwidth]{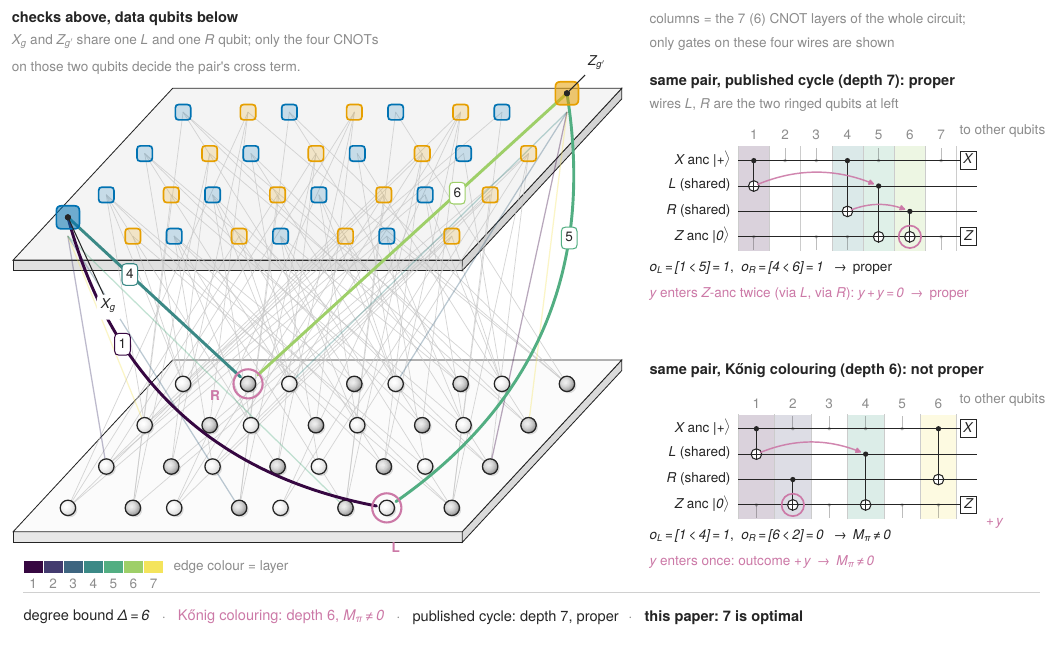}
\caption{\textbf{Degree bound versus correctness.} Left: a patch of the
$[[72,12,6]]$ Tanner graph drawn as two plates, checks above ($X$ blue,
$Z$ orange) and data qubits below; one $X$-check $X_g$ and one $Z$-check
$Z_{g'}$ sharing an $L$-qubit and an $R$-qubit are highlighted with their
twelve edges, the four on the shared qubits numbered by their layer in the
depth-$7$ schedule of Table~\ref{tab:depth7main} (edge colour = layer).
Right: the same four CNOTs on the four wires of the pair; faded gates lead
to the checks' other qubits. Top: in the published cycle the $X$-edge
precedes the $Z$-edge on both shared qubits, so the $X$-ancilla value $y$
enters the $Z$-ancilla twice and cancels ($o_L=o_R=1$). Bottom: a proper
$6$-edge-colouring of the full $[[72,12,6]]$ Tanner graph, restricted to
the pair, in which $y$ enters once, leaving $M_\pi\neq0$.}
\label{fig:setup}
\end{figure*}

The matrix $M_\pi$ is therefore the ancilla-to-ancilla contamination
matrix, and a schedule with $M_\pi=0$ is called \emph{proper}. The maximum
degree $\Delta$ of the Tanner graph bounds $D$ from below, and
K\H{o}nig's theorem gives a schedule of depth exactly $\Delta$ that need
not be proper (Fig.~\ref{fig:setup}; Supplementary Section~S1 records
this as a compiler with an exact depth certificate for commuting gate
sets, and Supplementary Section~S9 applies it to quantum memory).
\paragraph*{Two-block codes.}
Let $G$ be a finite abelian group of order $N$, written additively, and
$A=\{a_1,\dots,a_{\wA}\}$, $B=\{b_1,\dots,b_{\wB}\}\subset G$. The two-block
code $\mathrm{QC}(A,B)$~\cite{KovalevPryadko2013,PanteleevKalachev2021,
LinPryadko2024,WangLinPryadko2023} has data qubits $L_g,R_g$ and checks
$X_g,Z_g$, $g\in G$, with
\begin{equation}
X_g:\ L_{g+a_s},\,R_{g+b_t};\qquad Z_g:\ L_{g-b_t},\,R_{g-a_s};
\label{eq:twoblock}
\end{equation}
so $H_X=[A\,|\,B]$, $H_Z=[B^{\mathsf T}\,|\,A^{\mathsf T}]$ in the
notation of Ref.~\cite{Bravyi2024}, the BB codes being the case
$G=\mathbb{Z}_\ell\times\mathbb{Z}_m$ with $\wA=\wB=3$. The \emph{weights}
$\wA=|A|$ and $\wB=|B|$ are the numbers of monomials of the two
polynomials; every check has weight $\wA+\wB$, every $L$-qubit has $\wA$
$X$- and $\wB$ $Z$-neighbours, every $R$-qubit the reverse, and
$\Delta=\wA+\wB$. Unless stated otherwise $\wA=\wB=3$ and $\Delta=6$. We
write $H\le G$ for the subgroup generated by the differences $a_s-a_{s'}$
and $b_t-b_{t'}$; the connected components of the Tanner graph are the
cosets of $H$ (Section~\ref{sec:obstructions}). An edge carries the
\emph{label} $s$ or $t$ of the element $a_s$ or $b_t$ that defines it. $X_g$ and $Z_{g'}$ share $L_{g+a_s}$ and $R_{g+b_t}$ exactly
when $g'-g=a_s+b_t$; with $S(\delta)=\#\{(s,t):a_s+b_t=\delta\}$ they
overlap in $2S(g'-g)$ qubits. The code has \emph{distinct sums} (class D)
if $S\le1$ everywhere, so that every overlap has size $0$ or $2$; in the
polynomial notation this is the condition that the nine products $a_sb_t$
are distinct, and it holds for all seven codes of Ref.~\cite{Bravyi2024}.
It is of \emph{class C1} if $S(\delta_0)=2$ for exactly one $\delta_0$ and
$S\le1$ elsewhere; the two label pairs $(s,t),(s',t')$ with sum $\delta_0$
form the \emph{collision}, with \emph{collision element}
$\col=a_{s'}-a_s=b_t-b_{t'}$, and the $N$ pairs $(X_g,Z_{g+\delta_0})$
overlap in four qubits. Here $2^jG=\{2^jg:g\in G\}$, $\operatorname{ord}(\col)$
is the order of $\col$ in $G$, and for $\col$ of even order the
\emph{$2$-height} $h(\col)$ is the largest $h$ with $\col\in2^hG$; it is
finite exactly when the order is even, because the elements lying in every
$2^jG$ form the odd-order subgroup of $G$. Class C2 has two collisions or
one triple coincidence (Supplementary Section~S6).

\paragraph*{Corners and counting.}
A \emph{corner} is a data qubit $q$ with an $X$-edge $e$ and a $Z$-edge $e'$
at $q$; its orientation is $1$ if $\pi(e)<\pi(e')$ and $0$ otherwise. The
entry of $M_\pi$ for a pair is the orientation sum over its shared corners,
so $M_\pi=0$ reads
\begin{equation}
\sum_{q\in\operatorname{supp}X_g\cap\operatorname{supp}Z_{g'}}
o\bigl(q,e_g(q),e'_{g'}(q)\bigr)\equiv0\pmod 2
\label{eq:parity}
\end{equation}
for every overlapping pair, the constraint of
Refs.~\cite{Conrad2018,Zhang2026ASC,Nguyen2026}. Under distinct sums each
pair has one $L$-corner and one $R$-corner, which must have equal
orientation; this pairs the corners of the two blocks by a
label-preserving bijection $\Phi$ (Supplementary Section~S2). A four-qubit
overlap allows two $X$-first and two $Z$-first corners. For an $X$-check
$g$ let $P_g,V_g$ be the layer sets of its $L$- and $R$-edges; at depth
$\Delta$, $P_g\sqcup V_g=[\Delta]:=\{1,\dots,\Delta\}$ and every qubit and
check has one edge per layer. Let $\gamma_q(i)$ be the number of $Z$-edges
at $q$ in layers above $i$, and
\begin{equation}
F_L(g)=\sum_{i\in P_g}\gamma(i),\qquad F_R(g)=\sum_{i\in V_g}\gamma(i),
\label{eq:FLFR}
\end{equation}
evaluated at the qubit carrying each edge: the $X$-first corners of $X_g$ on
the two blocks. Each such corner lies in one overlapping pair, so
Eq.~\eqref{eq:parity} gives $F_L+F_R\equiv0\pmod 2$ for any overlap
pattern, and $F_L=F_R$ under distinct sums. A schedule is
\emph{translation-invariant} if $\pi$ depends only on label and edge type,
\emph{twisted} by a homomorphism $\varepsilon:G\to\Zq$ if it depends only
on label, type and $\varepsilon(g)$, and a layer is \emph{mixed} if it
holds $L$-edges of some $X$-checks and $R$-edges of others.

A few conventions are used throughout.
\emph{Layer $i$ is above layer $j$} means $i>j$; ``$X$-first'' means the
$X$-edge is in the lower layer. A \emph{layering} is any assignment of
layers with no qubit in two CNOTs of one layer; a \emph{schedule} is a
layering of depth $D$, and it is \emph{proper} if $M_\pi=0$.
\emph{Interleaved} means that $X$- and $Z$-ancillas share layers;
\emph{single-round} means one measurement of every check. An \emph{orbit}
of overlapping pairs is the set $\{(X_g,Z_{g+\delta}):g\in G\}$ for fixed
$\delta$. A schedule has \emph{period} $q$ if it is twisted by a
homomorphism onto $\Zq$ with $q>1$ dividing $N$; a found schedule has
\emph{no proper period} if it is not twisted by any such homomorphism, a
finite check performed on that schedule, which by itself says nothing
about other schedules of the same code. $\mathrm{QC}$ stands for
quasi-cyclic. The letters $k$ and $d$ are reserved for the code parameters
$[[n,k,d]]$ and $\ell,m$ for the BB group $\mathbb{Z}_\ell\times\mathbb{Z}_m$;
polynomial weights are $\wA,\wB$ and the collision element is $\col$.

\section{Results}
\label{sec:results}

Fig.~\ref{fig:mechanism} shows the mechanism; the full dependency map is
Supplementary Fig.~\ref{fig:fullmap}, and proofs not given in Methods are
in Supplementary Sections~S3--S5.

\subsection{Three general obstructions}
\label{sec:obstructions}

The first obstruction is a count of corners over a whole component of the
Tanner graph.

\begin{theorem}[mod-4 obstruction]
\label{thm:mod4}
Let every data qubit $v$ of a CSS code have $r_v$ $X$-neighbours and $s_v$
$Z$-neighbours with $r_v+s_v=\Delta$, and let every $X$-check have weight
$\Delta$. If $\sum_v r_vs_v\not\equiv0\pmod 4$ over some connected
component of the Tanner graph, no proper schedule of depth $\Delta$
exists.
\end{theorem}

Any depth-$\Delta$ layering of the component has exactly
$\frac12\sum_vr_vs_v$ $X$-first corners, since
$\sum_vN_v=\tfrac{\Delta-1}{2}\sum_vr_v-\sum_v\binom{r_v}{2}
=\tfrac12\sum_vr_vs_v$ (Methods), and properness makes the count of every
overlapping pair even. Only the weight of the $X$-checks and the degrees of
the data qubits enter; the weights of the $Z$-checks are unconstrained.

For a two-block code every data qubit has $r_vs_v=\wA\wB$, and the
components are explicit. Let $H\le G$ be the subgroup generated by the
differences $a_s-a_{s'}$ and $b_t-b_{t'}$. The component that contains
$X_g$ consists of the checks $X_{g+H}$ and $Z_{g+a_1+b_1+H}$ and the data
qubits $L_{g+a_1+H}$ and $R_{g+b_1+H}$ (Methods), so it has $|H|$ checks of
each type and $2|H|$ data qubits, and Theorem~\ref{thm:mod4} excludes depth
$\wA+\wB$ whenever $|H|\,\wA\wB$ is odd. For a connected Tanner graph
($H=G$) this means that $N$, $\wA$ and $\wB$ are all odd. For a BB code
with $2\ell m$ data qubits and a connected Tanner graph,
$\sum_vr_vs_v=18\ell m$, and the condition reads $\ell m$ odd. The Tanner
graphs of the seven codes of Ref.~\cite{Bravyi2024} are connected, and only
$[[90,8,10]]$, with $\ell m=45$, is excluded in this way. The
$\mathbb{Z}_6$ instance of Section~\ref{sec:sharpness}, $A=B=\{0,2,4\}$,
has $H=\{0,2,4\}$ and is excluded although $|G|$ is even, through
components of six data qubits with $\sum_vr_vs_v=54$. Beyond two-block
codes, suppose that every data qubit has $w$ $X$- and $w$ $Z$-neighbours
and every $X$-check has weight $2w$. A component with $n_0$ data qubits
then has $\sum_vr_vs_v=n_0w^2$, so depth $2w$ is impossible when $w$ is
odd and $n_0\equiv2\pmod 4$. The hypergraph product of a $(3,3)$-regular
classical code of odd length $n_c$ with itself is of this type, with
$2n_c^2$ data qubits and weight-$6$ checks. For the circulant code
$1+x+x^3$ of length $7$ it is a connected $[[98,18]]$ code, for length $15$
a connected $[[450,32]]$ code, and depth $6$ is impossible for both.

The other two obstructions concern schedules with a symmetry. If all
$X$-checks of a two-block code put their $L$-edges in one common layer set
$P$, every $L$-qubit has $X$-layers $P$ and $Z$-layers $[\Delta]\setminus
P$, every $R$-qubit the reverse, and
$F_L+F_R=|P|\,|[\Delta]\setminus P|=\wA\wB$, which properness makes even.
When $\wA\wB$ is odd, every proper schedule of depth $\wA+\wB$ therefore
contains a mixed layer, and in particular no translation-invariant proper
schedule of depth $\Delta$ exists (Supplementary Section~S3). This is why
the translation-invariant construction of Ref.~\cite{Nguyen2026} reaches
$\Delta$ only when one weight is even. The same count applies to a schedule
twisted by a homomorphism $\varepsilon:G\to\Zq$, for any overlap pattern.
Let $M=\varepsilon(H)\le\Zq$ be the subgroup generated by the images of the
differences $a_s-a_{s'}$ and $b_t-b_{t'}$. The classes in one coset of $M$
form a closed subsystem, in which $\sum_c(F_L(c)+F_R(c))=|M|\wA\wB$ while
every summand is even, so a proper twisted schedule of depth $\wA+\wB$
requires
\begin{equation}
|M|\,\wA\wB\equiv0\pmod 2
\label{eq:twisted}
\end{equation}
(Supplementary Section~S3). For $\wA=\wB=3$, $|M|$ must be even, and no
such schedule exists for odd $q$. None of these conditions excludes a
proper schedule that breaks all symmetry. That case is settled by the next
result.

\subsection{Exact depth seven}
\label{sec:main}

\begin{theorem}[exact depth seven]
\label{thm:main}
Let $G$ be a finite abelian group and $A,B\subset G$ three-element sets
whose nine sums $a_s+b_t$ are pairwise distinct. Then the minimum CNOT
depth of a proper single-round interleaved schedule of $\mathrm{QC}(A,B)$
is exactly $7$. In particular this holds for every bivariate bicycle code
with pairwise distinct products $a_sb_t$, including all seven codes of
Ref.~\cite{Bravyi2024}.
\end{theorem}

The upper bound is the translation-invariant schedule of
Table~\ref{tab:depth7main}, which is proper for every two-block code with
$|A|=|B|=3$, whatever its overlap pattern. The $L$-corner of $X_g$ with
labels $(s,t)$ has orientation $\mathbb{1}[\pi_{XL}(s)<\pi_{ZL}(t)]$ and
the $R$-corner $\mathbb{1}[\pi_{XR}(t)<\pi_{ZR}(s)]$. Since
$\pi_{XL}=(1,2,7)$ places $s=1,2$ below and $s=3$ above every entry of
$\pi_{ZL}=(5,3,4)$, and $\pi_{ZR}=(6,7,1)$ against $\pi_{XR}=(4,3,5)$ does
the same, both orientations equal $\mathbb{1}[s\le2]$ for all nine label
pairs. Every overlapping pair consists of corners indexed by label pairs,
so its orientation sum is even and Eq.~\eqref{eq:parity} holds. The
minimum depth of every two-block code with $|A|=|B|=3$ is therefore $6$ or
$7$. The schedule was also verified by $\F$ simulation on the seven codes of
Ref.~\cite{Bravyi2024} and on $40$ random instances (Supplementary
Section~S7).

\begin{table}[!tbp]
\caption{\textbf{The uniform depth-$7$ schedule used in
Theorem~\ref{thm:main}.} Layers by check type, block and label.}
\label{tab:depth7main}
\begin{ruledtabular}
\begin{tabular}{lccc}
 & label 1 & label 2 & label 3 \\
\colrule
$X$-check, $L$-edges & 1 & 2 & 7 \\
$X$-check, $R$-edges & 4 & 3 & 5 \\
$Z$-check, $L$-edges & 5 & 3 & 4 \\
$Z$-check, $R$-edges & 6 & 7 & 1 \\
\end{tabular}
\end{ruledtabular}
\end{table}

The lower bound takes three steps. For a proper depth-$\Delta$ schedule
let $a$ and $b$ be the lowest and highest layers in which any $X$-check has
an $L$-edge. The first step, support symmetry, holds for any $|A|$ and
$|B|$ with distinct sums: every proper schedule of depth $\Delta$ satisfies
\begin{equation}
a+b=\Delta+1,
\label{eq:support}
\end{equation}
and an $X$-check with an $L$-edge in layer $a$ also has one in layer $b$,
and conversely (Methods). For $|A|=|B|=3$ this leaves
$(a,b)\in\{(3,4),(2,5),(1,6)\}$. The first case is empty, since an
$X$-check needs three $L$-layers in $[a,b]$, and the second is excluded by
counting the corners of the layer-$2$ and layer-$5$ edges (Supplementary
Theorem~\ref{thm:S3}). In the remaining case $(a,b)=(1,6)$, the second
step, rigidity, shows that every data qubit carries exactly one $X$-edge
and one $Z$-edge in each of the layer pairs $\{1,2\}$, $\{3,4\}$ and
$\{5,6\}$. Its proof starts from an $X$-check with $L$-edges in layers $1$
and $6$, follows $\Phi$ through its $R$-edges, and shows layer by layer
that any other arrangement produces a corner whose partner cannot be placed
(Supplementary Section~S4). The third step is a count. Let $|A|=|B|=w$ and
$\Delta=2w$. If every data qubit has this pair structure, the number of
$Z$-edges above an $X$-edge depends only on its layer,
$\gamma(i)=w-\lceil i/2\rceil+\mathbb{1}[i\text{ odd}]$, and every
$X$-check $g$ has
\begin{equation}
F_L(g)+F_R(g)=\sum_{i=1}^{2w}\gamma(i)=w^2 .
\label{eq:pairsum}
\end{equation}
For $w=3$ the sum is $9$, which is odd, whereas distinct sums give
$F_L(g)=F_R(g)$ and hence an even sum. This contradiction proves the lower
bound of Theorem~\ref{thm:main}.

The argument is more general than the case it settles. Whenever a schedule
of a code with $\wA=\wB$ has the pair structure, Eq.~\eqref{eq:pairsum} and
$F_L=F_R$ contradict each other exactly when $\Delta\equiv2\pmod 4$; what
does not yet extend to $\wA=\wB\ge5$ is the rigidity step. Supplementary
Theorems~\ref{thm:S1} and~\ref{thm:S2} show that depth $\Delta$ is
unattainable when one weight is $1$ and the other odd, and when $\wA$ and
$\wB$ are odd and $a=(\wB+1)/2$.

\begin{figure*}[t]
\centering
\includegraphics[width=\textwidth]{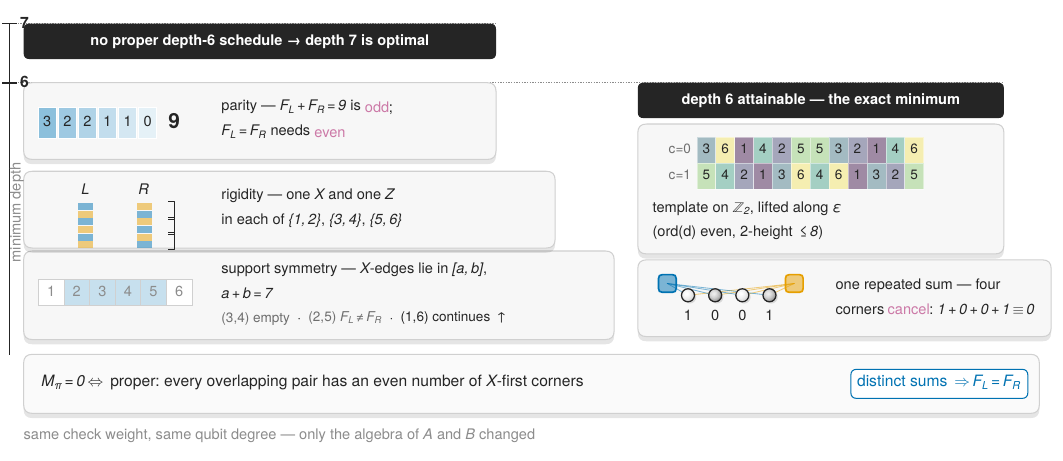}
\caption{\textbf{Why six layers fail, and when they do not.} Left
staircase, distinct sums: from $M_\pi=0$, support symmetry confines the
$X$-edges on $L$ to $[a,b]$ with $a+b=7$ (Eq.~\eqref{eq:support}) and
excludes the cases $(3,4)$ and $(2,5)$; rigidity forces one $X$- and one
$Z$-edge into each adjacent layer pair; the count of
Eq.~\eqref{eq:pairsum} then gives $F_L+F_R=9$, odd, while $F_L=F_R$
requires it even, so depth $7$ is optimal (Theorem~\ref{thm:main}). Right
staircase, one repeated sum: four shared corners can cancel, and a class
schedule on $\mathbb{Z}_2$ lifted along $\varepsilon$ attains depth $6$
whenever the collision element has even order and $2$-height at most $8$
(Theorem~\ref{thm:six}). The ruler marks the minimum depth reached by each
staircase; cell colours in the class strip are the layer scale of
Fig.~\ref{fig:setup}.}
\label{fig:mechanism}
\end{figure*}

\subsection{Exact depth six}
\label{sec:six}

When two sums coincide, one orbit of overlapping pairs shares four qubits
instead of two, and the four corners of such a pair can cancel in pairs.
We construct depth-six schedules for these codes from small periodic
templates. Fix $q$, a key
$\kappa=(\alpha_1,\alpha_2,\alpha_3;\beta_1,\beta_2,\beta_3)\in\Zq^6$ and a
designated pair $\{(s,t),(s',t')\}$ with
$\alpha_s+\beta_t=\alpha_{s'}+\beta_{t'}$, and write
$\colbar=\alpha_{s'}-\alpha_s$. A \emph{class schedule} assigns to every
class $c\in\Zq$, label and edge type a layer in $[6]$ such that (i)~the six
edges incident to the $X$-check of class $c$ occupy six distinct layers,
and so do the six edges incident to the $Z$-check of class $c$; (ii)~at the
$L$-qubit of class $c$, the $X$-edges of label $s$ from class $c-\alpha_s$
and the $Z$-edges of label $t$ from class $c+\beta_t$ occupy distinct
layers, and likewise on the $R$ block; (iii)~for every class and every
label pair other than the designated two, the $L$-corner of $X_c$ with
labels $(s,t)$ and the $R$-corner with the same labels, at the $Z$-check of
class $c+\alpha_s+\beta_t$, have equal orientation; (iv)~for the designated
pairs the four orientations have even sum. The constraints are indexed by
labels, not by which projected sums coincide in $\Zq$.

A class schedule lifts to every code that realises its key. Let
$\mathrm{QC}(A,B)$ be of class C1 with collision $\{(s,t),(s',t')\}$, let
$\varepsilon:G\to\Zq$ be a homomorphism, and let $T$ be a class schedule
for the key $(\varepsilon(a_s);\varepsilon(b_t))$ and that collision. Then
$\pi(e)=T(\varepsilon(g),\mathrm{label}(e),\mathrm{type}(e))$ is a proper
schedule of depth $6$ of $\mathrm{QC}(A,B)$. Indeed, every constraint of
$\mathrm{QC}(A,B)$ is indexed by a labelled object: a check (i), a data
qubit (ii), an overlapping pair together with its unique label pair (iii),
or the collision pair together with its two label pairs (iv). The edges at
$L_h$ come from $X_{h-a_s}$ and $Z_{h+b_t}$, of classes
$\varepsilon(h)-\alpha_s$ and $\varepsilon(h)+\beta_t$, so the conflict
constraints of the code are the class constraints (i)--(ii) at
$\varepsilon(g)$, $\varepsilon(h)$, $\varepsilon(r)$ evaluated on
$T(\varepsilon(\cdot))$. The overlapping pairs of a class-C1 code are
exactly the labelled pairs, the two collision pairs sharing one $Z$-check,
so Eq.~\eqref{eq:parity} for the pair $(X_g,Z_{g+a_s+b_t})$ is the class
constraint (iii) for $(\varepsilon(g),s,t)$, or (iv) for the collision.
Coincidences among projected sums $\alpha_s+\beta_t$ create no constraint
in $\mathrm{QC}(A,B)$, whose pairs are determined by the sums in $G$, and
remove none, because (iii) is imposed per class and label pair rather than
per class-level pair.

The templates themselves were found by computer. For $q\in\{2,4,8\}$ every
key with $\colbar\neq0$ admits a class schedule, and for $q=2^j$ with
$4\le j\le9$ every key with $\colbar=q/2$ admits one. For $q\in\{2,4,8\}$
this was established by exhaustive enumeration: every normalised key and
designated collision was decided by SAT, one class schedule was stored per
admissible key, and every stored schedule was checked against (i)--(iv)
and by $\F$ simulation of its pullback to a realising class-C1 code
(Supplementary Section~S5, which also records the $q=2$ template and its
pullback to a $[[28,6,4]]$ code over $\mathbb{Z}_{14}$). For $q\ge16$ the
keys with $\colbar=q/2$ reduce, by the symmetries of the class system and a
decomposition into subsystems of smaller modulus, to the one-parameter
family
\begin{equation}
\kappa_q(1,b)=(0,\,q/2,\,1;\;0,\,q/2,\,b),\qquad b\in\Zq
\label{eq:family}
\end{equation}
(Methods). Every member of the family was decided satisfiable by SAT for
$q\le512$, without any assumption for $q\le32$, and for $q\ge64$ under the
half-period symmetry
\begin{equation}
T(c+q/2,e)=\sigma_{\mathrm{blk}(e)}\bigl(T(c,e)\bigr),
\label{eq:halfperiod}
\end{equation}
with the fixed involutions $\sigma_L=(13)(24)(56)$ on $L$-edges and
$\sigma_R=(12)(36)(45)$ on $R$-edges, which turns instances that take up to
half an hour into instances decided in a second. Odd $q$ is excluded
analytically by Eq.~\eqref{eq:twisted}; $q=3$ and $q=5$ were also
enumerated and admit nothing, and $q=6$ was enumerated to test the
refinement discussed in Section~\ref{sec:sharpness}. In the enumeration,
no key with $\colbar=0$ admits a class schedule.

The modulus is chosen by the $2$-height of the collision element. If
$\col\in G$ has even order and $2$-height $h$, there is a homomorphism
$\varepsilon:G\to\mathbb{Z}_{2^{h+1}}$ with $\varepsilon(\col)=2^h$; the
proof (Supplementary Section~S5) splits $G/2^{h+1}G$ into cyclic
$2$-groups and picks a factor on which $\col$ survives. Together with the
templates and the lifting, this gives the second exact result.

\begin{theorem}[exact depth six]
\label{thm:six}
Let $\mathrm{QC}(A,B)$ be of class C1 with a collision element $\col$ of
even order and $2$-height $h(\col)\le8$, equivalently $\col\notin512G$.
Then the minimum CNOT depth of a proper single-round interleaved schedule
is exactly $6$.
\end{theorem}

The homomorphism given by the $2$-height has $q=2^{h+1}\le512$ and
$\varepsilon(\col)=q/2$, so the key $(\varepsilon(a_s);\varepsilon(b_t))$
has $\colbar=q/2$ and, after the normalisation of Methods, is one of the
keys for which a template exists. Lifting the template gives a proper
schedule of depth $6$, and the degree bound gives the lower bound.

Every element of even order has a finite $2$-height, so the only gap
between Theorem~\ref{thm:six} and the conjecture that even order suffices
is the modulus bound. By the reduction in Methods, the conjecture is
equivalent to the statement that every member of the
family~\eqref{eq:family} admits a class schedule for every $j$; it holds
for $j\le9$. Every class schedule found for the family has the half-period
symmetry~\eqref{eq:halfperiod}, with the same two involutions for every $j$
and $b$. The symmetry fixes the layers on the classes $c\ge q/2$ in terms
of those on $c<q/2$, and is the natural starting point for an analytic
template (Supplementary Section~S5). All $265$ class-C1 instances with
$\operatorname{ord}(\col)$ even in our data set have $h(\col)\le3$ and are
covered by Theorem~\ref{thm:six} (Supplementary Table~\ref{tab:twoheight}).

\subsection{The remaining regime}
\label{sec:sharpness}

\begin{figure*}[t]
\centering
\includegraphics[width=\textwidth]{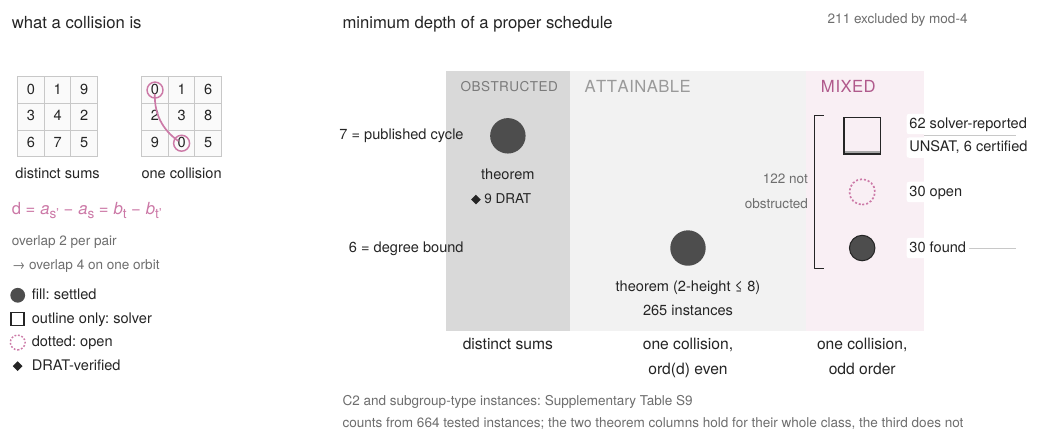}
\caption{\textbf{Where depth $6$ is attainable.} \textbf{(a)}~The nine
sums $a_s+b_t$ of a two-block code; a repeated sum (rose link) defines the
collision element $\col$ and turns one orbit of two-qubit overlaps into
four-qubit overlaps. \textbf{(b)}~Minimum depth of a proper schedule by
collision structure. Distinct sums: $7$ for every code
(Theorem~\ref{thm:main}; $\blacklozenge$ nine instances also
DRAT-certified). One collision with $\col$ of even order: $6$ for every
code with $h(\col)\le8$ (Theorem~\ref{thm:six}), which covers all $265$
tested instances, the two with $\col\in8G$ through the $q=16$ templates
(they were also found as twisted witnesses). Odd order: of $122$
instances not excluded by Theorem~\ref{thm:mod4}, $30$ reach depth $6$,
$62$ are refuted by the solver and $30$ are open; $211$ further instances
are excluded by Theorem~\ref{thm:mod4}. Fill encodes settled status
(theorem or certificate), outline the solver status, hatching a capped
run. The $56$ solver-reported refutations carry no proof certificate and
are marked as computational indication; $6$ further instances in that cell
are DRAT-certified. The plotted classes D and C1 are drawn from the
$664$-instance dataset; the $28$ class-C2 and $3$ subgroup-type instances
are in Supplementary Table~\ref{tab:h2}.}
\label{fig:sharpness}
\end{figure*}

Fig.~\ref{fig:sharpness} maps the depth-$6$ decision problem over $664$
instances with a recorded depth-$6$ outcome: the seven codes of
Ref.~\cite{Bravyi2024}, the $\mathbb{Z}_6$ instance of
Theorem~\ref{thm:mod4}, and $656$ sampled instances on cyclic groups of
order $10$--$24$, $28$, $30$, $32$ and $36$ (sampling and equivalence in
Supplementary Section~S6). For even $|G|\le8$ the nine sums cannot be
distinct, and in an exhaustive sweep of all $1441$ normalised instances
($33$ inequivalent) $1440$ admit a proper depth-$6$ schedule, the
exception being excluded by Theorem~\ref{thm:mod4}. Order $9$ is the
smallest order with distinct-sums instances: up to equivalence there are
three, one on $\mathbb{Z}_9$ and two on $\mathbb{Z}_3\times\mathbb{Z}_3$,
and all three are unsatisfiable, as Theorem~\ref{thm:main} requires; so is
every sampled distinct-sums instance of larger order. For $\wA=\wB=3$,
Eq.~\eqref{eq:twisted} excludes every twisted schedule of odd period, so
a proper schedule is either twisted with even period or has no period.
When $\operatorname{ord}(\col)$ is odd, the found schedules are of both
kinds: of $333$ such class-C1 instances, $211$ are excluded by
Theorem~\ref{thm:mod4}; of the remaining $122$, $30$ admit depth $6$, $62$
are refuted by the solver, $6$ of them with verified DRAT certificates and
$56$ without proof logging, and $30$ were undetermined within the cap. Of
the $30$ satisfiable instances, $24$ have a schedule twisted with an even
period, consistent with Eq.~\eqref{eq:twisted}: the relevant parity is
that of $|M|$, not of $\operatorname{ord}(\col)$; for the remaining $6$ no
twisted proper schedule exists for any $q\mid N$ (the search over all
proper divisors of $N$ completed without a cap), so every proper depth-$6$
schedule of those six codes is unperiodic (Supplementary
Fig.~\ref{fig:periods}). The refinement that a class schedule exists
exactly when $\operatorname{ord}(\colbar)$ is even is false, with $468$
admissible keys at $q=6$ having $\operatorname{ord}(\colbar)=3$. A
criterion for the odd-order regime is open; Section~\ref{sec:codes} shows
that it contains codes of interest.

\subsection{Depth-six codes with competitive parameters}
\label{sec:codes}

Theorem~\ref{thm:six} says nothing about distance, so we searched. For
every abelian group of order $N\in\{18,24,27,36,45,54,72\}$ we classified
the two-block codes with $\wA=\wB=3$---all pairs $(A,B)$ with $0\in A$,
$0\in B$ for $N\le36$, $4\times10^5$ random pairs per group for
$N\ge45$---by the multiset of sums, computed $k$, and estimated $d$ for the
orbit representatives under $\mathrm{Aut}(G)$, translations and block swap
by information-set decoding, confirming the bound by an exhaustive search
for a lighter logical operator whenever that was feasible (Methods).
Table~\ref{tab:depth6codes} lists the Pareto-optimal $(k,d)$ pairs by
length and collision structure; Supplementary Section~S8 lists the codes.

\begin{table*}[t]
\caption{\textbf{Best parameters found for two-block codes with
$\wA=\wB=3$, by length and collision structure.} Pareto-optimal $(k,d)$
pairs with $d\ge4$; each entry gives $(k,d)$ for the length $n$ in the first column. The search is over all abelian groups of order $N=n/2$ (exhaustive
over all $(A,B)$ for $N\le36$; $4\times10^5$ sampled pairs per group for
$N\ge45$). Distinct sums: depth $7$ by Theorem~\ref{thm:main}. Class C1
with $\operatorname{ord}(\col)$ even: depth $6$ by Theorem~\ref{thm:six}
(every listed code has $h(\col)\le2$). Class C1 with
$\operatorname{ord}(\col)$ odd: depth $6$ decided instance by instance
($\star$: depth-$6$ schedule found by the solver; $\dagger$: depth $6$
refuted with a verified DRAT certificate, for the
$\mathbb{Z}_{15}\times\mathbb{Z}_3$ and the $\mathbb{Z}_{45}$ instance
alike). Distances
without ``$\le$'' are exact (Methods); for odd $N$ no element has even
order and every distinct-sums code is also excluded by the two-block form
of Theorem~\ref{thm:mod4}.}
\label{tab:depth6codes}
\centering
\small
\setlength{\tabcolsep}{5pt}
\renewcommand{\arraystretch}{1.18}
\begin{tabular}{@{}rp{0.30\textwidth}p{0.30\textwidth}p{0.30\textwidth}@{}}
\toprule
$n$ & \paramcell{Distinct sums\\Depth $7$}
    & \paramcell{Class C1, $\operatorname{ord}(\col)$ even\\Depth $6$}
    & \paramcell{Class C1, $\operatorname{ord}(\col)$ odd\\Instance-dependent depth} \\
\midrule
36 & \paramcell{\kd{4,6}}
   & \paramcell{\kd{4,6}}
   & \paramcell{\kd{4,6}} \\
\addlinespace[3pt]
48 & \paramcell{\kd{8,4}, \kd{4,8}}
   & \paramcell{\kd{8,4}, \kd{4,8}}
   & \paramcell{\kd{8,4}, \kd{4,6}} \\
\addlinespace[3pt]
54 & \paramcell{\kd{8,6}, \kd{4,8}}
   & \paramcell{None ($N$ odd)}
   & \paramcell{\kd{4,8}} \\
\addlinespace[3pt]
72 & \paramcell{\kd{16,4}, \kd{12,6}, \kd{8,8}, \kd{4,\le10}}
   & \paramcell{\kd{12,4}, \kd{8,6}, \kd{4,8}}
   & \paramcell{\mbox{$(12,4)^\star$}, \mbox{$(8,6)^\star$}, \kd{4,8}} \\
\addlinespace[3pt]
90 & \paramcell{\kd{24,4}, \kd{12,6}, \kd{8,\le10}, \kd{4,\le12}}
   & \paramcell{None ($N$ odd)}
   & \paramcell{\mbox{$(12,6)^\dagger$}, \kd{4,\le10}} \\
\addlinespace[3pt]
108 & \paramcell{\kd{24,4}, \kd{16,6}, \kd{8,\le10}, \kd{4,\le12}}
    & \paramcell{\kd{12,6}, \kd{4,\le12}}
    & \paramcell{\kd{12,6}, \kd{8,\le8}, \kd{4,\le12}} \\
\addlinespace[3pt]
144 & \paramcell{\kd{32,4}, \kd{24,6}, \kd{16,\le8}, \kd{12,\le12}, \kd{8,\le10}, \kd{4,\le16}}
    & \paramcell{\kd{24,4}, \kd{16,6}, \kd{12,\le8}, \kd{8,\le10}, \kd{4,\le14}}
    & \paramcell{\kd{24,4}, \kd{16,6}, \kd{8,\le8}, \kd{4,\le14}} \\
\bottomrule
\end{tabular}
\end{table*}

Up to $n=48$ the depth-six codes match the best depth-seven two-block
codes; from $n=72$ on they trail them by one step in $k$ or in $d$, while
several points coincide. The $[[72,8,6]]$ code over
$\mathbb{Z}_6\times\mathbb{Z}_6$ with $A=1+x^4y^3+x^5y^3$ and
$B=1+xy^5+x^4y^3$ is the smallest example of interest: its only
coincidence is $a_1b_3=a_2b_1=x^4y^3$, so $\col=x^4y^3$ has order $6$ and
$2$-height $0$, and with $\varepsilon(x^iy^j)=j\bmod2$ the class schedule of
Table~\ref{tab:template72} pulls back to a proper depth-$6$ circuit,
verified by $\F$ simulation; the distance $6$ is exact. Compared with the
BB code $[[72,12,6]]$, one layer is traded for four logical qubits at
the same length and distance. The odd-order regime contains such codes
too: the solver found proper depth-$6$ schedules for a $[[72,8,6]]$ code
over $\mathbb{Z}_{36}$ with $\operatorname{ord}(\col)=9$ and for a
$[[72,12,4]]$ code over $\mathbb{Z}_6\times\mathbb{Z}_6$, and refuted depth
$6$ for a $[[90,12,6]]$ code over $\mathbb{Z}_{15}\times\mathbb{Z}_3$.
The distances are code distances; the circuit-level distances of the six-
and seven-layer circuits are not compared here.

\begin{table}[!tbp]
\caption{\textbf{Class schedule of the $[[72,8,6]]$ code.}
$G=\mathbb{Z}_6\times\mathbb{Z}_6$, $A=\{(0,0),(4,3),(5,3)\}$,
$B=\{(0,0),(1,5),(4,3)\}$, collision $a_1+b_3=a_2+b_1$,
$\varepsilon(i,j)=j\bmod2$, key $(0,1,1;0,1,1)$. Pulled back along
$\varepsilon$ it is a proper depth-$6$ schedule, verified by $\F$
simulation.}
\label{tab:template72}
\begin{ruledtabular}
\begin{tabular}{llccc}
class & edge type & label 1 & label 2 & label 3 \\
\colrule
$c=0$ & $X$-check, $L$-edges & 3 & 6 & 1 \\
      & $X$-check, $R$-edges & 2 & 5 & 4 \\
      & $Z$-check, $L$-edges & 5 & 4 & 2 \\
      & $Z$-check, $R$-edges & 3 & 1 & 6 \\
\colrule
$c=1$ & $X$-check, $L$-edges & 5 & 4 & 2 \\
      & $X$-check, $R$-edges & 3 & 6 & 1 \\
      & $Z$-check, $L$-edges & 3 & 1 & 6 \\
      & $Z$-check, $R$-edges & 2 & 4 & 5 \\
\end{tabular}
\end{ruledtabular}
\end{table}

\section{Discussion}
\label{sec:discussion}

Theorems~\ref{thm:main} and~\ref{thm:six} give two exact optima in one
model, seven and six, separated by a single repeated sum. The seven-layer
cycle is optimal for every BB code with distinct products and for every
two-block code over an abelian group with weight-three polynomials and
two-qubit overlaps; the proof is finite combinatorics on the Tanner graph
and applies to codes not yet constructed. A single collision, which changes
neither check weight nor qubit degree, creates four-qubit overlaps in which
the parity constraint is satisfied by cancellation, and when the collision
element has even order and $2$-height at most $8$ the template attains the
degree bound. The algebra of $A$ and $B$, not their weights, decides the
single-round depth. The freedom is compatible with distance: the depth-six
codes of Table~\ref{tab:depth6codes} match the best depth-seven two-block
codes up to $n=48$ and trail them by one step in $k$ or $d$ at
$n=72$--$144$, with $[[72,8,6]]$, $[[108,12,6]]$ and $[[144,16,6]]$ as
representatives. Whether the gap persists in an exhaustive search at
$n=144$, and how the six- and seven-layer circuits compare in circuit-level
distance and logical error rate, remain to be determined; the optimal
depth-seven circuit need not be the best fault-tolerant one, and the two
questions are separate.

The general obstructions reach beyond the family. Theorem~\ref{thm:mod4}
applies to any CSS code whose data qubits saturate the degree bound and
whose $X$-checks have weight $\Delta$, with hypergraph products of
$(3,3)$-regular codes of odd length as examples
(Section~\ref{sec:obstructions}); the twisted condition~\eqref{eq:twisted}
applies to any two-block code, and the absence of a common layer set
explains why symmetric constructions stop at $\Delta+1$ for two odd
weights. All three hold with
every Tanner-graph edge available as a coupler, unlike the
connectivity-driven bounds of Delfosse \emph{et al.}~\cite{Delfosse2021}.
Their scope is the single-round model: staggered, multi-round
schedules~\cite{Strikis2026,Yoder2025}, morphing
circuits~\cite{ShawTerhal2025,ShawTerhal2026} and flag
schemes~\cite{ChamberlandBeverland2018,ChaoReichardt2018} change the
model, and lower bounds there are open. Depth does not determine fault
tolerance; schedules of equal depth differ in circuit distance, computed
with tools such as \textsc{Stim}~\cite{Gidney2021}. The residual-error
metrics of Ref.~\cite{Strikis2026} provide complementary tools for
studying hook-error propagation and ranking left--right circuits.

The scheduling scheme is not limited to error correction. If no qubit is
the control of one CNOT and the target of another, all gates commute, every
colouring gives a correct circuit, and the degree bound is reached for the
given gates. A further application of this kind is Stab-QRAM
(Supplementary Section~S9), a QRAM for tables whose entries are affine
functions of the address. Its lookup uses only CNOT and $X$ gates, one
qubit per address and output bit and no ancillas, and its depth is at most
one more than the largest row or column weight of the data matrix. This
depth is optimal among schedules of these gates. General-purpose QRAM
stores arbitrary tables, with a number of routers or $T$ gates that grows
linearly with the number of
entries~\cite{Giovannetti2008PRA,Babbush2018,Jaques2023}. Stab-QRAM is
restricted to affine tables. For these the lookup is a linear reversible
circuit, and CNOT synthesis that allows other gates or ancillas can reach a
smaller depth~\cite{PMH2008,Jiang2020,MaslovZindorf2022}. Since Stab-QRAM
needs only CNOT and $X$ gates between two registers, a device built for the
syndrome-extraction circuits studied here can run it for any data matrix
whose support lies in that of $H_Z$ or $H_X^{\mathsf T}$. In syndrome
extraction the two lookups $H_Z\mathbf{x}$ and $H_X^{\mathsf T}\mathbf{y}$
share the data qubits, which are targets in one and controls in the other,
and this is where the parity obstruction comes from.

Four problems remain open. The rigidity step uses that a depth-$6$
schedule has four interior layers; for $\wA=\wB=5$ the count of
Eq.~\eqref{eq:pairsum} still applies, since $\Delta=10\equiv2\pmod 4$, but the cases
$(a,b)=(1,10),(2,9)$ resist the layer-by-layer argument of the rigidity
step, and we state the general odd case with $\wA+\wB\equiv2\pmod 4$ as a
conjecture supported by Theorem~\ref{thm:mod4}, the common-layer-set
argument and Supplementary Theorems~\ref{thm:S1} and~\ref{thm:S2}. When $\Delta\equiv0\pmod 4$, as for $(3,5)$, the parity
mechanism gives no contradiction and we have no evidence either way.
Between the necessary condition~\eqref{eq:twisted} and the sufficient
condition of Theorem~\ref{thm:six} lie the even-order collision
elements of $2$-height above $8$, where the conjecture is the statement
about the family~\eqref{eq:family}, and the odd-order elements, where the
regime is mapped but not characterised. Every member of the family admits
a class schedule with the half-period symmetry~\eqref{eq:halfperiod} for
two fixed involutions of the six layers, for all $q\le512$, but the symmetry does not extend to
$c+q/4$ and the layers on the classes $c<q/2$ depend on $c$ modulo
$2^{j-1}$ in an essential way for most $b$ (Supplementary Section~S5); an
analytic template has to prescribe those. For CSS codes without the
two-block structure,
attainability of the degree bound remains an instance-by-instance question
that the pipeline of Section~\ref{sec:methods} can answer but the theory
does not yet predict.

\section{Methods}
\label{sec:methods}

\paragraph*{Proof of Theorem~\ref{thm:mod4}.}
Fix a component with $n$ data qubits and $m_X$ $X$-checks; since every
$X$-check has weight $\Delta$, $\sum_vr_v=m_X\Delta$. At a data qubit $v$
with $X$-layers $P_v$, the layers above $i\in P_v$ are $Z$-layers except
for the $X$-layers above $i$, so the number of $X$-first corners at $v$ is
$N_v=\sum_{i\in P_v}(\Delta-i)-\binom{r_v}{2}$. Summing over $v$, the first
term is $\sum_e(\Delta-\pi(e))$ over all $X$-edges, which equals
$m_X\Delta(\Delta-1)/2$ because every $X$-check has one edge in each
layer. Hence
\begin{align}
\sum_vN_v
&=\frac{\Delta-1}{2}\sum_vr_v-\sum_v\frac{r_v(r_v-1)}{2}\nonumber\\
&=\frac12\sum_vr_v(\Delta-r_v)=\frac12\sum_vr_vs_v ,
\end{align}
for every valid depth-$\Delta$ layering. Each corner belongs to one
overlapping pair and properness makes each pair's count even, so
$\sum_vr_vs_v\equiv0\pmod 4$. \hfill$\square$

\paragraph*{Components of two-block codes.}
$X_g$ shares a data qubit with $Z_{g'}$ iff $g'-g\in\Sigma:=\{a_s+b_t\}$,
so the $X$-checks reachable from $X_g$ are $X_{g+H'}$ with
$H'=\langle\sigma-\sigma':\sigma,\sigma'\in\Sigma\rangle$; $H'=H$ because
$a_s-a_{s'}=(a_s+b_t)-(a_{s'}+b_t)$ and every $\sigma-\sigma'$ is the sum
of an $a$-difference and a $b$-difference. The $L$-qubits of $X_{g+H}$ are
$L_{g+a_s+H}$, a single coset since $a_s-a_{s'}\in H$; likewise for the
$R$-qubits $R_{g+b_t+H}$ and the $Z$-checks $Z_{g+a_s+b_t+H}$. Every data
qubit has $r_vs_v=\wA\wB$, so the sum over the component is
$2|H|\wA\wB\equiv2\pmod4$ iff $|H|\wA\wB$ is odd, and
Theorem~\ref{thm:mod4} applies. \hfill$\square$

\paragraph*{Proof of support symmetry.}
Let $x_i$ be the number of $X$-edges in layer $i$ on the $L$ block,
$\Lambda=\{i:x_i>0\}$, $a=\min\Lambda$, $b=\max\Lambda$. In layer $i$ the
$N$ $L$-qubits carry $x_i$ $X$-edges and $N-x_i$ $Z$-edges; the $N-x_i$
$X$-checks with no $L$-edge in layer $i$ have their layer-$i$ edge on the
$R$ block, so the $R$ block carries $N-x_i$ $X$-edges and $x_i$ $Z$-edges
there. Hence $Z$-edges on the $R$ block also lie in $[a,b]$, and every
$X$-check has $V_g\supseteq[1,a-1]\cup[b+1,\Delta]$. We use two properties
of $\Phi$ (Supplementary Section~S2): (P1)~the $\wB$ corners at an
$L$-edge of $X_g$ are mapped, one each, to the $\wB$ $R$-edges of $X_g$,
with equal orientations; (P2)~symmetrically, the $\wB$ corners at an
$R$-edge of a $Z$-check are mapped from its $\wB$ $L$-edges, one from
each.

Take $X_g$ with an $L$-edge in layer $b$. All layers above $b$ at its
$L$-qubit are $Z$-layers, so this edge has exactly $\Delta-b$ $X$-first
corners. By (P1), an $R$-edge of $X_g$ in a layer $v<a$ has its partner
$Z$-edge in $[a,b]$, orientation $1$, and there are $a-1$ such edges; if
$a\in V_g$, the $R$-qubit of the layer-$a$ edge has its $X$-edge in layer
$a$, so the partner lies above $a$, orientation $1$. Hence
$\Delta-b\ge a-1+\mathbb{1}[a\in V_g]$. Take $Z_{g'}$ with an $R$-edge in
layer $a$: it has $a-1$ $X$-first corners, and (P2) gives
$a-1\ge\Delta-b+\mathbb{1}[b\in U_{g'}]$ with $U_{g'}$ the $L$-layer set of
$Z_{g'}$. Adding forces both indicators to vanish and both inequalities to
be equalities: $a+b=\Delta+1$, and $a\in P_g$. Layer reversal
$i\mapsto\Delta+1-i$ preserves properness and fixes $(a,b)$, giving the
converse. \hfill$\square$

\paragraph*{Reduction of the template keys.}
Fix $q=2^j$ and a key with $\colbar=q/2$. Relabel so that the collision is
$\{(1,1),(2,2)\}$ and normalise $\alpha_1=\beta_1=0$ (Supplementary
Section~S5); the collision equation then forces $\alpha_2=\beta_2=q/2$, so
the key is $\kappa_q(a,b)=(0,q/2,a;0,q/2,b)$ with $a=\varepsilon(a_3)$,
$b=\varepsilon(b_3)$: $q^2$ keys per modulus. Three symmetries of the class
system identify further keys: $(a,b)\sim(b,a)$ ($A\leftrightarrow B$ with
$L\leftrightarrow R$), $(a,b)\sim(\omega a,\omega b)$ for every unit
$\omega$ of $\Zq$ (with $T'(c)=T(\omega^{-1}c)$), and
$(a,b)\sim(a+q/2,b+q/2)$ (swap labels $1\leftrightarrow2$ and
renormalise). Moreover the constraints (i)--(iv) only connect classes that
differ by key elements, so the class system of $\kappa_q(a,b)$ is a
disjoint union of copies of the class system of modulus $q/g$ with key
$\kappa_{q/g}(a/g,b/g)$, $g=\gcd(q/2,a,b)$, again with $\colbar$ equal to
half the modulus; it therefore suffices to treat keys with $\gcd(a,b)$
odd, which after a swap and a scaling have $a=1$, the
family~\eqref{eq:family}. The representatives under the three symmetries
number $2,5,11,21,39,73,139,269,527$ for $q=2,4,\dots,512$. For $q\le32$
every representative was decided satisfiable by SAT without further
assumptions ($39$ representatives at $q=32$ in $17$ minutes); for
$q\le512$ every
member of the family~\eqref{eq:family} was decided satisfiable under the
ansatz $T(c+q/2,e)=\sigma_{\mathrm{blk}(e)}(T(c,e))$ with
$\sigma_L=(13)(24)(56)$ on $L$-edges and $\sigma_R=(12)(36)(45)$ on
$R$-edges, whose solutions are class schedules and which makes the
instances easy (all $512$ keys at $q=512$ in twelve minutes); Supplementary
Section~S5 records how the symmetry was found and which stronger forms
fail.

\paragraph*{Class schedules with degenerate keys.}
For a key with repeated entries, $\alpha_s=\alpha_{s'}$ or
$\beta_t=\beta_{t'}$ (unavoidable at $q=2$ and the case for $648$ of the
$864$ admissible keys at $q=4$), the quotient system $\mathrm{QC}(\varepsilon(A),
\varepsilon(B))$ over $\Zq$ is not a two-block code with three-element
sets, so each stored class schedule is checked in two ways: directly
against (i)--(iv), and by pulling it back along $g\mapsto g\bmod q$ to a
\emph{realising} class-C1 code, with lifts $a_s=\alpha_s+qx_s$,
$b_t=\beta_t+qy_t$ in $\mathbb{Z}_{qM}$ chosen so that the designated
collision is the only coincidence, followed by the $\F$ circuit
simulation. All $72$ keys at $q=2$ ($M=7$, codes over $\mathbb{Z}_{14}$)
and all $864$ keys at $q=4$ (codes over $\mathbb{Z}_{28}$) pass.

\paragraph*{Table~\ref{tab:depth7main} and the published cycle.}
In our notation, rounds $1$--$7$ of the cycle of Ref.~\cite{Bravyi2024}
(Table~5 there; $A_s\leftrightarrow$ label $s$, $B_t\leftrightarrow$ label
$t$) are $X$ $L$-edges $(6,2,7)$, $X$ $R$-edges $(4,3,5)$, $Z$ $L$-edges
$(3,4,5)$ and $Z$ $R$-edges $(1,6,2)$. This is a proper
translation-invariant schedule of depth $7$ with
$o_L=o_R=\mathbb{1}[s=2]$, verified by $\F$ simulation on $[[72,12,6]]$ and
$[[144,12,12]]$, whereas Table~\ref{tab:depth7main} has
$o_L=o_R=\mathbb{1}[s\le2]$. The two are not equivalent under label
permutations, layer reversal, $X$/$Z$ duality and block swap, a group of
order $288$ acting on labelled schedules: the $8496$ translation-invariant
proper depth-$7$ schedules fall into $41$ orbits under it, and the two lie
in different orbits. Both attain the minimum of Theorem~\ref{thm:main}.

\paragraph*{Code search and distance estimation.}
For each group $G=\mathbb{Z}_\ell\times\mathbb{Z}_m$ of order $N$ the pairs
$A=\{0,a_2,a_3\}$, $B=\{0,b_2,b_3\}$ were enumerated (exhaustively for
$N\le36$; $4\times10^5$ random pairs per group for $N\in\{45,54,72\}$),
classified by the multiset of sums, and $k=n-\operatorname{rank}H_X-
\operatorname{rank}H_Z$ was computed over $\F$. Codes with $k>0$ were
reduced to orbit representatives under $\mathrm{Aut}(G)$, translations of
$A$ and $B$, and $(A,B)\mapsto(B,A)$. For each representative an upper
bound on $d$ ($=d_X=d_Z$ by the duality of Supplementary Section~S2) was
obtained by information-set decoding on a generator of $\ker H_Z$, a
vector being a nontrivial logical operator iff it has odd overlap with
some $Z$-logical representative (Prange with Lee--Brickell pairs; $300$
iterations per representative, $2000$ on the Pareto front, $10^4$ on the
codes with explicit schedules). Whenever
$\binom{n}{\lceil d/2\rceil}\le2.5\times10^6$ the bound was confirmed exact
by a meet-in-the-middle enumeration of all vectors of weight at most $d$
in $\ker H_Z$ outside the row space of $H_X$; entries of
Table~\ref{tab:depth6codes} without ``$\le$'' are exact. Depth-$6$
schedules for the listed class-C1 codes of even order were built from class
schedules of modulus $2^{h(\col)+1}$ and verified by $\F$ simulation
(Supplementary Section~S8); the odd-order instances were decided directly
by SAT at depth $6$, in $7$--$600$ seconds each.

\paragraph*{Encoding, solvers and verification.}
A schedule of depth $D$ is encoded with one Boolean variable per
Tanner-edge/layer pair, exactly-one clauses per edge, at-most-one clauses
for every Tanner vertex (data qubits and check ancillas) and layer, and
Eq.~\eqref{eq:parity} as XORs of layer comparisons with Tseitin
auxiliaries, in the manner of SAT-based circuit
synthesis~\cite{Peham2023,TanCong2020}; the encoding accepts any CSS code.
Unsatisfiability is decided by \textsc{kissat}~4.0.4~\cite{Biere2024Kissat}
with DRAT logging and every proof is checked by
\textsc{drat-trim}~\cite{Wetzler2014DRAT,HeuleHuntWetzler2013}, the
standard for computer-assisted refutations~\cite{Heule2016}; satisfiable
instances are decided by \textsc{CaDiCaL}~\cite{Biere2024CaDiCaL} via
PySAT~\cite{Ignatiev2018PySAT}; sweeps use
\textsc{Glucose}~\cite{AudemardSimon2009} under a $60$-second cap, and
capped instances are reported as undetermined, never as unsatisfiable.
Every schedule is verified by propagating the $\F$ register through the
layered CNOT list and comparing with Eq.~\eqref{eq:se-map} at $M_\pi=0$;
the same routine evaluates Eq.~\eqref{eq:parity} on arbitrary layerings.
Class schedules are enumerated by the same encoder on the class system,
stored per admissible key, and their pullbacks verified by the same
simulation. Layer reversal maps proper schedules to proper schedules; an
early encoding broke this symmetry by a clause (Supplementary Section~S7),
and the certificates of the seven codes of Ref.~\cite{Bravyi2024} were
regenerated without it. All runs were single-threaded on one workstation
(Intel Core Ultra 9 275HX, $15.6$~GiB); wall and CPU times agree within a
few per cent, and no cluster resources were used.

\paragraph*{Certificates.}
Eighteen instances carry DRAT certificates: the seven codes of
Ref.~\cite{Bravyi2024} (Table~\ref{tab:certificates}), the smallest and
largest certified distinct-sums instances ($|G|=10$ and $24$),
$G=\mathbb{Z}_6$, $A=B=\{0,2,4\}$, which Theorem~\ref{thm:mod4} excludes
through its six-qubit components and which is the hardest refutation here,
$558$ seconds and a $2.7$~GB proof with the reversal clause retained, and
six class-C1 instances of odd order not covered by a theorem (proofs
$123$--$911$~MB, $16$--$163$~s), and the two odd-order $[[90,12,6]]$ codes
of Supplementary Section~S8 (proofs $2.4$ and $2.8$~GB). As a cross-check, the $17$ certificate
CNFs of the first ten instances were re-decided by \textsc{CaDiCaL} and
\textsc{Glucose} with a $300$-second cap: $16$ are unsatisfiable under
both, and the $\mathbb{Z}_6$ file is decided by \textsc{CaDiCaL} in $233$
seconds but not by \textsc{Glucose} within the cap. All seven codes of
Ref.~\cite{Bravyi2024} are unsatisfiable at depth $6$ without
symmetry-breaking clauses, with solve times between $1.4$ and $6.3$
seconds and verified proofs between $18$ and $68$~MB, and satisfiable at
depth $7$ by the schedule of Table~\ref{tab:depth7main}. Of the $664$ instances of
Fig.~\ref{fig:sharpness}, $250$ are excluded by Theorems~\ref{thm:mod4}
or~\ref{thm:main}, $80$ of them beyond the solver's cap; $56$ are solver
refutations without proof logging and $6$ carry certificates; $30$ are
open; the remaining $322$ admit schedules verified by simulation, eight of
them found by twisted witnesses after the solver reached its cap.

\begin{table}[!tbp]
\caption{\textbf{Depth-6 refutations of the seven bivariate bicycle codes
of Ref.~\cite{Bravyi2024}.} Group $\mathbb{Z}_\ell\times\mathbb{Z}_m$,
\textsc{kissat} solve time and DRAT proof size, all without
symmetry-breaking clauses and all verified by \textsc{drat-trim}.
Polynomials in Supplementary Table~\ref{tab:instances}.}
\label{tab:certificates}
\begin{ruledtabular}
\begin{tabular}{lccc}
Code & $(\ell,m)$ & Time (s) & Proof (MB) \\
\colrule
$[[72,12,6]]$      & $(6,6)$   & 4.0 & 24 \\
$[[90,8,10]]$      & $(15,3)$  & 4.2 & 41 \\
$[[108,8,10]]$     & $(9,6)$   & 4.8 & 52 \\
$[[144,12,12]]$    & $(12,6)$  & 3.6 & 35 \\
$[[288,12,18]]$    & $(12,12)$ & 1.4 & 18 \\
$[[360,12,\le24]]$ & $(30,6)$  & 6.3 & 68 \\
$[[756,16,\le34]]$ & $(21,18)$ & 3.3 & 46 \\
\end{tabular}
\end{ruledtabular}
\end{table}

\paragraph*{Validation of the model and of the proof steps.}
The parity criterion~\eqref{eq:parity} and direct $\F$ simulation agree on
$300$ random valid layerings of each of two instances, the $[[72,12,6]]$
code and the $(\wA,\wB)=(2,3)$ code \texttt{oe23}; this checks the
implementation, the equivalence itself being the derivation of $M_\pi$.
The statements that hold for general $(\wA,\wB)$ with distinct
sums---support symmetry~\eqref{eq:support}, the corner bijection, $F_L=F_R$, and the
shell statements used in the rigidity proof---hold on all $417$ proper
depth-$\Delta$ schedules of four codes with one even weight,
$(\wA,\wB)\in\{(1,2),(2,1),(2,3),(4,3)\}$, $\Delta\in\{3,3,5,7\}$, of which
$5$ are translation-invariant, $72$ twisted and $340$ without a proper
period; the pair structure of the rigidity step cannot be tested
this way, since no proper depth-$6$ schedule of a weight-$(3,3)$ code
exists. The coset identity behind Eq.~\eqref{eq:twisted} holds on all
$12{,}931$ cosets formed from $289$ twisted witnesses and all $11{,}412$
admissible template keys, and on the $323$ twisted schedules found on the
satisfiable class-C2 instances, whose $|M|$ takes eleven values between
$2$ and $24$, all even; both subgroup instances are unsatisfiable at depth
$6$, so beyond classes C1 and C2 the theorem rests on its proof alone
(Supplementary Section~S7). Every number in
Sections~\ref{sec:six}--\ref{sec:methods} is anchored to a data file in
the repository.

\section*{Data availability}
All instance definitions, solver logs, DRAT certificates, schedules,
templates and the CSV files underlying every table and figure are
available at \url{https://doi.org/10.5281/zenodo.23407242}.

\section*{Code availability}
The SAT encoder, the simulation checker, the enumeration scripts, the
depth-certificate compiler of Supplementary Section~S1, the key-reduction,
code-search and distance-estimation scripts of Sections~\ref{sec:six}
and~\ref{sec:codes}, and a manifest mapping each figure to its data are
available at \url{https://github.com/Guangyi-Leo/parity-obstruction-bb-codes.git} (archived at
the Zenodo record above).

\section*{Acknowledgements}
We thank Frederic T.\ Chong, Yongshan Ding, Steve Girvin, Allen Mi, Kate
Smith and Youtao Zhang for helpful discussions. G.L.\ contributed to this 
work as a visiting student at the University of Pittsburgh; G.L., Y.G., 
Z.W.\ and J.L.\ are supported in part by the University of Pittsburgh School 
of Computing and Information, the Department of Computer Science, Pitt Cyber, 
the Pitt Momentum Fund, the PQI Community Collaboration Awards, the John 
C.\ Mascaro Faculty Scholar in Sustainability, NSF award 2535915, Thinking 
Machines Lab and Cisco Research. X.Z.\ acknowledges support from the AWS 
Center for Quantum Computing and the Fu Foundation School of Engineering 
and Applied Science of Columbia University. Z.Z.\ acknowledges NSF CAREER 
Award No.~2317471.

\section*{Author contributions}
G.L.\ conceived the project, proved the obstruction theorems, the main
theorem and the depth-six theorem, designed the SAT encoding and the
certificate pipeline, ran the computational study, and wrote the
manuscript. Y.G.\ and Z.W.\ contributed to the theoretical analysis and to
the validation of the proofs. X.Z.\ and Z.Z.\ contributed to the analysis
of the circuit model and its hardware assumptions and to the discussion of
hardware implementations of Stab-QRAM in Supplementary Section~S9. J.L.\
supervised the
project and revised the manuscript. All authors discussed the results and
reviewed the manuscript.

\section*{Competing interests}
J.L.\ has served as an Associate Editor of npj Quantum Information since
2024. The journal's editorial policies ensure that he was not involved in
the editorial handling or peer review of this manuscript. The other
authors declare no competing interests.

\onecolumngrid
\clearpage
\begin{center}\Large\bfseries Supplementary Information\end{center}
\raggedbottom
\providecommand{\SIwidefigurewidth}{0.94\textwidth}
\providecommand{\SIcolumnfigurewidth}{0.60\textwidth}
\setcounter{topnumber}{3}
\setcounter{bottomnumber}{2}
\setcounter{totalnumber}{5}
\setcounter{dbltopnumber}{2}
\renewcommand{\topfraction}{0.92}
\renewcommand{\bottomfraction}{0.85}
\renewcommand{\textfraction}{0.06}
\renewcommand{\floatpagefraction}{0.75}
\renewcommand{\dbltopfraction}{0.92}
\renewcommand{\dblfloatpagefraction}{0.72}
\setlength{\textfloatsep}{14pt plus 3pt minus 3pt}
\setlength{\floatsep}{9pt plus 3pt minus 2pt}
\setlength{\intextsep}{12pt plus 3pt minus 2pt}
\setcounter{section}{0}
\renewcommand{\thesection}{S\arabic{section}}
\setcounter{figure}{0}\renewcommand{\thefigure}{S\arabic{figure}}
\setcounter{table}{0}\renewcommand{\thetable}{S\arabic{table}}
\setcounter{equation}{0}\renewcommand{\theequation}{S\arabic{equation}}
\setcounter{theorem}{0}\renewcommand{\thetheorem}{S\arabic{theorem}}
\renewcommand{\theHsection}{S\arabic{section}}
\renewcommand{\theHfigure}{S\arabic{figure}}
\renewcommand{\theHtable}{S\arabic{table}}
\renewcommand{\theHequation}{S\arabic{equation}}
\renewcommand{\theHtheorem}{S\arabic{theorem}}

\begin{figure}[p]
\centering
\includegraphics[width=0.90\textwidth,height=0.82\textheight,keepaspectratio]{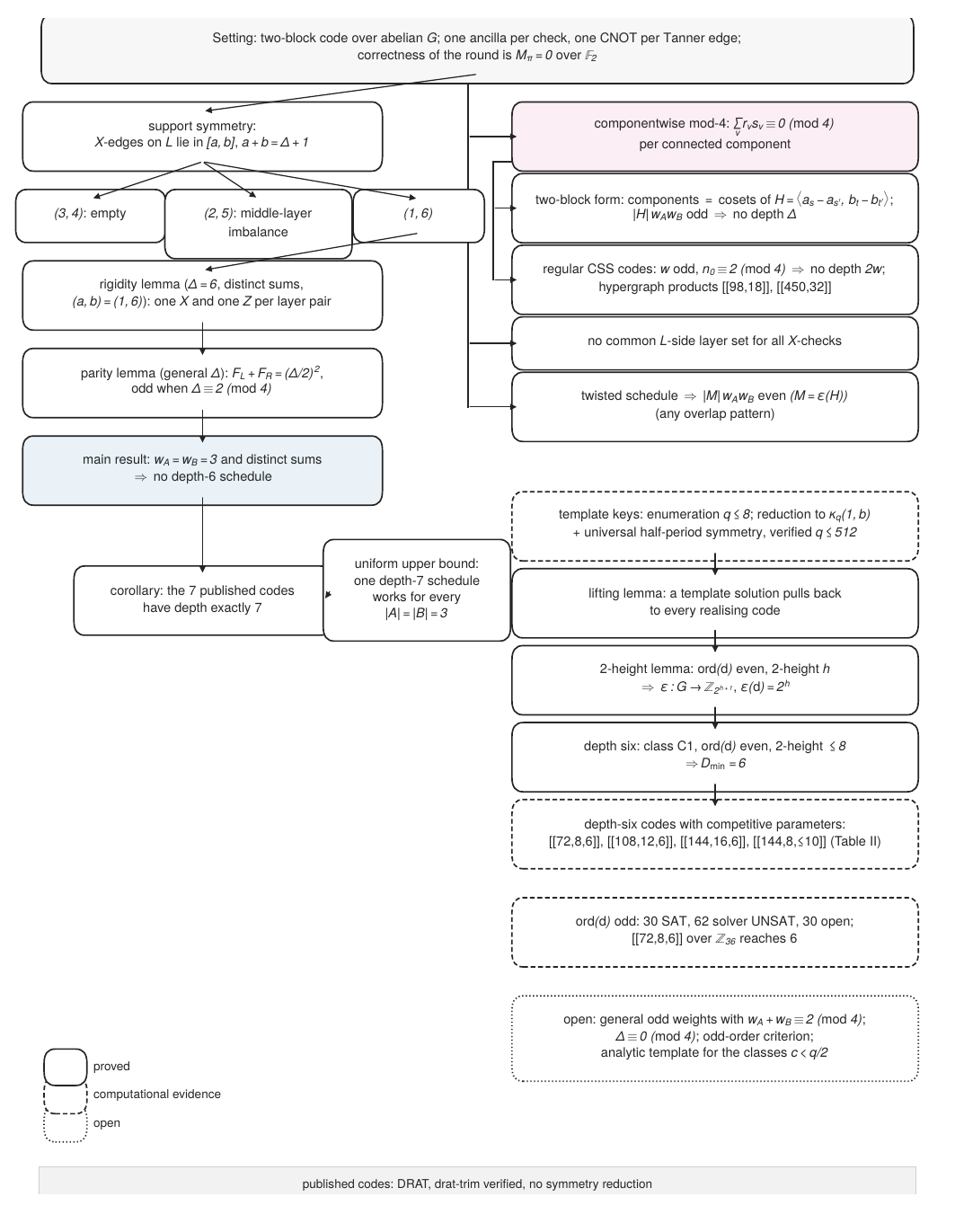}
\caption{\textbf{Full logical structure of the results.} Solid boxes are
proved statements, dashed boxes computational evidence, dotted boxes open
problems. The three general obstructions follow from $M_\pi=0$ alone. The
main chain runs from support symmetry, Eq.~\eqref{eq:support}, through the
rigidity step to the count of Eq.~\eqref{eq:pairsum}; together with the
schedule of Table~\ref{tab:depth7main} it gives Theorem~\ref{thm:main}. The
depth-six chain runs from the template enumeration and the reduction of
the keys through the lifting of class schedules and the $2$-height to
Theorem~\ref{thm:six}.}
\label{fig:fullmap}
\end{figure}

Theorems~1--3 are those of the main text, and Theorems~S1--S3 are proved
only here; the steps of the main proofs are referred to by name and by
equation number. Notation is that of Section~\ref{sec:model}; in a
depth-$\Delta$ schedule every data qubit and every check has exactly one
edge in each layer. Fig.~\ref{fig:fullmap} is the full dependency map.
Section~S9 applies the compiler of Section~S1 to quantum memory
(Stab-QRAM) and discusses possible hardware implementations.

\section{The depth certificate for commuting gate sets}
\label{si:compiler}

The degree bound of Section~\ref{sec:model} and the K\H{o}nig colouring
that attains it are instances of a depth-exact compiler for affine maps;
this section fixes what the certificate is exact for. It rests on a
standard fact about Clifford
circuits~\cite{DehaeneDeMoor2003,Gottesman1998,Aaronson2004}: a Clifford
circuit $C$ on $n+k$ qubits that maps every computational-basis input
$|\mathbf{x}\rangle|0\rangle^{\otimes k}$ to a computational-basis output
$|g(\mathbf{x})\rangle|h(\mathbf{x})\rangle$ acts by affine maps $g,h$ over
$\F$. To see this, write $CZ_iC^\dagger=(-1)^{\sigma_i}P_i$. A Pauli word
with an $X$ or $Y$ factor cannot stabilise a basis state, so
$P_i=Z^{\mathbf{v}_i}$; if $\sum_iu_i\mathbf{v}_i=0$ with
$\mathbf{u}\neq0$ then $\prod_{u_i=1}Z_i=\pm I$, which is false, so
$V=(\mathbf{v}_1\cdots\mathbf{v}_{n+k})$ is invertible. The output
$\mathbf{z}$ satisfies $V^{\mathsf T}\mathbf{z}=(\mathbf{x},0)+\bm{\sigma}$
and is therefore affine in $\mathbf{x}$. Conversely, a direct evaluation
shows that every affine map $f(\mathbf{x})=A\mathbf{x}+\mathbf{b}$ is
implemented on $n+m$ qubits by
\begin{equation}
C_f=\prod_j\Bigl[\Bigl(\prod_{k:A_{j,k}=1}\mathrm{CNOT}(x_k,d_j)\Bigr)
X_{d_j}^{b_j}\Bigr].
\label{eq:Cf}
\end{equation}

The gates of $C_f$ commute pairwise, so compiling it is a scheduling
problem. Read $(A,\mathbf{b})$ as a bipartite multigraph $G'_A$, with edge
set $E'$, that has one edge per nonzero $A_{j,k}$ and a pendant edge per
$b_j=1$; a layering is then an edge colouring of $G'_A$. With $c_k$ and
$w_j$ the column and row weights of $A$, the minimum depth among all
schedules of the gate multiset of $C_f$ is
\begin{equation}
D^\star=\max\Bigl\{\max_kc_k,\;\max_j(w_j+b_j)\Bigr\}=\Delta(G'_A),
\label{eq:Dstar}
\end{equation}
attained by any edge colouring of $G'_A$ with $D^\star$
colours~\cite{Konig1916}, which can be computed in $O(|E'|\log D^\star)$
time~\cite{ColeOstSchirra2001,GabowKariv1982}.

$D^\star$ is exact for this gate set and not a lower bound over all CNOT
circuits: with CNOTs inside the data register an all-ones column is
realised in depth $\lceil\log_2(m+1)\rceil$~\cite{MooreNilsson2001,
Jiang2020,MaslovZindorf2022,deBrugiere2021}. It depends only on the
degree sequence of $G'_A$, not on the rank of $A$ or on hardware
connectivity (for the relation of such compiled lookups to quantum memory
see Ref.~\cite{Jaques2023}). The two halves of an extraction round,
$A=H_X^{\mathsf T}$ and $A=H_Z$, are each scheduled exactly by
Eq.~\eqref{eq:Dstar}; interleaving them is an edge colouring of the Tanner
graph, attained with $\Delta$ colours, but the halves do not commute and
the interleaved circuit implements Eq.~\eqref{eq:se-map} with the cross
term $M_\pi$. Section~S9 applies this compiler to quantum memory.

\FloatBarrier

\section{Two-block codes and the corner bijection}
\label{si:corners}

\paragraph*{Proof of Eq.~\eqref{eq:heisenberg}.}
Under $\mathrm{CNOT}(c,t)$,
\[
X_c\mapsto X_cX_t,
\qquad
Z_t\mapsto Z_cZ_t,
\]
while $X_t$ and $Z_c$ are fixed. We propagate the Pauli operators
backwards through the scheduled circuit.

First consider $Z_{z_j}$. At a $Z$-check CNOT
$\mathrm{CNOT}(q,z_j)$, it acquires a factor $Z_q$. This factor is
subsequently carried onto $Z_{a_i}$ when the backward propagation reaches
an earlier $X$-check CNOT $\mathrm{CNOT}(a_i,q)$. Thus a shared data qubit
$q\in\operatorname{supp}X_i\cap\operatorname{supp}Z_j$ contributes
$Z_{a_i}$ precisely when its $X$-edge precedes its $Z$-edge. By
definition, the number of such shared qubits is $(M_\pi)_{ji}$ modulo
$2$. Therefore
\[
U^\dagger Z_{z_j}U
=
Z_{z_j}\,\overline{Z}_j
\prod_i Z_{a_i}^{(M_\pi)_{ji}}.
\]

Now consider $X_{a_i}$. At an $X$-check CNOT
$\mathrm{CNOT}(a_i,q)$, it acquires a factor $X_q$. Under backward
propagation, this factor is carried onto $X_{z_j}$ when the circuit
encounters an earlier $Z$-check CNOT $\mathrm{CNOT}(q,z_j)$. Hence a
shared data qubit contributes $X_{z_j}$ precisely when its $Z$-edge
precedes its $X$-edge.

Let
\[
r_{ij}
=
\bigl|
\operatorname{supp}X_i
\cap
\operatorname{supp}Z_j
\bigr|.
\]
Because the two incident CNOTs on a shared data qubit cannot occupy the
same layer, the number of $Z$-first shared qubits is
\[
r_{ij}-(M_\pi)_{ji}.
\]
The CSS commutation condition
\[
H_XH_Z^{\mathsf T}=0
\qquad\text{over }\F
\]
implies that every $X_i$--$Z_j$ overlap has even cardinality:
\[
r_{ij}\equiv0\pmod2.
\]
Consequently,
\[
r_{ij}-(M_\pi)_{ji}
\equiv
(M_\pi)_{ji}
\pmod2.
\]
It follows that
\[
U^\dagger X_{a_i}U
=
X_{a_i}\,\overline{X}_i
\prod_j X_{z_j}^{(M_\pi)_{ji}}.
\]
Thus Eq.~\eqref{eq:heisenberg} has the stated form. The two evolved
operators commute for every $i,j$, as required. \hfill$\square$

\paragraph*{Conventions.}
For BB codes, $G=\mathbb{Z}_\ell\times\mathbb{Z}_m$ and $A,B$ are the
exponent sets of the polynomials of Ref.~\cite{Bravyi2024}, a monomial
$x^iy^j$ being the element $(i,j)$. Table~\ref{tab:instances} lists the
seven codes of Ref.~\cite{Bravyi2024}; all have distinct sums.
Equivalently, the integer matrix $H_XH_Z^{\mathsf T}$ has entries in
$\{0,2\}$, since its $(g,g')$ entry equals $2S(g'-g)$.

\begin{table}[!htbp]
\caption{\textbf{The seven bivariate bicycle codes of
Ref.~\cite{Bravyi2024}.} Group, exponent sets $A$ and $B$ with the monomial
$x^iy^j$ written $(i,j)$, whether the nine sums are pairwise distinct, and
the parity of $\ell m$ relevant to Theorem~\ref{thm:mod4}.}
\label{tab:instances}
\begin{ruledtabular}
\begin{tabular}{lcllcc}
Code & $(\ell,m)$ & $A$ & $B$ & distinct sums & $\ell m$ \\
\colrule
$[[72,12,6]]$
  & $(6,6)$
  & $\{(3,0),(0,1),(0,2)\}$
  & $\{(0,3),(1,0),(2,0)\}$
  & yes & even \\
$[[90,8,10]]$
  & $(15,3)$
  & $\{(9,0),(0,1),(0,2)\}$
  & $\{(0,0),(2,0),(7,0)\}$
  & yes & odd \\
$[[108,8,10]]$
  & $(9,6)$
  & $\{(3,0),(0,1),(0,2)\}$
  & $\{(0,3),(1,0),(2,0)\}$
  & yes & even \\
$[[144,12,12]]$
  & $(12,6)$
  & $\{(3,0),(0,1),(0,2)\}$
  & $\{(0,3),(1,0),(2,0)\}$
  & yes & even \\
$[[288,12,18]]$
  & $(12,12)$
  & $\{(3,0),(0,2),(0,7)\}$
  & $\{(0,3),(1,0),(2,0)\}$
  & yes & even \\
$[[360,12,\le24]]$
  & $(30,6)$
  & $\{(9,0),(0,1),(0,2)\}$
  & $\{(0,3),(25,0),(26,0)\}$
  & yes & even \\
$[[756,16,\le34]]$
  & $(21,18)$
  & $\{(3,0),(0,10),(0,17)\}$
  & $\{(0,5),(3,0),(19,0)\}$
  & yes & even \\
\end{tabular}
\end{ruledtabular}
\end{table}

\paragraph*{Overlaps.}
$X_g$ acts on $L_{g+a_s}$ and $R_{g+b_t}$, while $Z_{g'}$ acts on
$L_{g'-b_t}$ and $R_{g'-a_s}$. They share $L_h$ if and only if
\[
h=g+a_s=g'-b_t,
\]
and share $R_r$ if and only if
\[
r=g+b_t=g'-a_s,
\]
both of which give
\[
g'-g=a_s+b_t.
\]
Each label pair satisfying $a_s+b_t=g'-g$ contributes one shared
$L$-qubit, $L_{g+a_s}$, where the $X$-edge has label $s$ and the
$Z$-edge has label $t$, and one shared $R$-qubit, $R_{g+b_t}$, where the
labels are $t$ and $s$, respectively.

\paragraph*{The corner bijection.}
Assume distinct sums. The map $\Phi$ sending the $L$-corner
\[
(L_{g+a_s},e^{X_g}_s,e^{Z_{g'}}_t)
\]
to the $R$-corner
\[
(R_{g+b_t},e^{X_g}_t,e^{Z_{g'}}_s),
\qquad
g'=g+a_s+b_t,
\]
is a bijection from $L$-corners to $R$-corners, and a schedule is proper
if and only if $\Phi$ preserves orientation. Indeed, an $L$-corner at $L_h$
pairs $X_g$ with $h=g+a_s$ and $Z_{g'}$ with $h=g'-b_t$, so $g'-g=a_s+b_t$,
and distinct sums determine $(s,t)$ from $g'-g$. The pair shares exactly
one corresponding $R$-qubit, namely $R_{g+b_t}$, with the stated labels,
and Eq.~\eqref{eq:parity} for the pair says that the two corners have equal
orientation. Two consequences of the labels are used repeatedly. (P1)~The
$\wB$ corners at the $L$-edge of label $s$ of $X_g$ are mapped to the $\wB$
$R$-edges of $X_g$, one to each, and the partner $Z$-edge at the $R$-edge
of label $t$ has label $s$. (P2)~The $\wB$ corners at the $R$-edge of
label $s$ of $Z_{g'}$ (one per $X$-check $X_{g'-a_s-b_t}$ at $R_{g'-a_s}$)
are mapped from the $\wB$ $L$-edges of $Z_{g'}$, one from each.

\paragraph*{Symmetries.}
Layer reversal $i\mapsto\Delta+1-i$ flips every orientation and preserves
properness; it maps $(a,b)$ to $(\Delta+1-b,\Delta+1-a)$, which is $(a,b)$
for proper schedules by Eq.~\eqref{eq:support}. The duality
\[
X_g\leftrightarrow Z_{-g},
\qquad
L_h\leftrightarrow R_{-h},
\]
maps $\mathrm{QC}(A,B)$ to itself with $X$ and $Z$ exchanged and
preserves properness.

\FloatBarrier

\section{General obstructions}
\label{si:obstructions}

\paragraph*{No common layer set.}
Suppose no layer is mixed. Then every $X$-check has the same $L$-layer set
$P$ and $R$-layer set $V=[\Delta]\setminus P$, $|P|=\wA$, $|V|=\wB$; every
$L$-qubit has $X$-layers $P$ and $Z$-layers $V$, every $R$-qubit the
reverse. For any $X$-check $F_L=\#\{(i,v)\in P\times V:i<v\}$ and
$F_R=\#\{(v,i)\in V\times P:v<i\}$, so $F_L+F_R=\wA\wB$, which
Eq.~\eqref{eq:parity} makes even. A translation-invariant schedule has no
mixed layer. \hfill$\square$

\paragraph*{Twisted schedules: proof of Eq.~\eqref{eq:twisted}.}
Let $\alpha_s=\varepsilon(a_s)$, $\beta_t=\varepsilon(b_t)$. A twisted
schedule assigns to the $X$-check of class $c$ layers $\pi_c(s)$,
$\nu_c(t)$ and to the $Z$-check of class $c$ layers $\mu_c(t)$,
$\kappa_c(s)$. The $L$-qubit of class $\lambda$ has $X$-edges from classes
$\lambda-\alpha_s$ and $Z$-edges from classes $\lambda+\beta_t$; the
$R$-qubit of class $\rho$ has $X$-edges from $\rho-\beta_t$ and $Z$-edges
from $\rho+\alpha_s$. Fix a coset $C$ of $M$ in $\varepsilon(G)$; since the
$\alpha_s$ are congruent modulo $M$, as are the $\beta_t$, the sets
$C+\alpha_s$, $C+\beta_t$, $C+\alpha_s+\beta_t$ are single cosets $C_L$,
$C_R$, $C_Z$ of size $|M|$, and the classes in $C\cup C_L\cup C_R\cup C_Z$
form a closed subsystem. For $c\in C$ let
$F_L(c)=\sum_{s,t}\mathbb{1}[\pi_c(s)<\mu_{c+\alpha_s+\beta_t}(t)]$ and
$F_R(c)=\sum_{s,t}\mathbb{1}[\nu_c(t)<\kappa_{c+\alpha_s+\beta_t}(s)]$;
these are the $F_L,F_R$ of any check of class $c$. Regrouping by qubit
class, at $\lambda\in C_L$ the $\wA$ $X$-edges and $\wB$ $Z$-edges fill
all $\Delta$ layers, so the $X$-first corners there number
$\sum_{i\in P_\lambda}(\Delta-i)-\binom{\wA}{2}$; summing over $C_L$ the
first term is $\sum_{c\in C}\sum_s(\Delta-\pi_c(s))$, and likewise on the
$R$ block with $\binom{\wB}{2}$. Since the $\Delta$ edges of each $X$-check
fill all layers, $\sum_s(\Delta-\pi_c(s))+\sum_t(\Delta-\nu_c(t))=
\binom{\Delta}{2}$, and
\begin{equation}
\begin{aligned}
\sum_{c\in C}\bigl(F_L(c)+F_R(c)\bigr)
&=|M|\Bigl[\tbinom{\Delta}{2}-\tbinom{\wA}{2}-\tbinom{\wB}{2}\Bigr]\\
&=|M|\,\wA\wB .
\end{aligned}
\end{equation}
Each summand is a sum over overlapping pairs of even numbers, hence even.
So $|M|\wA\wB$ is even; for $\wA=\wB=3$, $|M|$ is even, and a subgroup of
$\Zq$ with $q$ odd has odd order. \hfill$\square$

With $q=1$ this gives again that no translation-invariant proper
schedule of depth $\wA+\wB$ exists when $\wA\wB$ is odd. The proof uses
neither distinct sums nor the class of the code.

\FloatBarrier

\section{Proof of the main theorem}
\label{si:main}

The proof of support symmetry, Eq.~\eqref{eq:support}, is in Methods. It
says nothing about layers strictly between $a$ and $b$: a proper depth-$4$ schedule of
$G=\mathbb{Z}_5$, $A=\{0,1\}$, $B=\{0,2\}$ has every $X$-check's $L$-layer
set equal to $\{1,4\}$, with $a+b=5$ and the middle layers unused.
Throughout this section the code has distinct sums.

\begin{theorem}
\label{thm:S1}
Let $\mathrm{QC}(A,B)$ have distinct sums, $|A|=1$ and $|B|=\wB$ odd.
Then no proper schedule of depth $\Delta=\wB+1$ exists; by duality the
same holds for $|B|=1$, $|A|$ odd.
\end{theorem}

\begin{proof}
Let $p_g$, $q_{g'}$, $p_h$, $q_r$ be the layers of the single $L$-edge of
$X_g$, the single $R$-edge of $Z_{g'}$, the single $X$-edge at $L_h$ and
the single $Z$-edge at $R_r$. The $L$-edge of $X_g$ has $\Delta-p_g$
corners, all $X$-first, and by (P1)
$\Delta-p_g=\sum_t\mathbb{1}[v_t<q_{r_t}]$ over its $R$-edges in layers
$[\Delta]\setminus\{p_g\}$; the $R$-edge of $Z_{g'}$ has $q_{g'}-1$
corners, all $X$-first, and by (P2)
$q_{g'}-1=\sum_t\mathbb{1}[p_{h_t}<u_t]$. Layer counting gives
$\#\{g':q_{g'}=i\}=\#\{g:p_g=i\}$, so $p$ and $q$ take values in the same
$[a,b]$ with both ends attained. If $a<b$, take $p_g=b$: $R$-edges in
layers $v<a$ contribute $a-1$, and the $R$-edge in layer $a$ has
$q_r\neq a$, $q_r\ge a$, contributing $1$, so $\Delta-b\ge a$; take
$q_{g'}=a$: $L$-edges in layers $u>b$ contribute $\Delta-b$ and the
$L$-edge in layer $b$ has $p_h<b$, contributing $1$, so
$a-1\ge\Delta-b+1$, a contradiction. If $a=b$, the $Z$-side identity gives
$\Delta=2a-1$, odd; but $\Delta=\wB+1$ is even.
\end{proof}

\begin{theorem}
\label{thm:S2}
Let $\mathrm{QC}(A,B)$ have distinct sums, $|A|=\wA\ge3$ odd and
$|B|=\wB$ odd, and suppose a proper schedule of depth $\Delta=\wA+\wB$ has
$a=(\wB+1)/2$. Then no such schedule exists.
\end{theorem}

\begin{proof}
Here $b=\wA+(\wB+1)/2$, $[a,b]$ has $\wA+1$ layers, and the shells
$S_-=[1,a-1]$, $S_+=[b+1,\Delta]$ have $(\wB-1)/2$ layers each. Each
$X$-check uses $a$ or $b$, hence both by support symmetry:
$P_g=[a,b]\setminus\{c(g)\}$, $V_g=S_-\cup S_+\cup\{c(g)\}$. In layers $a,b$
every $X$-check is on $L$, so every $L$-qubit has $X$-layers
$[a,b]\setminus\{z_h\}$ and every $R$-qubit $Z$-layers
$[a,b]\setminus\{v_r\}$, with the interior $R$-edge of $X_g$ landing where
$v_r=c(g)$; layer counting gives $\#\{h:z_h=i\}=\#\{g:c(g)=i\}$, so the
$z_h$ lie in $[c_{\min},c_{\max}]$. Counting $Z$-layers above each edge,
$F_L=\wA(a-1)+1+W$ with $W=\#\{p\in(a,b)\setminus\{c\}:z_{h_p}>p\}$ and
$F_R=\wA(a-1)+(b-c)$, so $F_L=F_R$ gives $W(g)=b-c(g)-1$. If
$c_{\min}=c_{\max}=c$ then $z_h=c$ for all $h$, $W=c-a-1$ and
$2c=\Delta+1$, impossible for even $\Delta$. Otherwise $c(g)=c_{\max}$
gives $W\ge c_{\min}-a$, hence $c_{\min}+c_{\max}\le\Delta$, and
$c(g)=c_{\min}$ gives $W\le c_{\max}-a-2$, hence
$c_{\min}+c_{\max}\ge\Delta+2$.
\end{proof}

The parity of $\wA$ cannot be dropped from Theorem~\ref{thm:S2}: the last
step needs $\Delta$ even. For $G=\mathbb{Z}_{12}$, $A=\{0,1\}$,
$B=\{0,4,8\}$ ($\wA=2$, $\wB=3$, $\Delta=5$) a proper depth-$5$ schedule
with $a=2$, $b=4$ exists, in which every $X$-check has $L$-layers $\{2,4\}$,
so $c(g)=3$ for all $g$ and $2c=6=\Delta+1$ (found by SAT and verified by
$\F$ simulation).

\begin{theorem}[the case $(2,5)$]
\label{thm:S3}
Let $|A|=|B|=3$ with distinct sums. No proper schedule of depth $6$ has
$(a,b)=(2,5)$.
\end{theorem}

\begin{proof}
In layers $2$ and $5$ every $X$-check has its edge on the $L$ block
($2,5\in P_g$ for all $g$, by support symmetry and $|P_g|=3$ inside
$[2,5]$), so every $L$-qubit has $X$-edges in layers $2$ and $5$ and,
since its $X$-layers lie in $[a,b]=[2,5]$, a third $X$-layer
$c_L\in\{3,4\}$; its $Z$-layers are the complement $\{1,6,7-c_L\}$. On the
$R$ block layers $2$ and $5$ carry no $X$-edge, so every $R$-qubit has
$Z$-edges in layers $2$ and $5$ and a third $Z$-layer $c_R\in\{3,4\}$ (the
$Z$-edges on $R$ lie in $[a,b]$); its $X$-layers are $\{1,6,7-c_R\}$. Fix
$X_g$ with $P_g=\{2,5,c\}$, $V_g=\{1,6,7-c\}$. Its layer-$c$ $L$-edge lies
on an $L$-qubit whose $X$-layers contain $c$, so $c_L=c$ there; its
layer-$(7-c)$ $R$-edge lies on an $R$-qubit whose $X$-layers contain
$7-c$, so $c_R=c$ there. Counting $Z$-layers above each edge,
\begin{align*}
F_L(g)&=|\{6,7-c_L\}\cap(2,6]|+|\{6\}|+|\{6,7-c\}\cap(c,6]|\\
      &=2+1+(1+\mathbb{1}[c=3]),\\
F_R(g)&=|\{2,5,c_R\}|+0+|\{2,5,c\}\cap(7-c,6]|\\
      &=3+0+(1+\mathbb{1}[c=4]),
\end{align*}
using $7-c_L\in\{3,4\}>2$ in the first term, $7-c>c$ iff $c=3$ in the
third, and $5>7-c\ge3>2$, $c>7-c$ iff $c=4$ in the last. So
$F_L(g)=4+\mathbb{1}[c=3]$ and $F_R(g)=4+\mathbb{1}[c=4]$ differ by one
for either value of $c$, contradicting $F_L=F_R$.
\end{proof}

\paragraph*{Proof of the rigidity step.}
Let $\wA=\wB=3$, the sums distinct, $\pi$ a proper schedule of depth $6$
with $(a,b)=(1,6)$, and $J=\{2,3,4,5\}$. Let $x_i$ be the number of
$X$-edges in layer $i$ on the $L$ block; as in the proof of support
symmetry, layer $i$ then carries $N-x_i$ $X$-edges and
$x_i$ $Z$-edges on the $R$ block.

\emph{Types.} An $X$-check $g$ with $1\in P_g$ has $6\in P_g$ by support
symmetry; call it an $E$-check and write
$P_g=\{1,6,c_g\}$, $c_g\in J$, $V_g=J\setminus\{c_g\}$. An $X$-check with
$1\notin P_g$ has $6\notin P_g$ by the converse half of that statement, so
$P_g\subset J$, $|P_g|=3$: call it an $I$-check and write
$P_g=J\setminus\{d_g\}$, $V_g=\{1,6,d_g\}$. Let $E$, $I$ be the two sets
and $e=|E|$. Since $a=1$, some $X$-check has an $L$-edge in layer $1$, so
$e=x_1\ge1$. On the $R$ block layer $1$ carries $x_1=e$ $Z$-edges, one
per $Z$-check that has an $R$-edge in layer $1$; applying support
symmetry to the dual code ($X_g\leftrightarrow Z_{-g}$,
$L\leftrightarrow R$), whose lowest $X$-on-$L$ layer is therefore $1$,
these $Z$-checks also have an $R$-edge in layer $6$. Every $X$-check has
exactly one $L$-edge of each label $s$ and one $R$-edge of each label
$t$; write $s_i(g)$ for the label of the $L$-edge of $X_g$ in layer
$i\in P_g$ and $t_i(g)$ for the label of its $R$-edge in layer
$i\in V_g$.

\emph{(T1)} Let $g\in E$ and let $t$ be any label, $r=R_{g+b_t}$ the
$R$-qubit of the $R$-edge of $X_g$ of label $t$, in layer $\nu_g(t)$.
Then the $Z$-edge at $r$ of label $s_1(g)$, which belongs to
$Z_{g+a_{s_1(g)}+b_t}$, lies in a layer above $\nu_g(t)$, and the
$Z$-edge at $r$ of label $s_6(g)$, belonging to $Z_{g+a_{s_6(g)}+b_t}$,
lies below $\nu_g(t)$. Proof: the $L$-edge of $X_g$ in layer $1$, at
$h=L_{g+a_{s_1(g)}}$, has three corners, with the $Z$-edges at $h$ of
labels $t=1,2,3$; all are $X$-first. By (P1) the corner with label $t$
is mapped by $\Phi$ to the corner at $r=R_{g+b_t}$ formed by the $R$-edge
of $X_g$ of label $t$ and the $Z$-edge at $r$ of label $s_1(g)$, both
edges of the same pair $(X_g,Z_{g+a_{s_1(g)}+b_t})$; properness makes
this corner $X$-first, which is the first claim. The layer-$6$ $L$-edge
has all corners $Z$-first, and the same map gives the second claim.

\emph{(T2)} Let $g\in I$ and $s$ any label, $h=L_{g+a_s}$ the $L$-qubit
of the $L$-edge of $X_g$ of label $s$. Then the $Z$-edge at $h$ of label
$t_1(g)$ lies above that $L$-edge and the $Z$-edge at $h$ of label
$t_6(g)$ lies below it. Proof: the $R$-edge of $X_g$ in layer $1$ has
all three corners $X$-first; by (P1) applied in the reverse direction,
the corner with $Z$-label $s$ there is the image of the corner at $h$
formed by the $L$-edge of label $s$ and the $Z$-edge of label $t_1(g)$,
so that corner is $X$-first. Layer $6$ likewise.

\emph{Step 1: every $L$-qubit with an $X$-edge in layer $5$ has a
$Z$-edge in layer $6$.} Let $h$ carry the layer-$5$ $L$-edge of $X_g$,
of label $s^\ast=s_5(g)$. If $g\in I$, (T2) places the $Z$-edge of label
$t_1(g)$ at $h$ above layer $5$, hence in layer $6$. If $g\in E$, then
$c_g=5$ and $V_g=\{2,3,4\}$; let $t^\ast=t_2(g)$ and $r=R_{g+b_{t^\ast}}$.
The corner at $h$ with labels $(s^\ast,t^\ast)$ is mapped by $\Phi$ to the
corner at $r$ formed by the $R$-edge of $X_g$ in layer $2$ and the
$Z$-edge at $r$ of label $s^\ast$, both belonging to the pair
$(X_g,Z_{g+a_{s^\ast}+b_{t^\ast}})$. By (T1) the $Z$-edge at $r$ of label
$s_6(g)$ lies below layer $2$, i.e.\ in layer $1$. Since $s_6(g)\neq
s^\ast$ (they label different $L$-edges of $X_g$), the $Z$-edges at $r$
of labels $s^\ast$ and $s_6(g)$ are different edges, and $r$ has exactly
one edge in layer $1$; so the $Z$-edge of label $s^\ast$ at $r$ is not in
layer $1$, hence not below layer $2$, hence the corner at $r$ is
$X$-first. By properness the corner at $h$ is $X$-first: the $Z$-edge at
$h$ of label $t^\ast$ lies above layer $5$, in layer $6$. Distinct sums
enter through (P1): the corner at $h$ has a unique partner, which is the
corner at $r$ just described.

\emph{Step 2: every $R$-qubit with an $X$-edge in layer $5$ has a
$Z$-edge in layer $6$.} Let $r$ carry the layer-$5$ $R$-edge of $X_g$, of
label $t^\ast=t_5(g)$. If $g\in E$ (so $c_g\neq5$), (T1) places the
$Z$-edge of label $s_1(g)$ at $r$ above layer $5$, in layer $6$. If
$g\in I$, then $d_g=5$ and $P_g=\{2,3,4\}$; let $s^\ast=s_2(g)$ and
$h=L_{g+a_{s^\ast}}$. By (T2) the $Z$-edge at $h$ of label $t_6(g)$ is in
layer $1$ and the $L$-edge of $X_g$ occupies layer $2$ at $h$; the
$Z$-edge at $h$ of label $t^\ast\neq t_6(g)$ is a different edge, so it
lies above layer $2$, and the corner at $h$ with labels
$(s^\ast,t^\ast)$ is $X$-first. Its image under $\Phi$ is the corner at
$r=R_{g+b_{t^\ast}}$ formed by the layer-$5$ $R$-edge and the $Z$-edge at
$r$ of label $s^\ast$; by properness it is $X$-first, so that $Z$-edge is
in layer $6$.

\emph{Step 3: cardinalities.} Let $e_5=\#\{g\in E:c_g=5\}$ and
$\iota_5=\#\{g\in I:d_g=5\}$. The $X$-edges in layer $5$ on the $L$
block come from $E$-checks with $c_g=5$ and from $I$-checks with
$d_g\neq5$, so $x_5=e_5+(N-e-\iota_5)$, on as many distinct $L$-qubits.
The $L$-qubits with a $Z$-edge in layer $6$ number $N-x_6=N-e$, since
$x_6=|E|$. Step~1 gives an inclusion of the first set of qubits in the
second, hence
\begin{equation}
e_5+(N-e-\iota_5)\;\le\;N-e .
\end{equation}
The $X$-edges in layer $5$ on the $R$ block come from $E$-checks with
$c_g\neq5$ and $I$-checks with $d_g=5$, so there are $(e-e_5)+\iota_5$ of
them; the $R$-qubits with a $Z$-edge in layer $6$ number $x_6=e$, and
Step~2 gives
\begin{equation}
(e-e_5)+\iota_5\;\le\;e .
\end{equation}
The two inequalities read $e_5\le\iota_5$ and $\iota_5\le e_5$, so
$e_5=\iota_5$ and both inclusions are between sets of equal size, hence
equalities: on the $L$ block a qubit has an $X$-edge in layer $5$ if and
only if it has a $Z$-edge in layer $6$, and likewise on the $R$ block.
At any data qubit each layer holds exactly one edge, $X$ or $Z$. If the
layer-$5$ edge is $X$, the layer-$6$ edge is $Z$; if the layer-$5$ edge
is $Z$, the qubit has no $X$-edge in layer $5$, hence no $Z$-edge in
layer $6$, so its layer-$6$ edge is $X$. Thus the pair $\{5,6\}$ holds one
$X$- and one $Z$-edge at every data qubit.

\emph{Step 4: the other pairs.} Layer reversal $i\mapsto7-i$ maps $\pi$
to a proper schedule with $(a,b)=(1,6)$, maps $E$-checks to $E$-checks
($\{1,6,c\}\mapsto\{1,6,7-c\}$) and $I$-checks to $I$-checks, and
preserves distinct sums; Steps~1--3 applied to the reversed schedule
show that $\{1,2\}$ holds one $X$- and one $Z$-edge at every data qubit
of $\pi$. Each data qubit has three $X$-edges and three $Z$-edges; with
one of each in $\{1,2\}$ and one of each in $\{5,6\}$, the remaining
$X$-edge and $Z$-edge occupy $\{3,4\}$. \hfill$\square$

\paragraph*{Proof of Eq.~\eqref{eq:pairsum}.}
Under the pair structure the $Z$-edges above layer $i$ are one per pair
$\{2j-1,2j\}$ with $2j-1>i$, plus the partner within the pair of $i$ if $i$
is odd, so $\gamma(i)=w-\lceil i/2\rceil+\mathbb{1}[i\text{ odd}]$ and
$\sum_{i=1}^{2w}\gamma(i)=\sum_{j=1}^w(2w-2j+1)=w^2$. \hfill$\square$

\FloatBarrier

\section{Depth six: class schedules, lifting and templates}
\label{si:six}

\paragraph*{Lifting of class schedules.}
Conflicts: the six edges of $X_g$ receive the values
$T(\varepsilon(g),\cdot,\cdot)$, distinct by (i); the edges at $L_h$ come
from $X_{h-a_s}$ of class $\varepsilon(h)-\alpha_s$ with labels $s$ and from
$Z_{h+b_t}$ of class $\varepsilon(h)+\beta_t$ with labels $t$, distinct by
(ii); $R$-qubits and $Z$-checks likewise. Properness: a pair with
$g'-g=\delta\neq\delta_0$ has one label pair, $Z_{g'}$ has class
$\varepsilon(g)+\alpha_s+\beta_t$, and (iii) gives equal orientations; for
$\delta=\delta_0$ the two collision pairs share the $Z$-check of class
$\varepsilon(g)+\alpha_s+\beta_t=\varepsilon(g)+\alpha_{s'}+\beta_{t'}$
and (iv) gives an even sum. Coincidences among other projected sums play
no role, since every constraint of the code is one of (i)--(iv) for the
labelled pair that defines it. \hfill$\square$

\paragraph*{Normalisation of keys.}
Admissibility of a key is invariant under $(\alpha,\beta)\mapsto
(\alpha+u,\beta+v)$ for any $u,v\in\Zq$: if $T$ is a class schedule for
$(\alpha,\beta)$, then $T'$ defined by $T'(c,\cdot,X\cdot)=T(c,\cdot,X\cdot)$
and $T'(z,\cdot,Z\cdot)=T(z-u-v,\cdot,Z\cdot)$ is a class schedule for
$(\alpha+u,\beta+v)$, because the $L$-qubit of class $c$ in the shifted
system collects $X$-edges from classes $c-\alpha_s-u$ and $Z$-edges from
classes $c+\beta_t+v$, which is the original $L$-qubit of class $c-u$
after relabelling $Z$-classes by $z\mapsto z-u-v$, and the pair
constraints (iii)--(iv) pair $X$-class $c$ with $Z$-class
$c+\alpha_s+\beta_t+u+v$, which the same relabelling sends to the original
pairing; $R$-qubits are handled identically. We therefore fix
$\alpha_1=\beta_1=0$. A designated collision is an unordered pair
$\{(s,t),(s',t')\}$ with $s\neq s'$, $t\neq t'$: three choices of
$\{s,s'\}$, three of $\{t,t'\}$, and two ways to pair them, $18$ in all.
For each, the equality $\alpha_s+\beta_t=\alpha_{s'}+\beta_{t'}$ removes
one of the four free coordinates $(\alpha_2,\alpha_3,\beta_2,\beta_3)$,
leaving $q^3$ keys, hence $18q^3$ keys per modulus, the counts of
Table~\ref{tab:templates-count}. In that enumeration keys related by
label permutations, class translation or layer reversal are counted
separately; the reduction used for $q\ge16$ is described in Methods and
below. The enumerator iterates over the $18$
collisions and the $q^3$ solutions of the equality in lexicographic
order, and its output is checked to contain each normalised key exactly
once. In the proof of Theorem~\ref{thm:six} the key
$(\varepsilon(a_s);\varepsilon(b_t))$ of a code need not be normalised;
the invariance above reduces it to one that is.

\begin{figure}[!htbp]
\centering
\includegraphics[width=\SIcolumnfigurewidth]{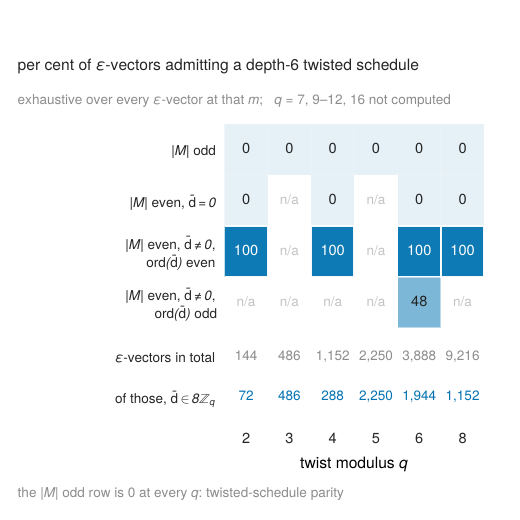}
\caption{\textbf{Class schedules by modulus.} For $q\in\{2,3,4,5,6,8\}$,
stacked counts of keys that admit a class schedule, do not admit one with
$\colbar\neq0$, and have $\colbar=0$.}
\label{fig:templates}
\end{figure}

\paragraph*{Enumeration.}
For $q\in\{2,3,4,5,6,8\}$, each designated pair and each normalised key,
the constraints (i)--(iv) were encoded on the class system and decided by
SAT; one class schedule was stored per admissible key, checked against
(i)--(iv) and by $\F$ simulation of its pullback to a realising class-C1
code (Methods). $q=7$ was not run; odd $q$ is excluded by
Eq.~\eqref{eq:twisted}. Table~\ref{tab:templates-count} and
Fig.~\ref{fig:templates} give the counts. Three regularities are observed
without exception: no admissible key has $\colbar=0$; every admissible key
has $|M|$ even, as Eq.~\eqref{eq:twisted} requires, which is why $q=3$
and $q=5$ admit nothing; and for $q\in\{2,4,8\}$ every key with
$\colbar\neq0$ is admissible, which is the first half of the template
statement of Section~\ref{sec:six}. The refinement that a key is admissible
exactly when $\operatorname{ord}(\colbar)$ is even is false: at $q=6$,
$468$ admissible keys have $\operatorname{ord}(\colbar)=3$.

\begin{table}[!htbp]
\caption{\textbf{Class schedules by modulus.} Normalised keys with one
designated collision, keys admitting a class schedule, and $|M|$ among
them; $q=7$ was not enumerated (odd, excluded by
Eq.~\eqref{eq:twisted}).}
\label{tab:templates-count}
\begin{ruledtabular}
\begin{tabular}{cccl}
$q$ & keys & admissible & $|M|$ among admissible \\
\colrule
2 & 144  & 72   & $\{2\}$ \\
3 & 486  & 0    & -- \\
4 & 1152 & 864  & $\{2,4\}$ \\
5 & 2250 & 0    & -- \\
6 & 3888 & 2412 & $\{2,6\}$ \\
8 & 9216 & 8064 & $\{2,4,8\}$ \\
\end{tabular}
\end{ruledtabular}
\end{table}

\FloatBarrier

\paragraph*{Keys with $\colbar=q/2$ for $q\ge16$.}
By the reduction in Methods the keys with $\colbar=q/2$ are
$\kappa_q(a,b)=(0,q/2,a;0,q/2,b)$, $(a,b)\in\Zq^2$, modulo
$(a,b)\sim(b,a)\sim(\omega a,\omega b)\sim(a+q/2,b+q/2)$, and every such
key reduces to a member of the family $\kappa_{q'}(1,b)$ at a modulus
$q'\mid q$. Table~\ref{tab:family} gives the numbers of representatives
and the outcome of the SAT decisions. For $q\le32$ all representatives
were decided without assumptions ($39$ at $q=32$ in $17$ minutes); single
instances at $q=64$ take up to half an hour, and the family was decided
there and beyond under the half-period symmetry described next, which the
solver handles in seconds.

\paragraph*{A universal half-period symmetry.}
Imposing $T(c+q/2,e)=\sigma(T(c,e))$ for a single solver-chosen
permutation $\sigma$ of the six layers is satisfiable only for $b=5$ at
$q=8$ and $b=9$ at $q=16$. Imposing it with a permutation
$\sigma_{\mathrm{type}(e)}$ depending on the edge type ($XL$, $XR$, $ZL$,
$ZR$) is satisfiable for every $b$ at $q=8,16,32,64,128$, with
fixed-point-free involutions in every solution inspected. One assignment
serves all $b$ at once: a
joint SAT instance over all $b$ at $q=16$ with the permutation depending
only on the block gives $\sigma_L=(13)(24)(56)$ on both $L$-edge types and
$\sigma_R=(12)(36)(45)$ on both $R$-edge types, and with these two
involutions imposed every member of the family at $q=2,4,\dots,512$ admits
a class schedule; this is the computation behind the templates of
Section~\ref{sec:six} for $q\ge64$ (all $512$ keys at $q=512$ in
twelve minutes). A second pair, $\sigma_L=(14)(23)(56)$,
$\sigma_R=(12)(35)(46)$, serves all $b$ for $q\le16$; a permutation shared
by $L$ and $R$, or one shared by $XL$ and $ZR$ and another by $XR$ and
$ZL$, does not exist at $q=8$ or $q=16$. The symmetry cannot be refined:
requiring in addition $T(c+q/4,e)=\sigma'_{\mathrm{type}(e)}(T(c,e))$ is
unsatisfiable for every $b$ at $q=8$ and $q=16$, and a schedule depending
on $c$ only through $(c\bmod2,\lfloor4c/q\rfloor)$ exists for $4$ of the
$16$ values of $b$ at $q=16$, one depending only on
$(c\bmod4,\mathbb{1}[c\ge q/2])$ for $2$ of them. An analytic template
therefore has to prescribe the layers on the classes $c<q/2$; the
half-period symmetry then determines the rest.

\paragraph*{The $2$-height homomorphism.}
Put $q=2^{h+1}$ and $\bar G=G/qG\cong\bigoplus_i\mathbb{Z}_{2^{e_i}}$ with
$e_i\le h+1$. The image $\bar{\col}$ of $\col$ in $\bar G$ is non-zero
because $\col\notin qG$, and $2\bar{\col}=0$ because $\col\in2^hG$; hence
$\bar{\col}=\sum_ic_i$ with $c_i\in\{0,2^{e_i-1}\}$ and some
$c_{i_0}\neq0$. Composing $G\to\bar G\to\mathbb{Z}_{2^{e_{i_0}}}$ with the
injection $x\mapsto2^{h+1-e_{i_0}}x$ into $\Zq$ sends $\col$ to
$2^{h+1-e_{i_0}}\cdot2^{e_{i_0}-1}=2^h$. \hfill$\square$

\begin{table}[!htbp]
\caption{\textbf{Keys with $\colbar=q/2$.} Number of keys
$\kappa_q(a,b)$, number of representatives under the three symmetries, and
how the keys were decided: all representatives by unconstrained SAT
($q\le32$), or every member of the family~\eqref{eq:family} under the
half-period symmetry ($q\ge64$; also done for $q\le32$).}
\label{tab:family}
\begin{ruledtabular}
\begin{tabular}{cccl}
$q$ & keys & reps. & decided \\
\colrule
2   & 4     & 2   & all, SAT \\
4   & 16    & 5   & all, SAT \\
8   & 64    & 11  & all, SAT \\
16  & 256   & 21  & all, SAT \\
32  & 1024  & 39  & all, SAT \\
64  & 4096  & 73  & family with symmetry, SAT \\
128 & 16384 & 139 & family with symmetry, SAT \\
256 & 65536 & 269 & family with symmetry, SAT \\
512 & 262144 & 527 & family with symmetry, SAT \\
\end{tabular}
\end{ruledtabular}
\end{table}

\FloatBarrier

\paragraph*{The $q=2$ template.}
The template displayed in Fig.~\ref{fig:mechanism} is the admissible
$q=2$ key $(0,1,0;0,1,0)$ with collision $\{(1,1),(2,2)\}$; its $24$ layer
values are listed in Table~\ref{tab:template-m2}. Pulled back along
$\varepsilon(g)=g\bmod2$ on $G=\mathbb{Z}_{14}$ with $A=\{0,1,10\}$,
$B=\{0,13,2\}$, it is a proper depth-$6$ schedule of a connected class-C1
code with parameters $[[28,6,4]]$ (the distance is exact: no logical
operator of weight below $4$ exists, by exhaustive search), verified by
$\F$ simulation.
Fifteen further pullbacks, across $q\in\{2,4,6,8\}$ and the groups
$\mathbb{Z}_{10}$, $\mathbb{Z}_{12}$, $\mathbb{Z}_{16}$, $\mathbb{Z}_{18}$,
$\mathbb{Z}_{24}$, $\mathbb{Z}_{32}$, $\mathbb{Z}_2\times\mathbb{Z}_6$ and
$\mathbb{Z}_2\times\mathbb{Z}_8$, with projections merging three to six
further sums, all pass.

\begin{table}[!htbp]
\caption{\textbf{The $q=2$ class schedule of Fig.~\ref{fig:mechanism}.}
Layers by class $c$, edge type and label, for the key $(0,1,0;0,1,0)$ with
collision $\{(1,1),(2,2)\}$; pulled back along $\varepsilon(g)=g\bmod2$ on
$\mathbb{Z}_{14}$ with $A=\{0,1,10\}$, $B=\{0,13,2\}$, it is a proper
depth-$6$ schedule of a $[[28,6,4]]$ code, verified by $\F$
simulation.}
\label{tab:template-m2}
\begin{ruledtabular}
\begin{tabular}{llccc}
class & edge type & label 1 & label 2 & label 3 \\
\colrule
$c=0$ & $X$-check, $L$-edges & 3 & 6 & 1 \\
      & $X$-check, $R$-edges & 4 & 2 & 5 \\
      & $Z$-check, $L$-edges & 5 & 3 & 2 \\
      & $Z$-check, $R$-edges & 1 & 4 & 6 \\
\colrule
$c=1$ & $X$-check, $L$-edges & 5 & 4 & 2 \\
      & $X$-check, $R$-edges & 1 & 3 & 6 \\
      & $Z$-check, $L$-edges & 4 & 6 & 1 \\
      & $Z$-check, $R$-edges & 3 & 2 & 5 \\
\end{tabular}
\end{ruledtabular}
\end{table}

\FloatBarrier

\paragraph*{The $2$-height in the data set.}
A homomorphism $G\to\mathbb{Z}_{2^j}$ nonzero on $\col$ exists iff
$\col\notin2^jG$, since $\mathrm{Hom}(G/2^jG,\mathbb{Z}_{2^j})$ separates
the points of $G/2^jG$; restricting Theorem~\ref{thm:six} to $q\le8$ would
therefore miss even-order elements inside $8G$, such as $\col=8$ in
$\mathbb{Z}_{32}$, which has $2$-height $3$ and is covered through $q=16$.
Table~\ref{tab:twoheight} splits the $598$ class-C1 instances by the
$2$-height; the two instances with $h(\col)=3$, both on $\mathbb{Z}_{32}$,
were also found to admit $\mathbb{Z}_{16}$-twisted proper schedules by
direct search, verified by simulation.

\begin{table}[!htbp]
\caption{\textbf{Class-C1 instances by the $2$-height of $\col$.} The two
instances with $h(\col)=3$ also admit $\mathbb{Z}_{16}$-twisted witnesses
found by direct search.}
\label{tab:twoheight}
\begin{ruledtabular}
\begin{tabular}{lccl}
condition on $\col$ & instances & witness & verdict \\
\colrule
$h(\col)\le2$ ($\col\notin8G$) & 263 & template, $q\le8$ & all SAT \\
$h(\col)=3$ ($\col\in8G$) & 2 & template, $q=16$ & both SAT \\
$\operatorname{ord}(\col)$ odd & 333 & -- & Table~\ref{tab:c1} \\
\end{tabular}
\end{ruledtabular}
\end{table}

\FloatBarrier

\section{Sharpness data}
\label{si:sharpness}

\paragraph*{Overlap taxonomy.}
Let $n_4$ and $n_6$ be the numbers of check pairs sharing exactly four and
six qubits; then $n_4+3n_6=N\sum_\delta\binom{S(\delta)}{2}$, and D is
$(0,0)$, C1 is $(N,0)$, C2 is $(2N,0)$ or $(0,N)$, any other pattern
``other''. The instance $G=\mathbb{Z}_8$, $A=\{0,1,4\}$, $B=\{0,7,4\}$ has
multiplicities $(1,1,1,1,2,3)$, hence $n_4=n_6=N$, and is ``other''
although exactly one $\delta$ has $S(\delta)=2$; this is why C1 requires
$S\le1$ elsewhere. The classifier reads the label from $H_XH_Z^{\mathsf T}$
and agrees with the sums on every classified instance and on this example.

\paragraph*{Small groups.}
The sweep over even orders $2$--$8$ is exhaustive over pairs with $0\in A$,
$0\in B$. Under $(A,B)\sim(\phi(A)+t_1,\phi(B)+t_2)$ for
$\phi\in\mathrm{Aut}(G)$ and $(A,B)\sim(B,A)$, the $1441$ pairs fall into
$33$ orbits, $32$ satisfiable and one unsatisfiable, the latter being the
$\mathbb{Z}_6$ instance $A=B=\{0,2,4\}$ alone (Tables~\ref{tab:sweep},
\ref{tab:sweep-census}); no orbit contains both outcomes. That instance has
components of six data qubits with $\sum_vr_vs_v=54\equiv2\pmod4$, so
Theorem~\ref{thm:mod4} excludes it, and its DRAT certificate confirms the
theorem. The odd orders $3$, $5$, $7$, $9$ were swept up to the same
equivalence: $1830$ pairs in $27$ orbits, decided under a $20$-second cap
(Table~\ref{tab:sweep-odd}). The three distinct-sums orbits, all of order
$9$ (recorded as O9-a, O9-b, O9-c), are unsatisfiable; of the $24$ orbits
with coinciding sums, $2$ are excluded by Theorem~\ref{thm:mod4} and $22$
reached the cap, so this part of the sweep is inconclusive rather than
negative.

\begin{table}[!htbp]
\caption{\textbf{Exhaustive sweep over even $|G|\le8$.}}
\label{tab:sweep}
\begin{ruledtabular}
\begin{tabular}{lc}
 & count \\
\colrule
normalised instances & 1441 \\
proper depth-$6$ schedule found and verified & 1440 \\
unsatisfiable, excluded by Theorem~\ref{thm:mod4} & 1 \\
\end{tabular}
\end{ruledtabular}
\end{table}

\begin{table}[!htbp]
\caption{\textbf{Census of the even-order sweep.} Orbits under
automorphisms, translations and block swap, and outcomes of the
representatives.}
\label{tab:sweep-census}
\begin{ruledtabular}
\begin{tabular}{lrrrrr}
group & $|\mathrm{Aut}(G)|$ & pairs & orbits & SAT & UNSAT \\
\colrule
$\mathbb{Z}_4$            & 2   & 9    & 1  & 1  & 0 \\
$\mathbb{Z}_6$            & 2   & 100  & 7  & 6  & 1 \\
$\mathbb{Z}_8$            & 4   & 441  & 13 & 13 & 0 \\
$\mathbb{Z}_2\times\mathbb{Z}_2$ & 6 & 9  & 1  & 1  & 0 \\
$\mathbb{Z}_2\times\mathbb{Z}_4$ & 8 & 441 & 9 & 9  & 0 \\
$\mathbb{Z}_2^3$          & 168 & 441  & 2  & 2  & 0 \\
\colrule
total                     &     & 1441 & 33 & 32 & 1 \\
\end{tabular}
\end{ruledtabular}
\end{table}

\begin{table}[!htbp]
\caption{\textbf{Odd-order sweep.} Orbits under the same equivalence and
outcomes of the representatives at a $20$-second cap.}
\label{tab:sweep-odd}
\begin{ruledtabular}
\begin{tabular}{lrrrrr}
group & orbits & pairs & SAT & UNSAT & capped \\
\colrule
$\mathbb{Z}_3$ & 1  & 1   & 0 & 0 & 1  \\
$\mathbb{Z}_5$ & 2  & 36  & 0 & 0 & 2  \\
$\mathbb{Z}_7$ & 5  & 225 & 0 & 1 & 4  \\
$\mathbb{Z}_9$ & 12 & 784 & 0 & 2 & 10 \\
$\mathbb{Z}_3\times\mathbb{Z}_3$ & 7 & 784 & 0 & 2 & 5 \\
\colrule
total & 27 & 1830 & 0 & 5 & 22 \\
\end{tabular}
\end{ruledtabular}
\end{table}

\begin{figure}[!htbp]
\centering
\includegraphics[width=\SIcolumnfigurewidth]{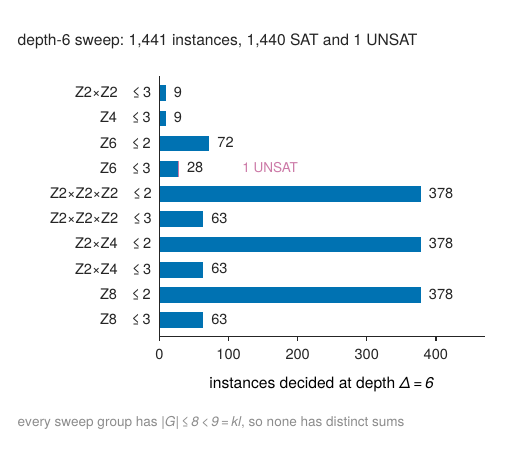}
\caption{\textbf{Depth-$6$ outcomes on groups of order at most nine.}
Per-group counts of normalised instances by outcome; odd orders are swept
up to equivalence.}
\label{fig:sweep}
\end{figure}

\FloatBarrier

\paragraph*{Instance generation for the sharpness map.}
The sampled instances of Table~\ref{tab:h2} are not enumerated. For cyclic
groups of order $N\in\{10,\dots,24,28,30,32,36\}$, $A$ and $B$ were each
drawn as $\{0\}$ plus two elements uniformly from $\{1,\dots,N-1\}$ (seed
$7$), classified, and kept up to a fixed quota per $(N,\text{class})$; the
subgroup instances $A=B=\{0,N/3,2N/3\}$ were added for $N\in\{12,18\}$.
Duplicates were removed by literal equality only. Under the equivalence
above, the $664$ instances fall into $475$ orbits ($416$ of $598$ in C1,
$30$ of $35$ in D, $26$ of $28$ in C2, all $3$ subgroup-type), so counts
over the $664$ are weighted by orbit size, most heavily at small $N$;
per-group counts are in the repository.

\paragraph*{Sampled groups and the seven codes of Ref.~\cite{Bravyi2024}.}
Table~\ref{tab:h2} lists the $664$ instances by class; the $\mathbb{Z}_6$
instance is the only one of order below $10$. Every distinct-sums instance
is unsatisfiable, as Theorem~\ref{thm:main} requires; the smallest and
largest, at $|G|=10$ and $24$, and the seven codes of
Ref.~\cite{Bravyi2024} carry DRAT certificates. Of the class-C2 instances
$27$ admit depth $6$ and one is excluded by Theorem~\ref{thm:mod4}; both
subgroup instances, in which $A\cup B$ generates a proper subgroup, are
excluded by Theorem~\ref{thm:mod4} applied to their components.

\begin{table}[!htbp]
\caption{\textbf{The $664$ depth-$6$ instances by class.} S: proper
depth-$6$ schedule found and verified; U$_{\rm t}$: excluded by
Theorem~\ref{thm:mod4} or~\ref{thm:main}; U$_{\rm s}$: refuted by the
solver (of the $62$, six carry DRAT certificates and $56$ have no proof
logging); O: solver capped, no theorem applies.}
\label{tab:h2}
\begin{ruledtabular}
\begin{tabular}{lrrrrr}
class & $n$ & S & U$_{\rm t}$ & U$_{\rm s}$ & O \\
\colrule
D ($|G|=10$--$24$)  & 28  & 0   & 28  & 0  & 0  \\
D (Ref.~\cite{Bravyi2024}) & 7 & 0 & 7  & 0  & 0  \\
C1                  & 598 & 295 & 211 & 62 & 30 \\
C2                  & 28  & 27  & 1   & 0  & 0  \\
subgroup type       & 2   & 0   & 2   & 0  & 0  \\
$\mathbb{Z}_6$, $A=B=\{0,2,4\}$ & 1 & 0 & 1 & 0 & 0 \\
\colrule
total               & 664 & 322 & 250 & 62 & 30 \\
\end{tabular}
\end{ruledtabular}
\end{table}

\begin{figure}[!htbp]
\centering
\begin{minipage}[t]{0.48\textwidth}
\vspace{0pt}
\centering
\includegraphics[width=\linewidth]{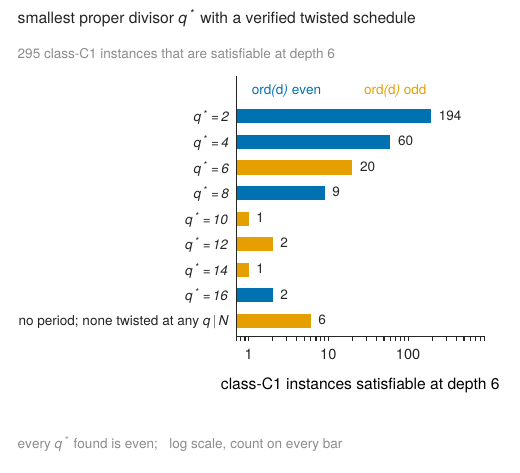}
\caption{\textbf{Periods of satisfiable class-C1 schedules.} Smallest
modulus $q^\ast$ under which a found proper schedule is twisted, split by
the parity of $\operatorname{ord}(\col)$; the last category collects the
six instances whose found schedule has no proper period and for which no
twisted schedule exists at any $q\mid N$.}
\label{fig:periods}
\end{minipage}\hfill
\begin{minipage}[t]{0.48\textwidth}
\vspace{0pt}
\centering
\includegraphics[width=\linewidth]{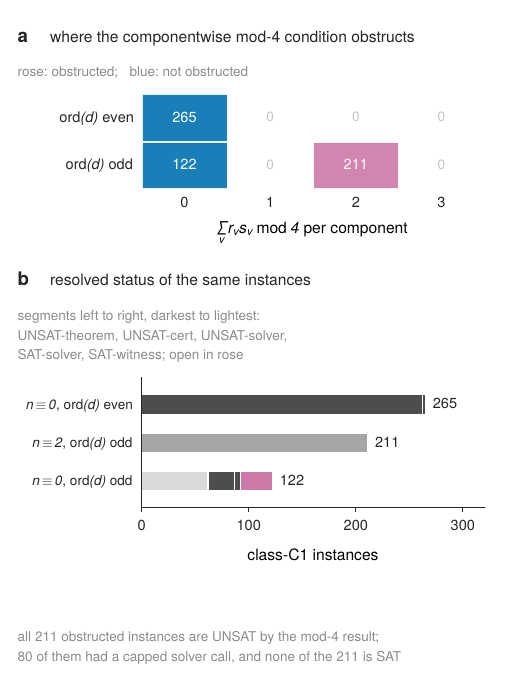}
\caption{\textbf{Theorem~\ref{thm:mod4} across the sweep.} Instances by
$\sum_vr_vs_v\bmod4$ per component, which for $r_v=s_v=3$ equals
$9n/2\bmod4$ with $n$ the component size, and by depth-$6$ outcome.}
\label{fig:thm1profile}
\end{minipage}
\end{figure}

\paragraph*{Class C1.}
Table~\ref{tab:c1} gives the $598$ class-C1 instances by the parity of
$\operatorname{ord}(\col)$, the $2$-height of $\col$,
Theorem~\ref{thm:mod4}, and outcome. All $265$ even-order instances admit
depth $6$ by Theorem~\ref{thm:six} ($h(\col)\le3$ throughout): $262$ were
also found by the solver, one by a template witness beyond the cap, and
the two with $h(\col)=3$ by $\mathbb{Z}_{16}$-twisted witnesses as well as
by the $q=16$ templates. Of the $333$ odd-order instances $211$ are
excluded by Theorem~\ref{thm:mod4}, the solver confirming $131$ and
reaching its cap on $80$; of the remaining $122$, $30$ admit depth $6$,
$62$ are refuted by the solver ($6$ with DRAT certificates, $56$ without
proof logging), and $30$ are open. Fig.~\ref{fig:periods} shows the
smallest modulus $q^\ast$ under which a found proper schedule is twisted:
$24$ of the $30$ satisfiable odd-order instances have an even
$q^\ast\in\{6,10,12,14\}$, and for the other $6$ the found schedule has no
proper period and the search for a twisted schedule over every $q\mid N$
completed without a cap and found none, so those codes admit no periodic
proper schedule. Fig.~\ref{fig:thm1profile} shows how
Theorem~\ref{thm:mod4} distributes over the sweep.

\begin{table}[!htbp]
\caption{\textbf{Class-C1 instances.} obs.: Theorem~\ref{thm:mod4}
obstructs; S$_{\rm s}$ satisfiable by solver; S$_{\rm w}$ by twisted
witness; U unsatisfiable (in the last row, $6$ with DRAT certificates and
$56$ solver-reported); O open. In the obstructed row the solver confirmed
$131$ and reached its cap on $80$.}
\label{tab:c1}
\begin{ruledtabular}
\begin{tabular}{lcrrrrr}
$\col$ & obs. & $n$ & S$_{\rm s}$ & S$_{\rm w}$ & U & O \\
\colrule
even, $h(\col)\le2$ & no  & 263 & 262 & 1 & 0   & 0  \\
even, $h(\col)=3$   & no  & 2   & 0   & 2 & 0   & 0  \\
odd              & yes & 211 & 0   & 0 & 211 & 0  \\
odd              & no  & 122 & 25  & 5 & 62  & 30 \\
\colrule
total            &     & 598 & 287 & 8 & 273 & 30 \\
\end{tabular}
\end{ruledtabular}
\end{table}

\FloatBarrier

\paragraph*{Coverage.}
Table~\ref{tab:coverage} partitions the $664$ instances by mathematical
and solver status. Mathematical status takes precedence: an instance
excluded by Theorem~\ref{thm:mod4} or~\ref{thm:main} counts as
unsatisfiable whether or not the solver finished, and ten of them also
carry DRAT certificates; six further instances not covered by a theorem
are certified; the eight capped satisfiable instances, three with
$\operatorname{ord}(\col)$ even and five odd, are settled by twisted
witnesses.

\begin{table}[!htbp]
\caption{\textbf{Coverage of the $664$ instances.} Rows: mathematical
status; columns: solver status.}
\label{tab:coverage}
\begin{ruledtabular}
\begin{tabular}{lcccc}
status & unsat & sat & capped & total \\
\colrule
unsatisfiable by theorem      & 170 & 0   & 80  & 250 \\
unsatisfiable by certificate  & 6   & 0   & 0   & 6   \\
unsatisfiable by solver only  & 56  & 0   & 0   & 56  \\
open                          & 0   & 0   & 30  & 30  \\
satisfiable (verified)        & 0   & 314 & 8   & 322 \\
\colrule
total                         & 232 & 314 & 118 & 664 \\
\end{tabular}
\end{ruledtabular}
\end{table}

\FloatBarrier

\section{Certificates and validation}
\label{si:certs}

\paragraph*{Encoding.}
Variables $x_{e,i}$ per Tanner edge $e$ and layer $i\in[D]$; exactly-one
clauses per edge; at-most-one clauses for every Tanner vertex (data qubits
and check ancillas) and layer; for each overlapping pair,
Eq.~\eqref{eq:parity} with the orientation of a corner $(q,e,e')$
represented by a Tseitin variable for
$\bigvee_{i<j}(x_{e,i}\wedge x_{e',j})$ and the parity by Tseitin variables
for the partial sums; $10^3$--$10^4$ variables for the instances here.
Class schedules use the same clauses on the class system with (i)--(iv)
in place of the code's constraints. The template encoder imposes
(iii)--(iv) per labelled pair, never regrouped by projected sum. An early
version added a symmetry-breaking clause that fixes the orientation of one
designated corner $(q_0,e_0,e_0')$ to $X$-first, i.e.\
$\bigvee_{i<j}(x_{e_0,i}\wedge x_{e_0',j})$. Layer reversal maps proper
schedules to proper schedules and flips every orientation, so every proper
schedule or its reversal satisfies the clause; the augmented formula is
therefore unsatisfiable if and only if the original is, and a certificate
for the augmented formula proves unsatisfiability of the original.
Certificates for the seven codes of Ref.~\cite{Bravyi2024} were
regenerated without the clause; the $\mathbb{Z}_6$ certificate and the two
class-D certificates retain it.

\begin{figure}[!htbp]
\centering
\includegraphics[width=\SIwidefigurewidth]{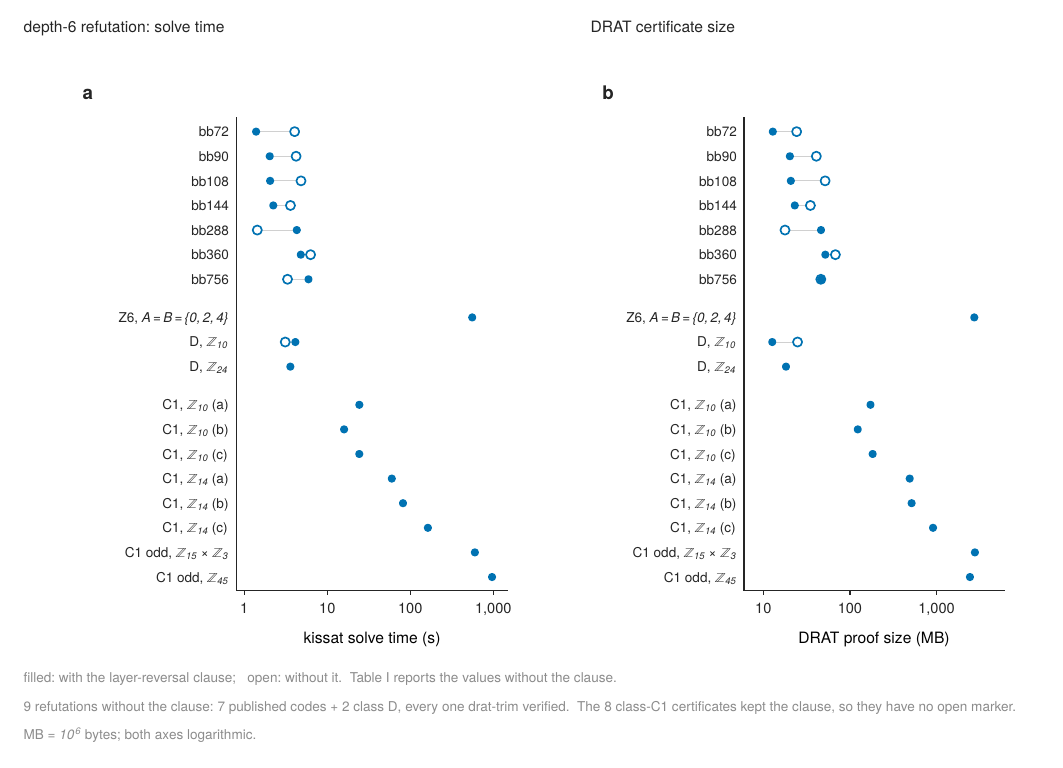}
\caption{\textbf{Certificates.} Solve time and DRAT proof size for each
certified instance, filled markers with the layer-reversal clause and open
markers without; Table~\ref{tab:certificates} reports the values without
the clause.}
\label{fig:certs}
\end{figure}

\paragraph*{Solvers and certificates.}
\textsc{kissat}~4.0.4~\cite{Biere2024Kissat} with DRAT logging, proofs
verified by \textsc{drat-trim}~\cite{Wetzler2014DRAT,HeuleHuntWetzler2013};
\textsc{CaDiCaL}~\cite{Biere2024CaDiCaL} and \textsc{Glucose}~4.2 through
PySAT~\cite{Ignatiev2018PySAT,AudemardSimon2009}; every satisfying
assignment decoded and verified by $\F$ simulation; every capped run
recorded as capped. Table~\ref{tab:certs-full} lists all eighteen certified
instances; Fig.~\ref{fig:certs} shows solve times and proof sizes. All
runs were single-threaded on one workstation (Intel Core Ultra 9 275HX,
$24$ logical cores, $15.6$~GiB, Linux/WSL2); wall and CPU time agree
within a few per cent.

\begin{table}[!htbp]
\caption{\textbf{DRAT certificates.} Instance, whether the
symmetry-breaking clause was present, \textsc{kissat} solve time, and
proof size; all certificates were verified by \textsc{drat-trim}.}
\label{tab:certs-full}
\centering
\small
\renewcommand{\arraystretch}{1.12}

\begin{ruledtabular}
\begin{tabular}{p{0.65\textwidth}ccc}
instance & clause & time (s) & proof \\
\colrule
$[[72,12,6]]$      & no  & 4.0 & 24 MB \\
$[[90,8,10]]$      & no  & 4.2 & 41 MB \\
$[[108,8,10]]$     & no  & 4.8 & 52 MB \\
$[[144,12,12]]$    & no  & 3.6 & 35 MB \\
$[[288,12,18]]$    & no  & 1.4 & 18 MB \\
$[[360,12,\le24]]$ & no  & 6.3 & 68 MB \\
$[[756,16,\le34]]$ & no  & 3.3 & 46 MB \\
D, $\mathbb{Z}_{10}$, $A=\{0,3,6\}$, $B=\{0,1,9\}$
  & yes & 4.1 & 13 MB \\
D, $\mathbb{Z}_{24}$, $A=\{0,1,8\}$, $B=\{0,5,14\}$
  & yes & 3.6 & 18 MB \\
$\mathbb{Z}_6$, $A=B=\{0,2,4\}$
  & yes & 558 & 2.7 GB \\
6 class-C1, odd order, not obstructed
  & yes & 16--163 & 123--911 MB \\
C1 odd, $\mathbb{Z}_{45}$, $A=\{0,12,33\}$, $B=\{0,24,39\}$
  & yes & 967 & 2.4 GB \\
C1 odd, $\mathbb{Z}_{15}\times\mathbb{Z}_3$, $A=\{(0,0),(8,2),(13,1)\}$,
$B=\{(0,0),(1,1),(14,2)\}$
  & yes & 600 & 2.8 GB \\
\end{tabular}
\end{ruledtabular}
\end{table}

\FloatBarrier

\paragraph*{Cross-check.}
The $17$ certificate CNFs of the first ten instances (seven codes of
Ref.~\cite{Bravyi2024} in both encodings, the two class-D instances, the
$\mathbb{Z}_6$ instance) were re-decided by \textsc{CaDiCaL} and
\textsc{Glucose} with a $300$-second cap: $16$ are unsatisfiable under
both; the $\mathbb{Z}_6$ file is decided by \textsc{CaDiCaL} in $233$
seconds and not by \textsc{Glucose} within the cap.

\paragraph*{Model ablation.}
The parity criterion~\eqref{eq:parity} and direct $\F$ simulation agree on
$300$ random valid depth-$\Delta$ layerings of each of $[[72,12,6]]$ and
\texttt{oe23}, $600$ in all; the identity
$\sum_vN_v=\frac12\sum_vr_vs_v$ of Theorem~\ref{thm:mod4} holds on $200$
random layerings of each, $400$ in all, proper or not.

\paragraph*{The uniform depth-$7$ schedule.}
The schedule used for the upper bound of Theorem~\ref{thm:main} is
Table~\ref{tab:depth7main} of the main text. Its nine corner orientations are
$o_L=o_R=\mathbb{1}[s\le2]$; it was simulated on the seven codes of
Ref.~\cite{Bravyi2024}, $20$ random distinct-sums instances and $20$
random class-C1/C2 instances, $47$ of $47$ proper. It is not equivalent to
the layer order of Ref.~\cite{Bravyi2024} under label permutations, layer
reversal, $X$/$Z$ duality and block swap (Methods): the published cycle has
$o_L=o_R=\mathbb{1}[s=2]$, and the $8496$ translation-invariant proper
depth-$7$ schedules fall into $41$ orbits under that group, the two
schedules lying in different orbits.

\paragraph*{Checks of the proof steps.}
Four codes with one even weight admit proper depth-$\Delta$ schedules
(Table~\ref{tab:lemmas}); all such schedules were enumerated up to $200$
per instance, $417$ in all ($5$ translation-invariant, $72$ twisted, $340$
without a proper period). On every schedule: $a+b=\Delta+1$; every
$X$-check with an $L$-edge in layer $a$ has one in layer $b$ and
conversely; $\Phi$ preserves orientation; $F_L=F_R$; (T1) and (T2) in
shell form; and $F_L+F_R$ even. The conditional statement
$F_L+F_R=\wA\wB$ when no layer is mixed applies to $140$ of the $417$ and
holds on all of them. \texttt{oe12} and \texttt{oe21} are not dual: the
dual of \texttt{oe12} has $A=\{0,4\}$, not equivalent to $\{0,3\}$ under
the units of $\mathbb{Z}_8$.

\begin{table}[!htbp]
\caption{\textbf{Codes used to check the proof steps.}}
\label{tab:lemmas}
\begin{ruledtabular}
\begin{tabular}{lcccccc}
instance & $G$ & $A$ & $B$ & weights & $\Delta$ & schedules \\
\colrule
\texttt{oe12} & $\mathbb{Z}_8$    & $\{0\}$       & $\{0,4\}$   & $(1,2)$ & 3 & 16  \\
\texttt{oe21} & $\mathbb{Z}_8$    & $\{0,3\}$     & $\{0\}$     & $(2,1)$ & 3 & 1   \\
\texttt{oe23} & $\mathbb{Z}_{12}$ & $\{0,1\}$     & $\{0,4,8\}$ & $(2,3)$ & 5 & 200 \\
\texttt{oe43} & $\mathbb{Z}_{16}$ & $\{0,1,2,3\}$ & $\{0,4,8\}$ & $(4,3)$ & 7 & 200 \\
\colrule
total & & & & & & 417 \\
\end{tabular}
\end{ruledtabular}
\end{table}

\FloatBarrier

\paragraph*{Coset identity.}
$\sum_{c\in C}(F_L(c)+F_R(c))=|M|\wA\wB$ was evaluated on every twisted
object in the repository: $12{,}931$ cosets, $295$ from the $289$ twisted
witnesses stored with their homomorphism and modulus, and $12{,}636$ from
the $11{,}412$ admissible template keys ($10{,}404$ with a single coset,
$1{,}008$ with several), with no failure and $|M|$ even throughout. On the
$27$ satisfiable class-C2 instances, twisted schedules were enumerated for
every $q\mid N$ under a $20$-second cap: $323$ found, $|M|$ taking eleven
even values between $2$ and $24$, $29$ combinations capped. Both subgroup
instances are unsatisfiable, so beyond classes C1 and C2 the theorem is
supported by its proof only.

\paragraph*{Reproducibility.}
The repository contains the encoder, the instance generator, all
certificates, schedules and witnesses with their homomorphisms, the
templates, the simulation checker, one CSV per table and figure, a
manifest mapping every figure to its columns, and a script that checks
every number of Sections~\ref{sec:six}--\ref{sec:methods} against those
files.

\FloatBarrier

\section{Depth-six codes: search data and explicit schedules}
\label{si:codes}

\paragraph*{Search data.}
Tables~\ref{tab:codes-a}--\ref{tab:codes-c} list, for every group and
every class (D: distinct sums; C1, even: class C1 with
$\operatorname{ord}(\col)$ even; C1, odd: class C1 with
$\operatorname{ord}(\col)$ odd), the Pareto-optimal $(k,d)$ pairs with
$d\ge4$ found by the search of Methods, one representative $(A,B)$ per
pair (elements of $\mathbb{Z}_\ell\times\mathbb{Z}_m$ written $(i,j)$, of
$\mathbb{Z}_N$ written $i$), the collision element, its order and its
$2$-height. A distance without ``$\le$'' is exact (meet-in-the-middle
confirmation); the search is exhaustive over all pairs with $0\in A$,
$0\in B$ for $N\le36$ and samples $4\times10^5$ pairs per group for
$N\in\{45,54,72\}$ (about $21\%$ of the pairs at $N=54$ and $6.5\%$ at
$N=72$), so the fronts at $N\ge45$ are lower bounds on what an exhaustive
search would find. Every class-C1 code of even order in the tables has
$h(\col)\le2$ and is covered by Theorem~\ref{thm:six}; the groups
$\mathbb{Z}_6\times\mathbb{Z}_6$, $\mathbb{Z}_2\times\mathbb{Z}_{18}$,
$\mathbb{Z}_9\times\mathbb{Z}_6$, $\mathbb{Z}_{18}\times\mathbb{Z}_3$ and
$\mathbb{Z}_{54}$ have $2G=4G$, so every even-order element there has
$2$-height $0$.

\FloatBarrier
\begingroup
\small
\setlength{\tabcolsep}{4pt}
\renewcommand{\arraystretch}{1.15}
\setlength{\LTcapwidth}{\textwidth}
\setlength{\LTpre}{10pt}
\setlength{\LTpost}{12pt}
\begin{longtable}{@{}llllllcc@{}}
\caption{\textbf{Pareto fronts, $N\in\{18,24,27\}$.}}
\label{tab:codes-a} \\
\toprule
group & class & $[[n,k,d]]$ & $A$ & $B$ & $\col$ & $\operatorname{ord}(\col)$ & $h(\col)$ \\
\midrule
\endfirsthead
\multicolumn{8}{l}{\tablename~\thetable\ (continued)} \\
\toprule
group & class & $[[n,k,d]]$ & $A$ & $B$ & $\col$ & $\operatorname{ord}(\col)$ & $h(\col)$ \\
\midrule
\endhead
\midrule
\multicolumn{8}{r}{\footnotesize Continued on next page} \\
\endfoot
\bottomrule
\endlastfoot
$\mathbb{Z}_{18}$ & C1, even & $[[36,4,6]]$ & $\{0,10,17\}$ & $\{0,1,14\}$ & 17 & 18 & 0 \\
$\mathbb{Z}_{18}$ & C1, odd & $[[36,4,6]]$ & $\{0,1,2\}$ & $\{0,5,16\}$ & 2 & 9 & -- \\
$\mathbb{Z}_{18}$ & D & $[[36,4,6]]$ & $\{0,8,13\}$ & $\{0,2,4\}$ & -- & -- & -- \\
\addlinespace[3pt]
$\mathbb{Z}_{2}\times\mathbb{Z}_{12}$ & C1, even & $[[48,8,4]]$ & $\{(0,0),(0,1),(1,2)\}$ & $\{(0,0),(0,5),(1,7)\}$ & (1,2) & 6 & 0 \\
$\mathbb{Z}_{2}\times\mathbb{Z}_{12}$ & C1, even & $[[48,4,6]]$ & $\{(0,0),(0,7),(1,2)\}$ & $\{(0,0),(0,7),(0,11)\}$ & (0,7) & 12 & 0 \\
$\mathbb{Z}_{2}\times\mathbb{Z}_{12}$ & C1, odd & $[[48,8,4]]$ & $\{(0,0),(0,8),(1,7)\}$ & $\{(0,0),(0,2),(0,10)\}$ & (0,8) & 3 & -- \\
$\mathbb{Z}_{2}\times\mathbb{Z}_{12}$ & D & $[[48,8,4]]$ & $\{(0,0),(1,1),(1,8)\}$ & $\{(0,0),(0,8),(0,10)\}$ & -- & -- & -- \\
$\mathbb{Z}_{2}\times\mathbb{Z}_{12}$ & D & $[[48,4,6]]$ & $\{(0,0),(0,1),(0,5)\}$ & $\{(0,0),(1,8),(1,10)\}$ & -- & -- & -- \\
\addlinespace[3pt]
$\mathbb{Z}_{4}\times\mathbb{Z}_{6}$ & C1, even & $[[48,8,4]]$ & $\{(0,0),(1,4),(2,5)\}$ & $\{(0,0),(0,5),(3,4)\}$ & (1,1) & 12 & 0 \\
$\mathbb{Z}_{4}\times\mathbb{Z}_{6}$ & C1, even & $[[48,4,6]]$ & $\{(0,0),(0,1),(2,2)\}$ & $\{(0,0),(1,4),(2,2)\}$ & (2,2) & 6 & 1 \\
$\mathbb{Z}_{4}\times\mathbb{Z}_{6}$ & C1, odd & $[[48,8,4]]$ & $\{(0,0),(0,1),(0,2)\}$ & $\{(0,0),(0,4),(2,5)\}$ & (0,2) & 3 & -- \\
$\mathbb{Z}_{4}\times\mathbb{Z}_{6}$ & D & $[[48,8,4]]$ & $\{(0,0),(2,5),(3,4)\}$ & $\{(0,0),(0,4),(2,2)\}$ & -- & -- & -- \\
$\mathbb{Z}_{4}\times\mathbb{Z}_{6}$ & D & $[[48,4,6]]$ & $\{(0,0),(2,1),(3,5)\}$ & $\{(0,0),(1,2),(2,4)\}$ & -- & -- & -- \\
\addlinespace[3pt]
$\mathbb{Z}_{24}$ & C1, even & $[[48,8,4]]$ & $\{0,10,20\}$ & $\{0,20,22\}$ & 20 & 6 & 2 \\
$\mathbb{Z}_{24}$ & C1, even & $[[48,4,8]]$ & $\{0,10,11\}$ & $\{0,2,16\}$ & 10 & 12 & 1 \\
$\mathbb{Z}_{24}$ & C1, odd & $[[48,8,4]]$ & $\{0,2,16\}$ & $\{0,4,8\}$ & 16 & 3 & -- \\
$\mathbb{Z}_{24}$ & C1, odd & $[[48,4,6]]$ & $\{0,1,8\}$ & $\{0,8,19\}$ & 8 & 3 & -- \\
$\mathbb{Z}_{24}$ & D & $[[48,4,8]]$ & $\{0,10,23\}$ & $\{0,17,22\}$ & -- & -- & -- \\
\addlinespace[3pt]
$\mathbb{Z}_{9}\times\mathbb{Z}_{3}$ & C1, odd & $[[54,4,8]]$ & $\{(0,0),(0,1),(7,0)\}$ & $\{(0,0),(4,2),(7,0)\}$ & (7,0) & 9 & -- \\
$\mathbb{Z}_{9}\times\mathbb{Z}_{3}$ & D & $[[54,8,6]]$ & $\{(0,0),(1,0),(8,0)\}$ & $\{(0,0),(0,2),(3,1)\}$ & -- & -- & -- \\
$\mathbb{Z}_{9}\times\mathbb{Z}_{3}$ & D & $[[54,4,8]]$ & $\{(0,0),(3,1),(4,2)\}$ & $\{(0,0),(0,1),(2,2)\}$ & -- & -- & -- \\
\addlinespace[3pt]
$\mathbb{Z}_{27}$ & C1, odd & $[[54,4,8]]$ & $\{0,4,26\}$ & $\{0,16,26\}$ & 26 & 27 & -- \\
$\mathbb{Z}_{27}$ & D & $[[54,4,8]]$ & $\{0,19,23\}$ & $\{0,5,7\}$ & -- & -- & -- \\
\end{longtable}
\endgroup
\FloatBarrier
\begingroup
\small
\setlength{\tabcolsep}{4pt}
\renewcommand{\arraystretch}{1.15}
\setlength{\LTcapwidth}{\textwidth}
\setlength{\LTpre}{10pt}
\setlength{\LTpost}{12pt}
\begin{longtable}{@{}llllllcc@{}}
\caption{\textbf{Pareto fronts, $N=36$ (exhaustive).}}
\label{tab:codes-b} \\
\toprule
group & class & $[[n,k,d]]$ & $A$ & $B$ & $\col$ & $\operatorname{ord}(\col)$ & $h(\col)$ \\
\midrule
\endfirsthead
\multicolumn{8}{l}{\tablename~\thetable\ (continued)} \\
\toprule
group & class & $[[n,k,d]]$ & $A$ & $B$ & $\col$ & $\operatorname{ord}(\col)$ & $h(\col)$ \\
\midrule
\endhead
\midrule
\multicolumn{8}{r}{\footnotesize Continued on next page} \\
\endfoot
\bottomrule
\endlastfoot
$\mathbb{Z}_{2}\times\mathbb{Z}_{18}$ & C1, even & $[[72,8,6]]$ & $\{(0,0),(0,2),(0,7)\}$ & $\{(0,0),(0,1),(0,14)\}$ & (0,5) & 18 & 0 \\
$\mathbb{Z}_{2}\times\mathbb{Z}_{18}$ & C1, even & $[[72,4,8]]$ & $\{(0,0),(0,13),(1,5)\}$ & $\{(0,0),(0,13),(1,2)\}$ & (0,13) & 18 & 0 \\
$\mathbb{Z}_{2}\times\mathbb{Z}_{18}$ & C1, odd & $[[72,12,4]]$ & $\{(0,0),(1,6),(1,12)\}$ & $\{(0,0),(0,3),(0,6)\}$ & (0,6) & 3 & -- \\
$\mathbb{Z}_{2}\times\mathbb{Z}_{18}$ & C1, odd & $[[72,8,6]]$ & $\{(0,0),(0,5),(0,10)\}$ & $\{(0,0),(0,1),(0,8)\}$ & (0,10) & 9 & -- \\
$\mathbb{Z}_{2}\times\mathbb{Z}_{18}$ & C1, odd & $[[72,4,8]]$ & $\{(0,0),(0,10),(0,17)\}$ & $\{(0,0),(0,10),(1,5)\}$ & (0,10) & 9 & -- \\
$\mathbb{Z}_{2}\times\mathbb{Z}_{18}$ & D & $[[72,8,6]]$ & $\{(0,0),(0,11),(1,4)\}$ & $\{(0,0),(0,17),(1,16)\}$ & -- & -- & -- \\
$\mathbb{Z}_{2}\times\mathbb{Z}_{18}$ & D & $[[72,4,8]]$ & $\{(0,0),(1,14),(1,16)\}$ & $\{(0,0),(1,1),(1,11)\}$ & -- & -- & -- \\
\addlinespace[3pt]
$\mathbb{Z}_{3}\times\mathbb{Z}_{12}$ & C1, even & $[[72,12,4]]$ & $\{(0,0),(0,5),(0,7)\}$ & $\{(0,0),(0,10),(0,11)\}$ & (0,2) & 6 & 1 \\
$\mathbb{Z}_{3}\times\mathbb{Z}_{12}$ & C1, even & $[[72,8,6]]$ & $\{(0,0),(0,2),(2,6)\}$ & $\{(0,0),(2,0),(2,10)\}$ & (0,2) & 6 & 1 \\
$\mathbb{Z}_{3}\times\mathbb{Z}_{12}$ & C1, even & $[[72,4,8]]$ & $\{(0,0),(1,3),(2,0)\}$ & $\{(0,0),(1,10),(2,1)\}$ & (1,3) & 12 & 0 \\
$\mathbb{Z}_{3}\times\mathbb{Z}_{12}$ & C1, odd & $[[72,12,4]]$ & $\{(0,0),(1,6),(2,6)\}$ & $\{(0,0),(1,9),(2,9)\}$ & (1,0) & 3 & -- \\
$\mathbb{Z}_{3}\times\mathbb{Z}_{12}$ & C1, odd & $[[72,4,6]]$ & $\{(0,0),(0,8),(2,1)\}$ & $\{(0,0),(2,7),(2,11)\}$ & (0,8) & 3 & -- \\
$\mathbb{Z}_{3}\times\mathbb{Z}_{12}$ & D & $[[72,16,4]]$ & $\{(0,0),(1,10),(2,2)\}$ & $\{(0,0),(0,8),(0,10)\}$ & -- & -- & -- \\
$\mathbb{Z}_{3}\times\mathbb{Z}_{12}$ & D & $[[72,8,8]]$ & $\{(0,0),(0,1),(0,8)\}$ & $\{(0,0),(1,8),(2,7)\}$ & -- & -- & -- \\
\addlinespace[3pt]
$\mathbb{Z}_{6}\times\mathbb{Z}_{6}$ & C1, even & $[[72,8,6]]$ & $\{(0,0),(4,3),(5,3)\}$ & $\{(0,0),(1,5),(4,3)\}$ & (4,3) & 6 & 0 \\
$\mathbb{Z}_{6}\times\mathbb{Z}_{6}$ & C1, even & $[[72,4,8]]$ & $\{(0,0),(0,5),(4,4)\}$ & $\{(0,0),(1,1),(1,2)\}$ & (0,5) & 6 & 0 \\
$\mathbb{Z}_{6}\times\mathbb{Z}_{6}$ & C1, odd & $[[72,12,4]]$ & $\{(0,0),(3,2),(3,4)\}$ & $\{(0,0),(3,1),(3,5)\}$ & (0,2) & 3 & -- \\
$\mathbb{Z}_{6}\times\mathbb{Z}_{6}$ & C1, odd & $[[72,4,8]]$ & $\{(0,0),(4,3),(4,4)\}$ & $\{(0,0),(1,3),(3,5)\}$ & (4,4) & 3 & -- \\
$\mathbb{Z}_{6}\times\mathbb{Z}_{6}$ & D & $[[72,16,4]]$ & $\{(0,0),(0,2),(3,4)\}$ & $\{(0,0),(1,0),(5,0)\}$ & -- & -- & -- \\
$\mathbb{Z}_{6}\times\mathbb{Z}_{6}$ & D & $[[72,12,6]]$ & $\{(0,0),(0,1),(3,5)\}$ & $\{(0,0),(1,0),(2,3)\}$ & -- & -- & -- \\
$\mathbb{Z}_{6}\times\mathbb{Z}_{6}$ & D & $[[72,4,8]]$ & $\{(0,0),(2,0),(4,5)\}$ & $\{(0,0),(1,3),(2,4)\}$ & -- & -- & -- \\
\addlinespace[3pt]
$\mathbb{Z}_{36}$ & C1, even & $[[72,12,4]]$ & $\{0,15,21\}$ & $\{0,3,33\}$ & 6 & 6 & 1 \\
$\mathbb{Z}_{36}$ & C1, even & $[[72,8,6]]$ & $\{0,2,28\}$ & $\{0,10,32\}$ & 26 & 18 & 1 \\
$\mathbb{Z}_{36}$ & C1, even & $[[72,4,8]]$ & $\{0,22,23\}$ & $\{0,13,29\}$ & 23 & 36 & 0 \\
$\mathbb{Z}_{36}$ & C1, odd & $[[72,12,4]]$ & $\{0,12,15\}$ & $\{0,6,12\}$ & 12 & 3 & -- \\
$\mathbb{Z}_{36}$ & C1, odd & $[[72,8,6]]$ & $\{0,8,22\}$ & $\{0,8,10\}$ & 8 & 9 & -- \\
$\mathbb{Z}_{36}$ & C1, odd & $[[72,4,8]]$ & $\{0,23,31\}$ & $\{0,7,8\}$ & 8 & 9 & -- \\
$\mathbb{Z}_{36}$ & D & $[[72,8,6]]$ & $\{0,16,20\}$ & $\{0,14,28\}$ & -- & -- & -- \\
$\mathbb{Z}_{36}$ & D & $[[72,4,\le10]]$ & $\{0,20,25\}$ & $\{0,1,29\}$ & -- & -- & -- \\
\end{longtable}
\endgroup
\FloatBarrier
\begingroup
\small
\setlength{\tabcolsep}{4pt}
\renewcommand{\arraystretch}{1.15}
\setlength{\LTcapwidth}{\textwidth}
\setlength{\LTpre}{10pt}
\setlength{\LTpost}{12pt}
\begin{longtable}{@{}llllllcc@{}}
\caption{\textbf{Pareto fronts, $N\in\{45,54,72\}$ (sampled).}}
\label{tab:codes-c} \\
\toprule
group & class & $[[n,k,d]]$ & $A$ & $B$ & $\col$ & $\operatorname{ord}(\col)$ & $h(\col)$ \\
\midrule
\endfirsthead
\multicolumn{8}{l}{\tablename~\thetable\ (continued)} \\
\toprule
group & class & $[[n,k,d]]$ & $A$ & $B$ & $\col$ & $\operatorname{ord}(\col)$ & $h(\col)$ \\
\midrule
\endhead
\midrule
\multicolumn{8}{r}{\footnotesize Continued on next page} \\
\endfoot
\bottomrule
\endlastfoot
$\mathbb{Z}_{15}\times\mathbb{Z}_{3}$ & C1, odd & $[[90,12,6]]$ & $\{(0,0),(8,2),(13,1)\}$ & $\{(0,0),(1,1),(14,2)\}$ & (13,1) & 15 & -- \\
$\mathbb{Z}_{15}\times\mathbb{Z}_{3}$ & C1, odd & $[[90,4,\le10]]$ & $\{(0,0),(1,1),(11,2)\}$ & $\{(0,0),(6,1),(10,2)\}$ & (11,2) & 15 & -- \\
$\mathbb{Z}_{15}\times\mathbb{Z}_{3}$ & D & $[[90,24,4]]$ & $\{(0,0),(7,0),(9,0)\}$ & $\{(0,0),(11,0),(12,0)\}$ & -- & -- & -- \\
$\mathbb{Z}_{15}\times\mathbb{Z}_{3}$ & D & $[[90,12,6]]$ & $\{(0,0),(2,0),(13,1)\}$ & $\{(0,0),(7,1),(8,0)\}$ & -- & -- & -- \\
$\mathbb{Z}_{15}\times\mathbb{Z}_{3}$ & D & $[[90,8,\le10]]$ & $\{(0,0),(10,0),(14,0)\}$ & $\{(0,0),(8,1),(13,2)\}$ & -- & -- & -- \\
\addlinespace[3pt]
$\mathbb{Z}_{45}$ & C1, odd & $[[90,12,6]]$ & $\{0,12,33\}$ & $\{0,24,39\}$ & 21 & 15 & -- \\
$\mathbb{Z}_{45}$ & C1, odd & $[[90,4,\le10]]$ & $\{0,25,41\}$ & $\{0,25,38\}$ & 25 & 9 & -- \\
$\mathbb{Z}_{45}$ & D & $[[90,24,4]]$ & $\{0,6,27\}$ & $\{0,9,12\}$ & -- & -- & -- \\
$\mathbb{Z}_{45}$ & D & $[[90,8,\le8]]$ & $\{0,11,29\}$ & $\{0,2,39\}$ & -- & -- & -- \\
$\mathbb{Z}_{45}$ & D & $[[90,4,\le12]]$ & $\{0,14,19\}$ & $\{0,1,17\}$ & -- & -- & -- \\
\addlinespace[3pt]
$\mathbb{Z}_{9}\times\mathbb{Z}_{6}$ & C1, even & $[[108,12,6]]$ & $\{(0,0),(4,4),(5,5)\}$ & $\{(0,0),(7,1),(8,2)\}$ & (1,1) & 18 & 0 \\
$\mathbb{Z}_{9}\times\mathbb{Z}_{6}$ & C1, even & $[[108,4,\le12]]$ & $\{(0,0),(5,4),(7,1)\}$ & $\{(0,0),(2,0),(7,3)\}$ & (2,3) & 18 & 0 \\
$\mathbb{Z}_{9}\times\mathbb{Z}_{6}$ & C1, odd & $[[108,12,6]]$ & $\{(0,0),(1,1),(2,2)\}$ & $\{(0,0),(2,5),(7,4)\}$ & (2,2) & 9 & -- \\
$\mathbb{Z}_{9}\times\mathbb{Z}_{6}$ & C1, odd & $[[108,8,\le8]]$ & $\{(0,0),(1,2),(4,0)\}$ & $\{(0,0),(0,2),(4,2)\}$ & (4,0) & 9 & -- \\
$\mathbb{Z}_{9}\times\mathbb{Z}_{6}$ & C1, odd & $[[108,4,\le12]]$ & $\{(0,0),(4,5),(5,4)\}$ & $\{(0,0),(1,0),(5,4)\}$ & (5,4) & 9 & -- \\
$\mathbb{Z}_{9}\times\mathbb{Z}_{6}$ & D & $[[108,24,4]]$ & $\{(0,0),(0,1),(0,2)\}$ & $\{(0,0),(3,5),(6,4)\}$ & -- & -- & -- \\
$\mathbb{Z}_{9}\times\mathbb{Z}_{6}$ & D & $[[108,16,6]]$ & $\{(0,0),(6,2),(6,4)\}$ & $\{(0,0),(2,2),(4,4)\}$ & -- & -- & -- \\
$\mathbb{Z}_{9}\times\mathbb{Z}_{6}$ & D & $[[108,8,\le10]]$ & $\{(0,0),(3,1),(3,5)\}$ & $\{(0,0),(2,3),(7,0)\}$ & -- & -- & -- \\
$\mathbb{Z}_{9}\times\mathbb{Z}_{6}$ & D & $[[108,4,\le12]]$ & $\{(0,0),(2,2),(4,3)\}$ & $\{(0,0),(1,4),(5,4)\}$ & -- & -- & -- \\
\addlinespace[3pt]
$\mathbb{Z}_{18}\times\mathbb{Z}_{3}$ & C1, even & $[[108,12,6]]$ & $\{(0,0),(10,1),(17,2)\}$ & $\{(0,0),(11,2),(16,1)\}$ & (7,1) & 18 & 0 \\
$\mathbb{Z}_{18}\times\mathbb{Z}_{3}$ & C1, even & $[[108,4,\le12]]$ & $\{(0,0),(10,2),(11,1)\}$ & $\{(0,0),(4,1),(17,1)\}$ & (1,2) & 18 & 0 \\
$\mathbb{Z}_{18}\times\mathbb{Z}_{3}$ & C1, odd & $[[108,12,6]]$ & $\{(0,0),(8,0),(13,0)\}$ & $\{(0,0),(7,0),(17,0)\}$ & (8,0) & 9 & -- \\
$\mathbb{Z}_{18}\times\mathbb{Z}_{3}$ & C1, odd & $[[108,8,\le8]]$ & $\{(0,0),(12,2),(16,1)\}$ & $\{(0,0),(0,1),(2,2)\}$ & (16,1) & 9 & -- \\
$\mathbb{Z}_{18}\times\mathbb{Z}_{3}$ & C1, odd & $[[108,4,\le12]]$ & $\{(0,0),(0,2),(16,2)\}$ & $\{(0,0),(13,2),(16,0)\}$ & (16,0) & 9 & -- \\
$\mathbb{Z}_{18}\times\mathbb{Z}_{3}$ & D & $[[108,24,4]]$ & $\{(0,0),(3,2),(6,1)\}$ & $\{(0,0),(3,0),(15,0)\}$ & -- & -- & -- \\
$\mathbb{Z}_{18}\times\mathbb{Z}_{3}$ & D & $[[108,16,6]]$ & $\{(0,0),(6,1),(6,2)\}$ & $\{(0,0),(8,1),(16,2)\}$ & -- & -- & -- \\
$\mathbb{Z}_{18}\times\mathbb{Z}_{3}$ & D & $[[108,8,\le10]]$ & $\{(0,0),(7,2),(17,1)\}$ & $\{(0,0),(0,1),(15,2)\}$ & -- & -- & -- \\
$\mathbb{Z}_{18}\times\mathbb{Z}_{3}$ & D & $[[108,4,\le12]]$ & $\{(0,0),(1,1),(17,1)\}$ & $\{(0,0),(4,1),(11,2)\}$ & -- & -- & -- \\
\addlinespace[3pt]
$\mathbb{Z}_{54}$ & C1, even & $[[108,12,6]]$ & $\{0,6,39\}$ & $\{0,39,51\}$ & 39 & 18 & 0 \\
$\mathbb{Z}_{54}$ & C1, even & $[[108,4,\le12]]$ & $\{0,1,29\}$ & $\{0,37,38\}$ & 1 & 54 & 0 \\
$\mathbb{Z}_{54}$ & C1, odd & $[[108,12,6]]$ & $\{0,3,33\}$ & $\{0,30,42\}$ & 30 & 9 & -- \\
$\mathbb{Z}_{54}$ & C1, odd & $[[108,8,\le8]]$ & $\{0,2,22\}$ & $\{0,22,50\}$ & 22 & 27 & -- \\
$\mathbb{Z}_{54}$ & C1, odd & $[[108,4,\le12]]$ & $\{0,19,32\}$ & $\{0,32,40\}$ & 32 & 27 & -- \\
$\mathbb{Z}_{54}$ & D & $[[108,12,6]]$ & $\{0,6,30\}$ & $\{0,21,42\}$ & -- & -- & -- \\
$\mathbb{Z}_{54}$ & D & $[[108,8,\le8]]$ & $\{0,2,4\}$ & $\{0,14,22\}$ & -- & -- & -- \\
$\mathbb{Z}_{54}$ & D & $[[108,4,\le12]]$ & $\{0,34,53\}$ & $\{0,10,38\}$ & -- & -- & -- \\
\addlinespace[3pt]
$\mathbb{Z}_{12}\times\mathbb{Z}_{6}$ & C1, even & $[[144,24,4]]$ & $\{(0,0),(7,0),(11,3)\}$ & $\{(0,0),(7,3),(8,3)\}$ & (4,3) & 6 & 0 \\
$\mathbb{Z}_{12}\times\mathbb{Z}_{6}$ & C1, even & $[[144,16,6]]$ & $\{(0,0),(10,0),(10,4)\}$ & $\{(0,0),(2,0),(8,2)\}$ & (10,0) & 6 & 1 \\
$\mathbb{Z}_{12}\times\mathbb{Z}_{6}$ & C1, even & $[[144,8,\le10]]$ & $\{(0,0),(1,0),(7,5)\}$ & $\{(0,0),(4,3),(9,4)\}$ & (7,5) & 12 & 0 \\
$\mathbb{Z}_{12}\times\mathbb{Z}_{6}$ & C1, even & $[[144,4,\le12]]$ & $\{(0,0),(7,1),(11,2)\}$ & $\{(0,0),(5,3),(7,1)\}$ & (7,1) & 12 & 0 \\
$\mathbb{Z}_{12}\times\mathbb{Z}_{6}$ & C1, odd & $[[144,24,4]]$ & $\{(0,0),(0,2),(6,1)\}$ & $\{(0,0),(0,1),(0,2)\}$ & (0,2) & 3 & -- \\
$\mathbb{Z}_{12}\times\mathbb{Z}_{6}$ & C1, odd & $[[144,8,\le8]]$ & $\{(0,0),(6,2),(10,4)\}$ & $\{(0,0),(0,1),(8,5)\}$ & (4,2) & 3 & -- \\
$\mathbb{Z}_{12}\times\mathbb{Z}_{6}$ & D & $[[144,32,4]]$ & $\{(0,0),(8,2),(10,4)\}$ & $\{(0,0),(2,0),(10,0)\}$ & -- & -- & -- \\
$\mathbb{Z}_{12}\times\mathbb{Z}_{6}$ & D & $[[144,24,6]]$ & $\{(0,0),(2,4),(10,5)\}$ & $\{(0,0),(0,5),(6,1)\}$ & -- & -- & -- \\
$\mathbb{Z}_{12}\times\mathbb{Z}_{6}$ & D & $[[144,16,\le8]]$ & $\{(0,0),(3,2),(3,4)\}$ & $\{(0,0),(4,2),(11,4)\}$ & -- & -- & -- \\
$\mathbb{Z}_{12}\times\mathbb{Z}_{6}$ & D & $[[144,12,\le12]]$ & $\{(0,0),(7,4),(11,5)\}$ & $\{(0,0),(2,1),(7,2)\}$ & -- & -- & -- \\
$\mathbb{Z}_{12}\times\mathbb{Z}_{6}$ & D & $[[144,4,\le14]]$ & $\{(0,0),(5,5),(8,1)\}$ & $\{(0,0),(1,4),(7,5)\}$ & -- & -- & -- \\
\addlinespace[3pt]
$\mathbb{Z}_{72}$ & C1, even & $[[144,24,4]]$ & $\{0,60,66\}$ & $\{0,30,42\}$ & 60 & 6 & 2 \\
$\mathbb{Z}_{72}$ & C1, even & $[[144,16,6]]$ & $\{0,40,68\}$ & $\{0,8,52\}$ & 28 & 18 & 2 \\
$\mathbb{Z}_{72}$ & C1, even & $[[144,12,\le8]]$ & $\{0,3,51\}$ & $\{0,51,66\}$ & 51 & 24 & 0 \\
$\mathbb{Z}_{72}$ & C1, even & $[[144,4,\le14]]$ & $\{0,29,64\}$ & $\{0,17,37\}$ & 35 & 72 & 0 \\
$\mathbb{Z}_{72}$ & C1, odd & $[[144,24,4]]$ & $\{0,42,66\}$ & $\{0,48,60\}$ & 24 & 3 & -- \\
$\mathbb{Z}_{72}$ & C1, odd & $[[144,16,6]]$ & $\{0,4,8\}$ & $\{0,44,64\}$ & 8 & 9 & -- \\
$\mathbb{Z}_{72}$ & C1, odd & $[[144,8,\le8]]$ & $\{0,28,68\}$ & $\{0,14,46\}$ & 40 & 9 & -- \\
$\mathbb{Z}_{72}$ & C1, odd & $[[144,4,\le14]]$ & $\{0,10,50\}$ & $\{0,40,65\}$ & 40 & 9 & -- \\
$\mathbb{Z}_{72}$ & D & $[[144,16,6]]$ & $\{0,32,52\}$ & $\{0,56,64\}$ & -- & -- & -- \\
$\mathbb{Z}_{72}$ & D & $[[144,12,\le8]]$ & $\{0,42,48\}$ & $\{0,39,60\}$ & -- & -- & -- \\
$\mathbb{Z}_{72}$ & D & $[[144,8,\le10]]$ & $\{0,38,58\}$ & $\{0,44,70\}$ & -- & -- & -- \\
$\mathbb{Z}_{72}$ & D & $[[144,4,\le16]]$ & $\{0,25,62\}$ & $\{0,8,34\}$ & -- & -- & -- \\
\end{longtable}
\endgroup

\FloatBarrier

\paragraph*{Explicit depth-six schedules.}
Table~\ref{tab:template72} of the main text gives the class schedule of the
$[[72,8,6]]$ code. Table~\ref{tab:templates-more} gives class schedules,
found by SAT on the class system and verified by $\F$ simulation of the
pullback, for the $[[108,12,6]]$ code over $\mathbb{Z}_{54}$, the
$[[108,4,\le12]]$ code over $\mathbb{Z}_9\times\mathbb{Z}_6$, the
$[[144,12,\le8]]$ code over $\mathbb{Z}_{72}$ and the $[[144,8,\le10]]$
code over $\mathbb{Z}_{12}\times\mathbb{Z}_6$, all with $h(\col)=0$ and
$q=2$, and Table~\ref{tab:template144-16} the $q=4$ class schedule of the
$[[144,16,6]]$ code over $\mathbb{Z}_{12}\times\mathbb{Z}_6$, whose
collision element $(10,0)$ has $2$-height $1$. The $[[144,16,6]]$ code
over $\mathbb{Z}_{72}$ ($A=\{0,40,68\}$, $B=\{0,8,52\}$, $\col=28$ of
$2$-height $2$) was scheduled at $q=8$ in the same way; its key
$(0,0,4;0,0,4)$ generates the subgroup $\{0,4\}$ of $\mathbb{Z}_8$, so its
class system is four copies of a $q=2$ system, as described in Methods.
Information-set decoding with $10^4$ iterations did not lower any of the
distance bounds of these codes.

\begin{table}[!htbp]
\caption{\textbf{Class schedules ($q=2$) of four depth-six codes.} Layers
by class $c$ and edge type ($XL$/$XR$: $X$-check $L$-/$R$-edges, $ZL$/$ZR$:
$Z$-check $L$-/$R$-edges), labels $1,2,3$ in order. Collisions are given
as label pairs $(s,t)$; $\varepsilon$ reduces the last coordinate modulo
$2$ ($g\bmod2$ on cyclic groups).}
\label{tab:templates-more}
\begin{ruledtabular}
\begin{tabular}{lcccc}
code, $c$ & $XL$ & $XR$ & $ZL$ & $ZR$ \\
\colrule
\multicolumn{5}{l}{$[[108,12,6]]$: $\mathbb{Z}_{54}$, $A=\{0,6,39\}$, $B=\{0,39,51\}$,}\\
\multicolumn{5}{l}{\quad collision $(1,2),(3,1)$, key $(0,0,1;0,1,1)$}\\
$c=0$ & $(3,4,1)$ & $(6,2,5)$ & $(1,4,3)$ & $(2,5,6)$ \\
$c=1$ & $(6,2,5)$ & $(3,1,4)$ & $(5,6,2)$ & $(1,4,3)$ \\
\colrule
\multicolumn{5}{l}{$[[108,4,\le12]]$: $\mathbb{Z}_9\times\mathbb{Z}_6$, $A=\{(0,0),(5,4),(7,1)\}$,}\\
\multicolumn{5}{l}{\quad $B=\{(0,0),(2,0),(7,3)\}$, collision $(2,1),(3,3)$, key $(0,0,1;0,0,1)$}\\
$c=0$ & $(4,2,5)$ & $(3,6,1)$ & $(6,1,4)$ & $(5,2,3)$ \\
$c=1$ & $(1,6,3)$ & $(2,5,4)$ & $(3,2,5)$ & $(6,4,1)$ \\
\colrule
\multicolumn{5}{l}{$[[144,12,\le8]]$: $\mathbb{Z}_{72}$, $A=\{0,3,51\}$, $B=\{0,51,66\}$,}\\
\multicolumn{5}{l}{\quad collision $(1,2),(3,1)$, key $(0,1,1;0,1,0)$}\\
$c=0$ & $(4,2,5)$ & $(1,3,6)$ & $(5,3,2)$ & $(4,6,1)$ \\
$c=1$ & $(6,1,3)$ & $(5,2,4)$ & $(4,6,1)$ & $(2,3,5)$ \\
\colrule
\multicolumn{5}{l}{$[[144,8,\le10]]$: $\mathbb{Z}_{12}\times\mathbb{Z}_6$, $A=\{(0,0),(1,0),(7,5)\}$,}\\
\multicolumn{5}{l}{\quad $B=\{(0,0),(4,3),(9,4)\}$, collision $(1,2),(3,3)$, key $(0,0,1;0,1,0)$}\\
$c=0$ & $(4,2,5)$ & $(6,3,1)$ & $(1,4,6)$ & $(3,5,2)$ \\
$c=1$ & $(6,1,3)$ & $(5,2,4)$ & $(2,5,3)$ & $(1,6,4)$ \\
\end{tabular}
\end{ruledtabular}
\end{table}

\begin{table}[!htbp]
\caption{\textbf{Class schedule ($q=4$) of the $[[144,16,6]]$ code.}
$G=\mathbb{Z}_{12}\times\mathbb{Z}_6$, $A=\{(0,0),(10,0),(10,4)\}$,
$B=\{(0,0),(2,0),(8,2)\}$, collision $(1,1),(2,2)$, $\col=(10,0)$,
$\varepsilon(i,j)=i\bmod4$, key $(0,2,2;0,2,0)$, $\colbar=2$.}
\label{tab:template144-16}
\begin{ruledtabular}
\begin{tabular}{lcccc}
$c$ & $XL$ & $XR$ & $ZL$ & $ZR$ \\
\colrule
0 & $(4,1,6)$ & $(2,5,3)$ & $(1,4,6)$ & $(5,2,3)$ \\
1 & $(4,5,2)$ & $(6,1,3)$ & $(5,3,2)$ & $(1,4,6)$ \\
2 & $(2,5,3)$ & $(6,4,1)$ & $(5,2,3)$ & $(4,6,1)$ \\
3 & $(6,3,1)$ & $(2,4,5)$ & $(1,6,4)$ & $(3,2,5)$ \\
\end{tabular}
\end{ruledtabular}
\end{table}

\FloatBarrier

\paragraph*{Odd-order instances decided directly.}
Depth $6$ was decided by SAT for the best odd-order class-C1 codes of
Tables~\ref{tab:codes-b} and~\ref{tab:codes-c}: the $[[72,8,6]]$ code over
$\mathbb{Z}_{36}$ ($A=\{0,8,22\}$, $B=\{0,8,10\}$, $\col=8$ of order $9$)
is satisfiable ($60$~s), as is the $[[72,12,4]]$ code over
$\mathbb{Z}_6\times\mathbb{Z}_6$ ($A=\{(0,0),(3,2),(3,4)\}$,
$B=\{(0,0),(3,1),(3,5)\}$, $\col=(0,2)$ of order $3$; $7$~s), both
schedules verified by $\F$ simulation; the $[[90,12,6]]$ code over
$\mathbb{Z}_{15}\times\mathbb{Z}_3$ ($A=\{(0,0),(8,2),(13,1)\}$,
$B=\{(0,0),(1,1),(14,2)\}$, $\col=(13,1)$ of order $15$) is unsatisfiable:
first by \textsc{CaDiCaL} in $523$~s without proof logging, then re-decided
by \textsc{kissat} with DRAT logging in $600$~s, giving a $2.8$~GB proof
verified by \textsc{drat-trim}. The $[[90,12,6]]$ code over
$\mathbb{Z}_{45}$ ($A=\{0,12,33\}$, $B=\{0,24,39\}$, $\col=21$ of order
$15$), whose earlier PySAT run crashed and which was therefore left
undecided, is also unsatisfiable: \textsc{kissat} refutes it in $967$~s and
\textsc{drat-trim} verifies the $2.4$~GB proof. Both certificates are in
Table~\ref{tab:certs-full} and Fig.~\ref{fig:certs}.

\FloatBarrier

\section{A further application: Stab-QRAM}
\label{si:qram}

The compiler of Section~S1 applies to any circuit whose CNOTs commute. In
this section we use it to build a quantum random access memory (QRAM) for
affine Boolean data, which we call Stab-QRAM. We describe the memory and
its resources, its relation to the syndrome-extraction circuits of the main
text, and possible hardware implementations. Here $n$ and $m$ are the
numbers of address and output qubits, $j$ and $k$ label these qubits, and
$A$ and $\mathbf{b}$ are the matrix and offset of the stored table; these
symbols do not have their main-text meanings. We write $\mathrm{wt}(\cdot)$
for the number of nonzero entries.

\paragraph*{The memory.}
A QRAM with $2^n$ cells of $m$ bits stores a table $f:\F^n\to\F^m$ and
answers a superposition of addresses
coherently~\cite{Giovannetti2008,Giovannetti2008PRA,Jaques2023},
\begin{equation}
U_f\sum_{\mathbf{x}}\alpha_{\mathbf{x}}|\mathbf{x}\rangle|\mathbf{d}\rangle
=\sum_{\mathbf{x}}\alpha_{\mathbf{x}}|\mathbf{x}\rangle
|\mathbf{d}+f(\mathbf{x})\rangle ,
\label{eq:qram}
\end{equation}
where $\mathbf{x}$ is the address register and $\mathbf{d}$ the output
register. General-purpose designs can store any table, but their resources
grow linearly with the number of cells: the bucket-brigade architecture
uses $O(2^n)$ quantum routers~\cite{Giovannetti2008PRA}, and table lookup
by unary iteration uses $O(2^n)$ $T$ gates~\cite{Babbush2018}. By the fact
recalled in Section~S1, a Clifford circuit can implement Eq.~\eqref{eq:qram} only if
$f$ is affine, $f(\mathbf{x})=A\mathbf{x}+\mathbf{b}$ with
$A\in\F^{m\times n}$ and $\mathbf{b}\in\F^m$. There are $2^{m(n+1)}$ such
tables out of $2^{m2^n}$. Examples are the state update
$\mathbf{x}_{t+1}=A\mathbf{x}_t$ of a linear feedback shift
register~\cite{Golomb1967}, or its affine version with an added
$\mathbf{b}$, and the syndrome map of a classical linear code.

\begin{figure}[!htbp]
\centering
\includegraphics[width=\SIwidefigurewidth]{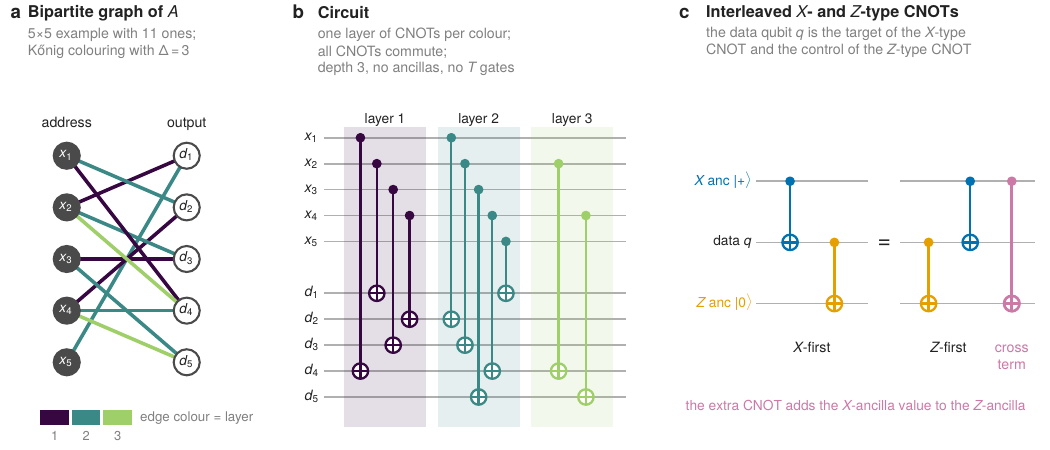}
\caption{\textbf{Stab-QRAM and syndrome extraction.} \textbf{(a)}~Bipartite
graph of a $5\times5$ matrix $A$, with address qubits $x_k$, output qubits
$d_j$ and one edge for each $A_{j,k}=1$, coloured by a K\H{o}nig edge
colouring with $\Delta=3$ colours. \textbf{(b)}~The corresponding circuit,
with one layer of CNOTs per colour. No qubit is both a control and a
target, so all gates commute, and the circuit maps
$|\mathbf{x}\rangle|\mathbf{d}\rangle$ to
$|\mathbf{x}\rangle|\mathbf{d}+A\mathbf{x}\rangle$ in depth $3$ (checked by
$\F$ simulation). $X$ gates for $\mathbf{b}$ would be placed in idle slots
of the output qubits or in one more layer. \textbf{(c)}~In syndrome
extraction a data qubit $q$ of the code is the target of an $X$-type CNOT
and the control of a $Z$-type CNOT. Applying the $X$-type CNOT first adds a
CNOT from the $X$-ancilla to the $Z$-ancilla, which contributes to the
cross term $M_\pi$ of Eq.~\eqref{eq:se-map}. The layer colours are taken
from the scale of Fig.~\ref{fig:setup}, from dark (early) to light
(late).}
\label{fig:qram}
\end{figure}

\paragraph*{Construction.}
For affine $f$ the lookup is the circuit $C_f$ of Eq.~\eqref{eq:Cf}, with one
$\mathrm{CNOT}(x_k,d_j)$ for every $A_{j,k}=1$ and one $X$ gate on $d_j$
for every $b_j=1$ (Fig.~\ref{fig:qram}a,b). It acts on $n+m$ qubits, uses
no ancillas and no $T$ gates, and contains $\mathrm{wt}(A)\le mn$ CNOTs and
$\mathrm{wt}(\mathbf{b})$ $X$ gates. Every CNOT is controlled by an address
qubit and targets an output qubit. Two CNOTs commute unless the target of
one is the control of the other, and an $X$ gate on an output qubit
commutes with every CNOT, because no output qubit is used as a control. All
gates of $C_f$ therefore commute, and compiling the circuit only requires a
schedule. By Eq.~\eqref{eq:Dstar}, the minimum depth among
schedules of these gates is $D^\star=\Delta(G'_A)$, and a K\H{o}nig
colouring attains it in time $O(|E'|\log D^\star)$, where $E'$ is the edge
set of $G'_A$. Since a column of $A$ has at most $m$ ones and a row at most
$n$, $D^\star\le\max(m,n+1)$. If the $X$ gates are instead applied in a
separate last layer, the depth is $\Delta+1$, with $\Delta$ the largest row
or column weight of $A$. The depth depends on the degrees of $G'_A$ and not
on the rank of $A$, so matrices with small row and column weights give
shallow circuits. Loading a new table changes only which CNOTs are applied.

\paragraph*{Relation to syndrome extraction.}
Without its cross term, Eq.~\eqref{eq:se-map} consists of two such lookups
with $\mathbf{b}=0$. The $Z$-half, $\mathbf{z}\mapsto\mathbf{z}+H_Z\mathbf{x}$,
is the lookup for $A=H_Z$, with the data register $\mathbf{x}$ of the code
as address register and the $Z$-ancillas $\mathbf{z}$ as output register.
The $X$-half, $\mathbf{x}\mapsto\mathbf{x}+H_X^{\mathsf T}\mathbf{y}$, is the
lookup for $A=H_X^{\mathsf T}$, with the $X$-ancillas $\mathbf{y}$ as
address register and the data register as output register. The gates
within each half commute, and Section~S1 schedules each half exactly. For
the BB codes every check has weight six and every data qubit lies in three
checks of each type, so each half needs six layers, and the two halves
applied one after the other need twelve. Interleaving the halves is an edge
colouring of the Tanner graph, which again has degree six. A data qubit of
the code is then the target of its $X$-type CNOTs and the control of its
$Z$-type CNOTs, and these gates do not commute. An $X$-first ordering
equals the $Z$-first ordering followed by a CNOT from the $X$-ancilla to
the $Z$-ancilla (Fig.~\ref{fig:qram}c), and these additional CNOTs form the
cross term $M_\pi\mathbf{y}$. No such terms occur in the memory, and the
degree bound is always reached for its gates. In syndrome extraction they
must cancel. For two-block codes with weights $\wA=\wB=3$ and distinct sums
this costs one layer (Theorem~\ref{thm:main}), while a single collision whose element has
even order and $2$-height at most $8$ removes the cost
(Theorem~\ref{thm:six}).

\paragraph*{Scope.}
Stab-QRAM stores only affine tables. The correspondence between affine
maps and Clifford circuits is standard~\cite{DehaeneDeMoor2003,
Gottesman1998,Aaronson2004}, and $D^\star$ is optimal only for the gate set
of $C_f$. The lookup is a linear reversible map, and synthesis with CNOTs
inside the registers or with ancillas can give a smaller
depth~\cite{PMH2008,MooreNilsson2001,Jiang2020,MaslovZindorf2022}, as noted
in Section~S1. Relative to the number of cells the resources are
logarithmic when $m=O(n)$, with $n+m$ qubits and depth at most
$\max(m,n+1)$. This reflects the size of the description
$(A,\mathbf{b})$, which has $m(n+1)$ bits, and is not an advantage over
general-purpose QRAM, which stores arbitrary tables. For the same reason,
questions about the table alone can be answered classically by linear
algebra over $\F$, and the memory is useful only as a $T$-free subroutine
in algorithms that also contain non-affine steps. We do not develop such
applications here. Non-affine tables would require non-Clifford gates. On
hardware with limited connectivity the CNOTs between the two registers must
be routed; provable depth bounds for CNOT circuits on a line and on a grid
are given in Refs.~\cite{Kutin2007,GoubaultMartiel2023}.

\paragraph*{Possible hardware implementations.}
Stab-QRAM is practical on current hardware because of its specialised
design; the points below concern the hardware requirements for the affine
tables it stores. Unlike a general-purpose QRAM, it relies only on CNOT and
$X$ gates, with specific connectivity requirements for the CNOTs. This
allows implementations on widely used platforms, including trapped
ions~\cite{Bruzewicz2019}, Rydberg atom arrays~\cite{Bluvstein2024},
photonics~\cite{ZhangS2024}, and superconducting qubits coupled to
microwave or mechanical
modes~\cite{WangZ2025,LiZ2025,Miao2025,Shen2025,ZhangS2025,Hann2019,Hann2021}.
For superconducting circuits in particular, Stab-QRAM does not need
controlled-iSWAP gates, which require native support on specially
engineered hardware~\cite{LiZ2025} or quantum
routers~\cite{Miao2025,Shen2025,ZhangS2025}; a general-purpose processor
with optimised CNOT gates can run it. It can also benefit from recent
superconducting architectures with enhanced connectivity, either through a
shared bus~\cite{ZhangX2023,Song2017} or across separate
modules~\cite{WuX2024,Heya2025,ZhouC2023}. On photonic platforms for
measurement-based quantum computing~\cite{Raussendorf2003,WuB2020}, the
address and output qubits can be encoded as time-bin or dual-rail qubits
in cluster states. The Clifford layers of the lookup are then carried out
as parallel measurements in fixed bases, which implement the CNOT and $X$
gates with Pauli-frame tracking, so no adaptive measurement angles are
needed. The photonic implementation requires only short optical delays,
using fibre loops~\cite{Asavanant2019,Larsen2019} or low-loss on-chip
waveguides~\cite{Jia2025}, and no long-lived quantum storage, while
photonic switching and multiplexing provide the parallelism within a
round. Because the construction has no $T$ gates, it is suitable for
demonstrations on current noisy intermediate-scale quantum
devices~\cite{Preskill2018} and can be scaled up as the hardware improves.

Hardware built for the circuits of the main text has couplers along the
Tanner graph. It can therefore run the lookup for any matrix $A$ whose
support lies in that of $H_Z$, with the data qubits of the code as address
register and the $Z$-ancillas as output register, or in that of
$H_X^{\mathsf T}$, with the $X$-ancillas as address register and the data
qubits as output register.

\FloatBarrier

\end{document}